\documentclass[preprint,3p,authoryear]{elsarticle}

\journal{Icarus}

\usepackage{amssymb}
\usepackage{amsmath}

\usepackage{lineno}

\usepackage{doi}
\usepackage{textcomp}
\usepackage{color}
\usepackage{ulem}
\usepackage{hyperref}
\hypersetup{colorlinks=true, urlcolor=blue}

\begin{document}

\begin{frontmatter}



\title{Resolved Spectrophotometric Properties of the Ceres Surface from Dawn Framing Camera Images\tnoteref{label1}\tnoteref{label2}}
\tnotetext[label1]{\doi{10.1016/j.icarus.2017.01.026}}
\tnotetext[label2]{\copyright 2017. This manuscript version is made available under the CC-BY-NC-ND 4.0 licence:\\ \url{https://creativecommons.org/licenses/by-nc-nd/4.0/}}


\author[DLR]{S.~E.~Schr\"oder\corref{cor1}}
\author[DLR]{S.~Mottola}
\author[DLR]{U.~Carsenty}
\author[INAF]{M.~Ciarniello}
\author[DLR]{R.~Jaumann}
\author[PSI]{J.-Y.~Li}
\author[INAF]{A.~Longobardo}
\author[PSI]{E.~Palmer}
\author[Brown]{C.~Pieters}
\author[DLR]{F.~Preusker}
\author[JPL]{C.~A.~Raymond}
\author[UCLA]{C.~T.~Russell}

\cortext[cor1]{Corresponding author}

\address[DLR]{Deutsches Zentrum f\"ur Luft- und Raumfahrt (DLR), 12489 Berlin, Germany}
\address[INAF]{Istituto di Astrofisica e Planetologia Spaziali, Istituto Nazionale di Astrofisica (INAF), 00133 Rome, Italy}
\address[PSI]{Planetary Science Institute, Tucson, AZ 85719, U.S.A.}
\address[Brown]{Department of Geological Sciences, Brown University, Providence, RI 02912, U.S.A.}
\address[JPL]{Jet Propulsion Laboratory (JPL), California Institute of Technology, Pasadena, CA 91109, U.S.A.}
\address[UCLA]{Institute of Geophysics and Planetary Physics (IGPP), University of California, Los Angeles, CA 90095-1567, U.S.A.}

\begin{abstract}

We present a global spectrophotometric characterization of the Ceres surface using Dawn Framing Camera (FC) images. We identify the photometric model that yields the best results for photometrically correcting images. Corrected FC images acquired on approach to Ceres were assembled into global maps of albedo and color. Generally, albedo and color variations on Ceres are muted. The albedo map is dominated by a large, circular feature in Vendimia Planitia, known from HST images \citep{L06}, and dotted by smaller bright features mostly associated with fresh-looking craters. The dominant color variation over the surface is represented by the presence of ``blue'' material in and around such craters, which has a negative spectral slope over the visible wavelength range when compared to average terrain. We also mapped variations of the phase curve by employing an exponential photometric model, a technique previously applied to asteroid Vesta \citep{S13b}. The surface of Ceres scatters light differently from Vesta in the sense that the ejecta of several fresh-looking craters may be physically smooth rather than rough. High albedo, blue color, and physical smoothness all appear to be indicators of youth. The blue color may result from the desiccation of ejected material that is similar to the phyllosilicates/water ice mixtures in the experiments of \citet{P16}. The physical smoothness of some blue terrains would be consistent with an initially liquid condition, perhaps as a consequence of impact melting of subsurface water ice. We find red terrain (positive spectral slope) near Ernutet crater, where \citet{dS17} detected organic material. The spectrophotometric properties of the large Vendimia Planitia feature suggest it is a palimpsest, consistent with the \citet{Ma16} impact basin hypothesis. The central bright area in Occator crater, Cerealia Facula, is the brightest on Ceres with an average visual normal albedo of about 0.6 at a resolution of 1.3 km per pixel (six times Ceres average). The albedo of fresh, bright material seen inside this area in the highest resolution images (35 m per pixel) is probably around unity. Cerealia Facula has an unusually steep phase function, which may be due to unresolved topography, high surface roughness, or large average particle size. It has a strongly red spectrum whereas the neighboring, less-bright, Vinalia Faculae are neutral in color. We find no evidence for a diurnal ground fog-type haze in Occator as described by \citet{N15}. We can neither reproduce their findings using the same images, nor confirm them using higher resolution images. FC images have not yet offered direct evidence for present sublimation in Occator.

\end{abstract}


\end{frontmatter}


\section{Introduction}
\label{sec:introduction}

The NASA Dawn spacecraft has been in orbit around dwarf planet Ceres since March 6, 2015. The observations by the onboard framing camera (FC) and visible and infrared spectrometer (VIR) reveal a dark world with little color variation and several enigmatic bright areas in the Occator crater \citep{dS15,N15}. The surface shows evidence for aqueous alteration; ammoniated phyllosilicates are widespread \citep{A16}. Small, isolated patches of water ice lie exposed on the surface \citep{Co16}. Data for a third instrument, a gamma ray and neutron detector (GRaND) are best acquired in the close proximity to the surface that is attained in the low altitude orbit at the end of the Ceres mission. Its first results point at the widespread presence of water ice below the surface, mostly at higher latitudes \citep{PY16}. In this paper, we use FC images for a global spectrophotometric characterization of Ceres. The FC is a CCD camera with one clear (F1) and seven narrow band filters (F2-F8) and a $5.5^\circ \times 5.5^\circ$ field of view (FOV) \citep{Si11}. On approach to Ceres, during a phase of the mission named {\it RC3}\footnote{Mission phase names are indicated in italics.} (the third rotational characterization) the FC mapped the entire visible surface through all filters at a resolution of 1.3~km per pixel. The {\it RC3} images are well suited for our analysis.

Our approach is similar to that adopted by \citet{S13b} for Dawn's previous target, asteroid Vesta. The first step in the spectrophotometric characterization is the derivation of photometric models for each FC filter. The models are valid for the phase angle range covered by the FC image data but not necessarily beyond that. Like at Vesta, the data extend to very large phase angles ($>100^\circ$), but do not cover the smallest ones ($<5^\circ$). This means that we cannot characterize the opposition surge, defined as a non-linear increase towards zero phase angle of the surface reflectance, when displayed on a logarithmic scale. Neither can we provide an independent estimate of Ceres' visual geometric albedo, which is defined for the integrated disk at zero phase angle and estimated to be $0.10 \pm 0.01$ \citep{T89}. Photometric models that represent ``Ceres average'' can be used to photometrically correct the images. Photometric correction is a technique for decoupling the intrinsic reflectivity of the surface from apparent reflectivity variations due to varying illumination and viewing geometry conditions. In other words, local topography is no longer evident in an image after photometric correction and intrinsic reflectivity variations over the surface are revealed. We evaluate a variety of photometric models (Sec.~\ref{sec:best_fit_model}) and aim to identify the one that best represents average Ceres, where we are mostly concerned with the ability of the models to reproduce the light scattering properties of the surface rather than their physical correctness. We present visual albedo maps, assembled from photometrically corrected clear filter images (Sec~\ref{sec:albedo}). Combining albedo maps for different narrow-band filters into color composites reveals spectral diversity over the surface. We present such composite maps, both in enhanced and false color (Sec.~\ref{sec:color_maps}). Other than color, we also show maps of several photometric parameters. To increase the reliability of the results, we choose a very simple model for the phase curve, an exponential model, of which we map the amplitude and slope. Our aim is to relate these photometric parameters to the physical properties of the regolith and thereby uncover the physical processes shaping the surface of Ceres. Other investigations of Ceres' spectrophotometric properties using Dawn data are presented by \citet{C16}, \citet{L16b}, and \citet{Lo16}.

The first albedo and color maps of the Ceres surface were constructed by \citet{L06} from Hubble Space Telescope (HST) observations at a spatial resolution of 30~km per pixel. Their albedo maps at three different wavelengths revealed 11 discrete albedo features ranging in scale from 40 to 350~km. The authors also derived a first photometric model for the resolved surface, but as the HST data were restricted to the very narrow phase angle range of $6.2^\circ$-$7.4^\circ$ their photometric characterization was necessarily incomplete. A preliminary color map made from Dawn FC {\it RC3} images was included in \citet{N15}. Ceres does not have strongly diagnostic absorption features in the wavelength range of the FC (400-1000 nm), but several regions are relatively blue (brighter at shorter wavelengths) and a few areas are relatively red (brighter at longer wavelengths). The initial FC map shows a number of localized bright areas (also called spots). One of these, Cerealia Facula, is located in the center of Occator crater\footnote{The coordinates of Occator and that of other craters discussed in this paper are provided in Table~\ref{tab:craters}.}, and is by far the brightest feature on Ceres. Here, \citet{dS16} reported the detection of sodium carbonate, thought to derive from a brine transported to the surface. Surprisingly, \citet{N15} reported the discovery of an enigmatic diurnal haze in Occator associated with Cerealia Facula. The existence of haze on such a small world without atmosphere would be extraordinary and have important implications. However, the evidence as presented by the authors does not appear extraordinary. We critically evaluate their arguments and see how they hold up in light of post-{\it RC3} observations (Sec.~\ref{sec:haze}). The large brightness of the Occator area took many by surprise. It can be identified with feature \#5 in the HST maps \citep{L06}, but appears unremarkable there due to the small size of the area. Because it is so much brighter than anything else on Ceres, we discuss the spectrophotometric properties of the Occator area separately (Sec.~\ref{sec:Occator}). \citet{L16} looked for changes in brightness in the decade between the HST and FC observations, but did not find any, neither for feature \#5 nor Ceres integrated. We extend their analysis to the entire surface by comparing the HST albedo maps with our FC albedo maps at HST resolution (Sec.~\ref{sec:albedo}). Color maps from VIR data are presented by \citet{C16}. The VIR map in the visual wavelength range is broadly consistent with the FC map in \citet{N15}. The nature and origin of the widely distributed blue material is yet unknown. We map its distribution in clarity and detail, study its spectral properties, and propose a plausible explanation for its origin (Sec~\ref{sec:blue_material}). Note that a full mineralogical characterization of blue material, which requires input from other instruments onboard Dawn, is beyond the scope of this paper. In a recent paper, \citet{N16} presented additional global color maps derived from FC images. Partly similar in its aims, their paper focused on mineralogical interpretation, linking various color units to different types of carbonaceous chondrites, whereas in this paper we focus more on photometric characterization of the surface.

\section{Data and methods}

There are actually two Framing Cameras onboard Dawn: FC1 and FC2 \citep{Si11}. FC2 was the prime camera during the Ceres mission and it acquired all images discussed in this paper. The mission at Ceres was split into different phases. Three rotational characterizations ({\it RC1}, {\it RC2}, {\it RC3}) were performed on approach to Ceres, and generally involved imaging the dwarf planet over a full rotation with all filters. Following approach, Dawn went into successively lower mapping orbits. In this paper we mostly analyze images from {\it RC3} (1.3~km per pixel), the {\it Survey} orbit (0.41~km per pixel), and the high altitude mapping orbit ({\it HAMO}; 0.14~km per pixel). Images acquired in the low altitude mapping orbit ({\it LAMO}) have the highest spatial resolution of the entire mission (about 35~m per pixel). {\it RC3} images are well-suited for evaluating the photometric model performance as they have the entire disk of Ceres in the FOV and span the largest range of phase angles for any mission phase. The {\it RC3} phase was different from the identically named phase at Dawn's previous target, asteroid Vesta. At Ceres, Dawn came in over the north pole to acquire images at very high phase angle to enable a visual search for a rarified extended atmosphere \citep{Li15}. This manoeuvre led to an uneven phase angle coverage over the surface later in {\it RC3}, which complicated our analysis as described below. The {\it RC3} images cover the entire illuminated surface. Due to Ceres' current small obliquity ($4^\circ$) some polar areas do not receive sunlight \citep{S16}. {\it RC3} took place during summer on the northern hemisphere, and therefore part of the south pole was in the dark. The {\it RC3} images selected for evaluating the photometric model performance are listed in Table~\ref{tab:images}. The clear filter images we chose are all those acquired in the last stage of the {\it RC3} phase, which are most similar in character to the Vesta {\it RC3} images. We did not include the limb images acquired at extreme phase angles in early stages of {\it RC3}. The narrow-band images we selected for determining the color photometric model parameters are all those acquired in {\it RC3}.

The first step in the image processing chain was to calibrate the images to radiance factor ($I/F$) following \citet{S13a} (see Sec.~\ref{sec:models} for definitions). {\it HAMO} color images required a correction for stray light according to \citet{S14a}. {\it RC3} color images were not corrected as the method is not accurate if Ceres only fills part of the FOV. For many applications, we projected the images with the USGS Integrated Software for Imagers and Spectrometers ISIS3 \citep{A04,B12}. This software also proved invaluable for calculating the local photometric angles (incidence, emission, phase) for each image pixel. Projection accuracy was improved by image registration, which involved shifting the observed image to match a simulated image prior to projection through cross-correlation and cubic convolution interpolation\footnote{Using the SolarSoft IDL routines tr\_get\_disp and shift\_img (\url{http://www.lmsal.com/solarsoft/})}. The simulated image was calculated from the shape model assuming the Akimov disk function (see Sec~\ref{sec:models}). We employed the {\it HAMO} shape model described by \citet{Pr16}. This model is a result of sophisticated stereo-photogrammetric analysis (e.g.\ bundle block adjustment) of stereo images from {\it HAMO} orbital mapping, and is presented as a raster digital elevation model (DEM) with a lateral grid space of 60~pixels per degree (135~m per pixel). It covers about 98\% of Ceres' surface (few permanently shadowed areas near the poles required interpolation), and has a vertical accuracy of about 10~m. The model is oriented in the Ceres reference frame, which is defined by the tiny crater Kait \citep{Ro16} and the Ceres rotation state derived by \citet{Pr16}.

\section{Model description}
\label{sec:models}

The surface reflectance of a planetary body depends on the wavelength $\lambda$ and the angles of observation: the local angle of incidence $\iota$ of sunlight, the local angle of emergence $\epsilon$, and the phase angle $\alpha$. For convenience we define $\mu_0 = \cos\iota$ and $\mu = \cos\epsilon$. The {\it bidirectional reflectance} of the surface is defined as
\begin{equation}
r({\mu_0, \mu}, \alpha, \lambda) = I({\mu_0, \mu}, \alpha, \lambda) / J(\lambda),
\label{eq:reflectance}
\end{equation}
where $I$ is the radiance in W~m$^{-2}$~\textmu m$^{-1}$~sr$^{-1}$ and $J$ is the normal solar irradiance in W~m$^{-2}$~\textmu m$^{-1}$, which depends on the distance of the planet to the Sun. The {\it radiance factor} \citep{H81}, also known as $I/F$ (with $F = J / \pi$), is
\begin{equation}
r_{\rm F} = \pi r.
\label{eq:rad_fac}
\end{equation}
We use the term ``reflectance'' for the radiance factor in the remainder of this paper. A photometric model for the surface provides an analytical or tabular expression for $r_{\rm F}$. Most commonly, a single model is derived and assumed to be representative for the ``average'' surface (``global photometric model''). Images can then be ``photometrically corrected'' by dividing the observed reflectance of each image pixel by the reflectance predicted by the global model for the photometric angles associated with that pixel. The average photometrically corrected reflectance is therefore unity, but is often multiplied with a factor to represent the normal albedo at a certain phase angle.

In this paper we apply two kinds of models to the surface of Ceres. In the first kind (Type~I) the reflectance can be separated into two parts, the {\it equigonal albedo} and the {\it disk function} \citep{K01,S11}:
\begin{equation}
r_{\rm F} = A_{\rm eq}(\alpha) D(\mu_0, \mu, \alpha).
\label{eq:type_I}
\end{equation}
In this paper, we refer to the equigonal albedo as the {\it phase function}. Plotting the phase function as a function of phase angle produces the {\it phase curve}. It describes the phase dependence of the brightness \citep{S11}:
\begin{equation}
A_{\rm eq}(\alpha) = A_{\rm N} f(\alpha),
\label{eq:equigonal_albedo}
\end{equation}
where $A_{\rm N}$ is the {\it normal albedo}, and $f(\alpha)$ is the phase function normalized to unity at $\alpha = 0^\circ$. The latter depends on the choice of disk function $D$, which describes how the reflectance varies over the planetary disk at constant phase angle. An equigonal albedo image has no brightness trend from limb to terminator for a surface of constant albedo. Note that $A_{\rm N}$ as defined in Eq.~\ref{eq:equigonal_albedo} is not equal to the \citet{H81} normal albedo, which is defined at zero phase but arbitrary incidence angle ($\iota = \epsilon$), and thus depends on the local topography. The normal albedo in Eq.~\ref{eq:equigonal_albedo} is independent of local topography, as such brightness variations are contained in the disk function in Eq.~\ref{eq:type_I}. For most of this paper we assume that the reflectance over the whole surface is described by a single combination of disk and phase function and attempt to find the one that best describes average Ceres. For the phase function we adopt a polynomial of degree $d$:
\begin{equation}
A_{\rm eq}(\alpha) = \sum\limits_{i=0}^d C_i \alpha^i,
\label{eq:poly_phase}
\end{equation}
where $C_0 = A_{\rm N}$ is the normal albedo. One of our main aims is to find a phase function that best reproduces the observations in order to have the optimum photometric correction, and a polynomial offers the flexibility required. All models in this paper required $d = 4$ to achieve a satisfactory fit to the data. In Sec.~\ref{sec:exp_model} we adopt a different approach. We choose a disk function and allow the phase function to vary over the surface in an attempt to link such variations to the physical properties of the regolith. To simplify our analysis we adopt an exponential phase function of the form
\begin{equation}
A_{\rm eq}(\alpha) = A_{\rm N} f(\alpha) = A_{\rm N} e^{-\nu \alpha},
\label{eq:exp_model}
\end{equation}
which has two parameters: the amplitude $A_{\rm N}$ (the normal albedo) and slope $\nu$.

We evaluate several disk functions, each normalized at $\iota = \epsilon = \alpha = 0^\circ$. Most of these have a parameter that usually depends on phase angle. The well-known Lommel-Seeliger law (L-S) has no parameters:
\begin{equation}
D({\mu_0, \mu}) = 2 \frac{\mu_0}{\mu_0 + \mu}.
\label{eq:L-S}
\end{equation}
This law naturally arises from the radiative transfer theory of a particulate medium when considering only single scattering \citep{H81}. A peculiar aspect of the L-S law is that it predicts strong limb brightening when the incidence angle is small and emission angle large or vice versa. This makes it an unrealistic disk function for general use, and we do not consider it in this paper. The Lambert law $D = \mu_0$ for an isotropically scattering surface is not well suited either as a disk function for atmosphereless solar system bodies. However, a Lambert term can be added to the L-S term to counter the limb brightening and improve performance (e.g.\ \citealt{BV83}):
\begin{equation}
D_{\rm L}({\mu_0, \mu}, \alpha) = c_{\rm L}(\alpha) \frac{2 \mu_0}{\mu_0 + \mu} + [1 - c_{\rm L}(\alpha)] \mu_0,
\label{eq:L-S+Lam}
\end{equation}
The free parameter $c_{\rm L}$ governs the balance between the L-S and Lambert contributions. A third model is that of \citet{M41}:
\begin{equation}
D_{\rm M}({\mu_0, \mu, \alpha}) = \mu_0^{c_{\rm M}(\alpha)} \mu^{{c_{\rm M}(\alpha)} - 1},
\label{eq:Minnaert}
\end{equation}
with free parameter $c_{\rm M}$. The fourth model is the Akimov disk function \citep{S11}:
\begin{equation}
D_{\rm A}({\alpha, \beta, \gamma}) = \cos \frac{\alpha}{2} \cos \left[ \frac{\pi}{\pi - \alpha} \left( \gamma - \frac{\alpha}{2} \right) \right] \frac{(\cos \beta)^{c_{\rm A}(\alpha) \alpha / (\pi - \alpha)}}{\cos \gamma},
\label{eq:Akimov}
\end{equation}
in which the photometric latitude $\beta$ and longitude $\gamma$ depend on the incidence, emergence, and phase angles as follows:
\begin{equation}
\begin{split}
\mu_0 & = \cos \beta \cos (\alpha - \gamma) \\
\mu & = \cos \beta \cos \gamma.
\end{split}
\end{equation}
The model parameter $c_{\rm A}$ is called $\eta$ by \citet{S11}. We separately evaluate the case of $c_{\rm A} = 1$ (``Akimov'') and the case for which $c_{\rm A}$ is allowed to vary (``parameterized Akimov''). We consider the model-specific parameters ($c_{\rm L}$, $c_{\rm M}$, $c_{\rm A}$) to be linear functions of phase angle with two fit parameters:
\begin{equation}
c_{\rm X}(\alpha) = a_{\rm X} + b_{\rm X} \alpha.
\label{eq:model_parameter}
\end{equation}

The second kind of photometric model we include in this paper (Type~II) is not explicitly separated into a disk and phase function. We evaluate one such model, that of \citet{H81,H84,H86}:
\begin{equation}
r_{\rm F}(\mu_0^\prime, \mu^\prime, \alpha) = \frac{w}{4} \frac{\mu_0^\prime}{\mu_0^\prime + \mu^\prime} [B_{\rm SH}(\alpha) P(\alpha) + H(\mu_0^\prime,w)H(\mu^\prime,w) - 1] f(\bar{\theta}),
\label{eq:Hapke}
\end{equation}
with $w$ the so-called single scattering albedo. The $H$-function is given by
\begin{equation}
H(\mu,w) = \frac{1 + 2\mu}{1 + 2 \mu \gamma(w)},
\end{equation}
with $\gamma = \sqrt{1-w}$. The shadow hiding opposition effect $B_{\rm SH}$ is described by
\begin{equation}
B_{\rm SH}(\alpha) = 1 + B_{\rm S0} B_{\rm S}(\alpha) = 1 + \frac{B_{\rm S0}}{1 + \tan(\alpha/2) / h_{\rm S}},
\label{eq:SHOE}
\end{equation}
with $B_{\rm S0}$ and $h_{\rm S}$ the amplitude and width of the opposition effect, respectively. $P(\alpha)$ is the phase function of a single particle, for which we adopt the double Henyey-Greenstein function as formulated by \citet{SH07}:
\begin{equation}
P(\alpha) = \frac{1+c}{2} \frac{1-b^2}{(1 + 2b \cos \alpha + b^2)^{3/2}} + \frac{1-c}{2} \frac{1-b^2}{(1 - 2b \cos \alpha + b^2)^{3/2}},
\label{eq:dHG}
\end{equation}
with two parameters $b$ and $c$. Scattering is backward\footnote{\citet{SH07} erroneously stated the opposite.} for $c<0$ and forward for $c>0$, with $|c|<1$. The model includes the term $f(\bar{\theta})$ that describes the effects of ``macroscopic roughness'', a measure of the roughness of the regolith with $\bar{\theta}$ the mean slope angle of the surface facets. This term is rather complicated and we do not reproduce it here; details can be found in \citet{H84}. The inclusion of macroscopic roughness also changes $\mu_0$ and $\mu$ to $\mu_0^\prime$ and $\mu^\prime$ in a way we also do not reproduce here. Note that we do not include the coherent backscatter opposition effect in Eq.~\ref{eq:Hapke}, as the phase angles of our data are far from zero.

\section{Global photometric models}
\label{sec:best_fit_model}

\subsection{Clear filter}
\label{sec:best_fit_clear_model}

Our approach for determining the best-fit photometric models for the clear filter images is slightly different from that in \citet{S13b}. Here we fit the modeled reflectance to the reflectance of all pixels of all images simultaneously, rather than fitting the model to image averages. Fitting was performed using the Levenberg-Marquardt algorithm with constrained search spaces for the model parameters\footnote{Using the MPFITFUN library retrieved from \url{http://cow.physics.wisc.edu/~craigm/idl/idl.html}} \citep{M78,M09}. The wide phase angle range covered by the 350 images in Table~\ref{tab:images} is well suited for a photometric characterization of the surface, but dealing with the large amount of data associated with this number of images was too cumbersome. But since there is a substantial amount of overlap among the images we can do with only half of them without sacrificing quality of the fit. We therefore selected only the images with odd IDs ($n = 175$) for model fitting. Note that the images between {\bf 36356} and {\bf 36517} were acquired with alternating short and long exposures, and the odd images represent the relatively long exposures. However, we do employ all images ($n = 350$) for evaluating the best-fit model performance. Only the north pole was observed at very high phase angles, but we assume that this area is representative for the entire surface in this range.

The two classes of photometric models (Type~I: models with separate phase and disk function, Type~II: Hapke) require different approaches. We determined the Type~1 model parameters in two steps. {\bf Step~1}: Fit the reflectance of the full model (Eq.~\ref{eq:type_I}) to all pixels that meet a certain limit on the incidence and emission angles of all images combined. We evaluated three limits: $(\iota, \epsilon) < 70^\circ$, $80^\circ$, and $85^\circ$, for which the total number of pixels was 31,204,034, 39,434,249, and 42,374,163, respectively. The retrieved photometric model parameters (Eq.~\ref{eq:model_parameter}) are listed in Table~\ref{tab:c2_coef}. While this step gives the best-fit model parameters for the entire data set, low-phase angle images have more pixels than high-phase angle images and thereby dominate the fit. To ensure similar weight for high-phase angle images we perform an additional step. {\bf Step~2}: Adopt the disk function parameters from step~1 (Table~\ref{tab:c2_coef}) and determine the parameters of the polynomial phase function (Eq.~\ref{eq:poly_phase}) by fitting the average reflectances of the model images to those of the observed images (number of images $n = 175$). The parameterless Akimov model did not require this step. Table~\ref{tab:phase_coef_clear} lists the best-fit parameters of the Type~I models for $(\iota, \epsilon) < 80^\circ$. We omit the results for the $70^\circ$ and $85^\circ$ limits, the former because the loss of surface coverage was not compensated by notable improvement of model performance, and the latter because the model performance was unsatisfactory (more details below).

Evaluating the Type~II Hapke model (Eq.~\ref{eq:Hapke}) required only a single step: the six model parameters ($w$, $B_{\rm S0}$, $h_{\rm S}$, $\bar{\theta}$, $b$, $c$) were determined by fitting the model reflectance to all pixels that meet the limit on the incidence and emission angles of all images combined. The Hapke parameters are known to be interdependent to some degree (e.g.\ \citealt{SH07}, \citealt{S12}), and cannot be estimated reliably if the phase angle range of the data is limited. In our case the minimum average phase angle is too large ($7.5^\circ$) to reliably estimate the width ($h_{\rm S}$) and amplitude ($B_{\rm S0}$) of the opposition effect and the macroscopic roughness ($\bar{\theta}$) simultaneously. Therefore we first tried several combinations and let all other parameters vary freely in the fitting process. For the best-fit solutions the Henyey-Greenstein parameters always converged to $b = 0.30$ and $c = -0.65$. We fixed these parameters at their respective values to explore the ($h_{\rm S}$, $B_{\rm S0}$) space more in-depth. \citet{SH89} determined $h_{\rm S} = 0.06$ and $B_{\rm S0} = 1.6$ from $V$-band images with a phase angle range of $1^\circ$-$21^\circ$, which was subsequently adopted by \citet{L06}. These values do not fit the FC data well, perhaps because of the different phase angle and/or wavelength ranges covered by both data sets. Neither fixing $h_{\rm S}$ at 0.06 nor fixing $B_{\rm S0}$ at 1.6 would lead to satisfactory fits. Instead, the data are most compatible with smaller $h_{\rm S}$ and larger $B_{\rm S0}$ values. The best-fit Hapke model for $(\iota, \epsilon) < 80^\circ$ has $w = 0.113$, $B_{\rm S0} = 4.0$, $h_{\rm S} = 0.02$ (fixed), $\bar{\theta} = 22^\circ$, $b = 0.30$ (fixed), and $c = -0.65$ (fixed). We stress that the SHOE parameter values we derive are not necessarily the ``correct'' values for Ceres, because our data do not cover the opposition range. They are merely the values that fit our data best and thereby provide the highest quality photometric correction (excluding SHOE from the model, $B_{\rm S0} = 0$, enlarged the residuals almost fourfold). Our best-fit value for the photometric roughness of $\bar{\theta} = 22^\circ \pm 2^\circ$ is consistent with the $20^\circ$ adopted by \citet{SH89}, who pointed out that their restricted phase angle range prevented them from finding a unique solution for this parameter. It is much smaller than the $44^\circ$ derived from low phase angle HST images by \citet{L06}, who speculated on the physical significance of such a high value. But the photometric roughness of Ceres is probably not unusually high. \citet{R15} and \citet{L16} also find values around $20^\circ$. It seems that analyzing full-disk images at low phase angles does not suffice to reliably derive the roughness parameter; high phase angle images are required in accordance with \citet{SH89}. \citet{C16} find a photometric roughness of $\bar{\theta} = 29^\circ \pm 6^\circ$ at 0.55~\textmu m using VIR data, a little higher than the other values but probably not significantly so. We also implemented the \citet{H02} model, but found its performance lacking. Compared to the \citet{H81,H84,H86} model, the optimization routine took an extremely long time to converge, and the optimum model solution was much worse than that of the earlier model. \citet{L13} had a similar experience when applying this model to Vesta.

We first evaluate the performance of our photometric models in a qualitative way by comparing their ability to photometrically correct images. Figure~\ref{fig:artifacts} shows photometrically corrected images (observed image divided by model image) using the least restrictive limit for the photometric angles, $(\iota, \epsilon) < 85^\circ$, for images acquired at three different average phase angles: low, intermediate, and high. Differences between the models mostly occur at larger photometric angles, and are most pronounced for the high phase angle images. For the low phase angle images (top row) the differences are rather subtle: the Hapke model image is darker towards the outer parts of the image, and the poles of the Lambert/L-S, Minnaert, and Hapke images are brighter than those of the Akimov images. At intermediate average phase angle (middle row) the most pronounced difference is the fact that the Minnaert and Hapke images have bright image poles (where $\iota$ and $\epsilon$ are close to $90^\circ$). The high average phase angle images (bottom row) show dramatic differences. The brightness distribution over the disk can be completely opposite depending on the photometric models, the extremes being Akimov (bright poles, dark limb) and Hapke (dark poles, bright limb). Most applications discussed in this paper use the low phase angle images for which the artifacts of the photometric correction are relatively minor. To avoid the aforementioned bright pole artifacts we adopt $(\iota, \epsilon) < 80^\circ$ for the remainder of the paper. The photometric model images are nearly indistinguishable for $(\iota, \epsilon) < 70^\circ$, but using such a low limit restricts the surface coverage too much. Furthermore, some photometric models may perform better than others and not display any artifacts in the $70^\circ < (\iota, \epsilon) < 80^\circ$ range (more about this in Sec.~\ref{sec:albedo}).

We quantitatively evaluate the performance of all photometric models by calculating the coefficient of variation of the root-mean-square error, a measure of the goodness-of-fit (GOF). It is calculated from the sum-of-squares of the difference between the measured reflectance and that modeled, and is expressed as the coefficient of variation of the root-mean-square error:
\begin{equation}
{\rm CV(RMSE)} = \frac{1}{\bar{r}_{\rm F}} \sqrt{\frac{1}{n} \sum\limits_{i=1}^n (\hat{r}_{{\rm F},i} - r_{{\rm F},i})^2},
\label{eq:GOF}
\end{equation}
where $r_{{\rm F},i}$ and $\hat{r}_{{\rm F},i}$ are the observed and modeled reflectance of pixel $i$, $n$ the total number of pixels included in the analysis, and $\bar{r}_{\rm F}$ the average observed reflectance of these pixels. By defining the GOF like this, rather than by simply the sum-of-squares, it accounts for the difference in number of illuminated pixels between the images and may be compared between campaigns. The model fit performance for $(\iota, \epsilon) < 80^\circ$ is shown in Fig.~\ref{fig:phot_mod_comp}. For high phase angle images, the parameterless Akimov model stands out as performing worse than the others. For the group of images with intermediate phase angle ($30^\circ < \alpha < 50^\circ$), the Minnaert model performs slightly worse than the others. For low phase angle images, all models perform very similarly except for the Hapke model, which is sometimes better, sometimes worse than the others. As said, most applications discussed here use the low phase angle images for which the parameterized Akimov and Hapke models perform best, and therefore we adopt these two models for the remainder of the paper. Figure~\ref{fig:phase_curves} shows the clear filter phase curve associated with the parameterized Akimov disk function. As the Hapke model is so widely used for photometric correction it is important to investigate in detail the consequences of its application with respect to alternative models like Akimov.

Our {\it a priory} assumption was that the disk function parameters are linear functions of phase angle (Eq.~\ref{eq:model_parameter}). To assess whether this assumption is justified, we determined the optimum value of the disk function parameter for each image individually after correcting for the phase angle variation over the disk using the phase functions in Table~\ref{tab:phase_coef_clear}. In Fig.~\ref{fig:photmod_parameters} we compare the ``data'', the parameters derived from fitting the individual images, with the linear models derived from the fit to all images simultaneously. The results are different for each of the three disk functions evaluated (Akimov, Lambert/L-S, Minnaert). For Akimov (Eq.~\ref{eq:Akimov}), the linear behavior appears to break down at low phase angles. However, the large scatter in the data indicates that the disk function is not sensitive to $c_{\rm A}$, and adopting the linear model value should give satisfactory results. For the Lambert/Lommel-Seeliger combination (Eq.~\ref{eq:L-S+Lam}), the linear model aligns very well with the data. Interestingly, the disk function becomes ``super-Lambertian'' at large phase angles ($c_{\rm L} < 0$), something that \citet{S13b} also noticed for Vesta. For Minnaert (Eq.~\ref{eq:Minnaert}), the linear model does not align well with the data at high phase angles. Here, the data appear to make a non-linear upturn, again becoming ``super-Lambertian'' ($c_{\rm M} > 1$). Thus, the linear model may not be satisfactory for $c_{\rm M}$ at large phase angles. Figure~\ref{fig:photmod_parameters} also shows some of the models used for other solar system bodies. Perhaps surprisingly, as the geometric albedo of Ceres is one fourth that of Vesta, the dependence of the disk function parameters on phase angle is quite similar for both bodies. The Akimov parameter model has a different slope, but as the Akimov disk function is not very sensitive to its parameter, the difference may not be significant. \citet{S11} give this parameter (constant) values between 0.34 and 0.52 for the Moon. The Lambert/L-S parameter behaves in similar fashion for Ceres, Vesta, and the Moon. Scattering is mostly Lommel-Seeliger-like towards zero phase angle and even beyond Lambert-like at large phase angles ($> 100^\circ$). The Minnaert parameter is around 0.55 towards zero phase angle for all bodies considered, and increases with phase angle. For the very low-albedo nucleus of comet Tempel~1, $c_{\rm M}$ increases less rapidly with phase angle than for Ceres and Vesta, although the associated uncertainties were rather large \citep{Li13}. For the asteroid Lutetia, we found two sources for the Minnaert parameter in the literature. \citet{M15} found $c_{\rm M}$ to drop off at higher phase angle, but the linear relation of \citet{H16} shows the familiar increase.

\subsection{Color filters}
\label{sec:best_fit_color_model}

A total of 525 images is available for determining the photometric model parameters (Table~\ref{tab:images}), which means $n = 75$ for each narrow-band filter. They can be divided in two groups of different average phase angle: $7^\circ$-$11^\circ$ and $32^\circ$-$49^\circ$. This covers a phase angle range for which all clear filter photometric models perform about equally well (see Sec.~\ref{sec:best_fit_clear_model}). This means that we are free to choose the most convenient model for the color data, and we select the parameterless Akimov disk function in combination with a second-order polynomial phase function (parabola). To find the phase function for each filter we applied the method for Type~I models described in the previous section, where step~1 was not necessary. Figure~\ref{fig:phase_curves} shows the best-fit phase curves for all filters over their range of validity ($7^\circ < \alpha < 49^\circ$), with the parameters listed in Table~\ref{tab:phase_coef_color}. The FC narrow band images are contaminated by up to 15\% in-field stray light, depending on filter \citep{Si11,S14a}. We did not correct the {\it RC3} images for stray light since our method is not guaranteed to work if a substantial part of the image is empty space. The best-fit phase curves are therefore valid only for {\it RC3} images uncorrected for stray light for the purpose of photometric correction.

With the curves in hand we should, in principle, be able to retrieve the degree of phase reddening of the Ceres spectrum. Phase reddening, the increase of the visible spectral slope with increasing phase angle, was first reported for Ceres by \citet{T83} and confirmed by \citet{R15}, \citet{C16}, and \citet{Lo16}. \citet{L16} made an attempt to quantitatively compare the spectral slope derived from ground based observations with that derived from FC images acquired on approach to Ceres. Unfortunately, the visible spectral slope as defined by the authors is very sensitive to errors in the calibration, in particular the in-field stray light correction (or lack thereof). Of the two filters involved in the slope equation (F2 and F6; see Table~\ref{tab:phase_coef_color} for definition), especially F6 is strongly contaminated by stray light \citep{S14a}. The spectral slope derived by \citet{L16} from ground based data changes from $-1$\%/0.1~\textmu m at $\alpha = 0^\circ$ to $+1$\%/0.1~\textmu m at $\alpha = 20^\circ$. Stray light present in FC images could easily decrease the spectral slope by about 2-4\%/0.1~\textmu m, which is larger than the proposed amount of phase reddening. This is the case for the {\it RC3} images analyzed here, but also for the {\it RC1} and {\it RC2} images analyzed by \citet{L16}. Not only will the amount of stray light be different (to an unknown degree) for each {\it RC}, but it will also be different within {\it RC3} for different phase angles. We therefore conclude that the FC images acquired on approach are not suitable for determining the degree of Ceres phase reddening.

\section{Resolved spectrophotometric properties}

\subsection{Albedo}
\label{sec:albedo}

\begin{figure}
\centering
\includegraphics[width=\textwidth,angle=0]{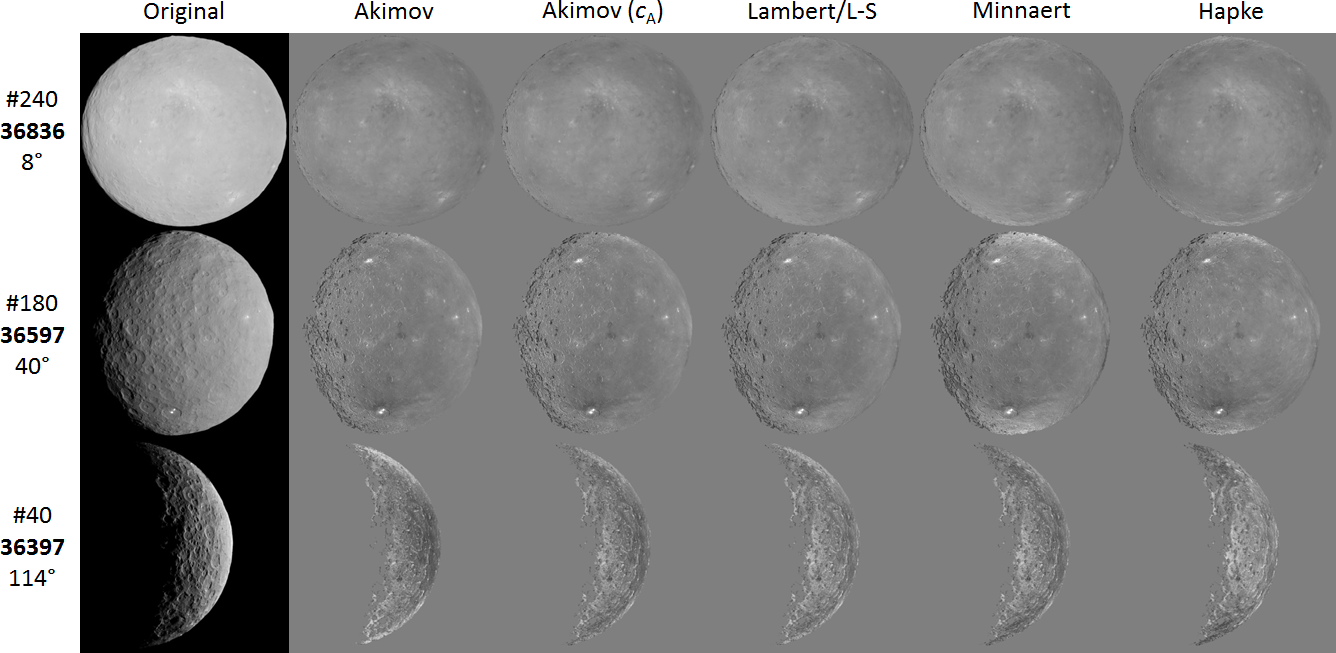}
\caption{Photometric correction properties of all models compared for three clear filter images of different average phase angle with ($\iota, \epsilon) < 85^\circ$. The labels on the left list the image number in Fig.~\ref{fig:phot_mod_comp}, the image ID (bold), and the average phase angle.}
\label{fig:artifacts}
\end{figure}

\begin{figure}
\centering
\includegraphics[width=8cm,angle=0]{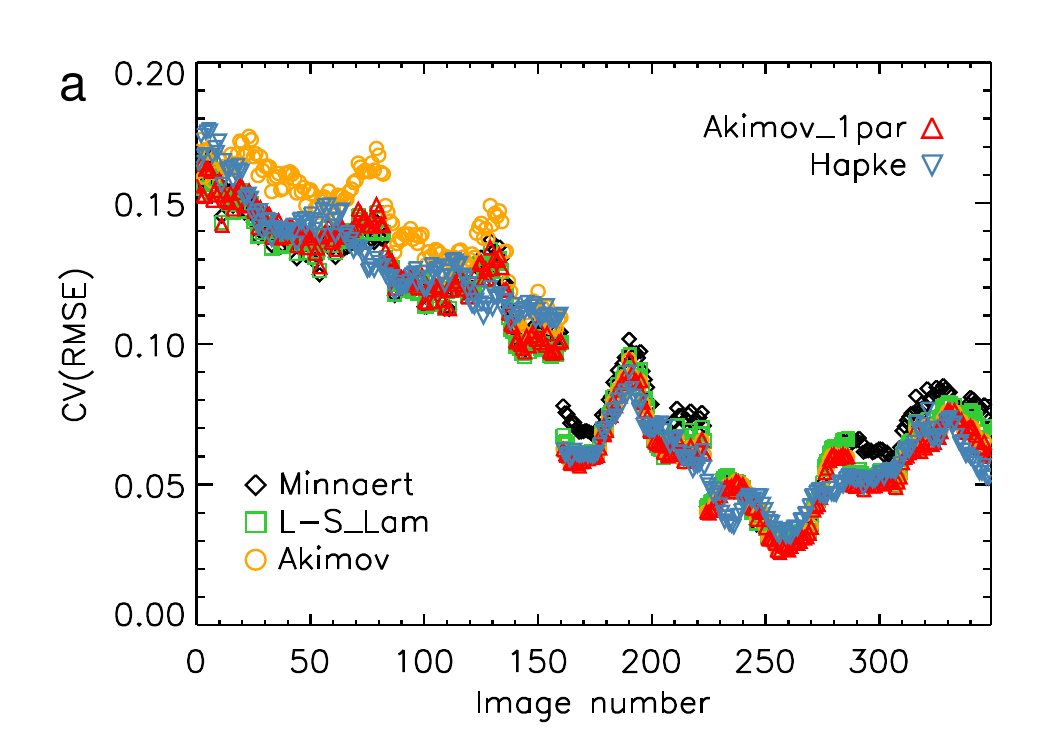}
\includegraphics[width=8cm,angle=0]{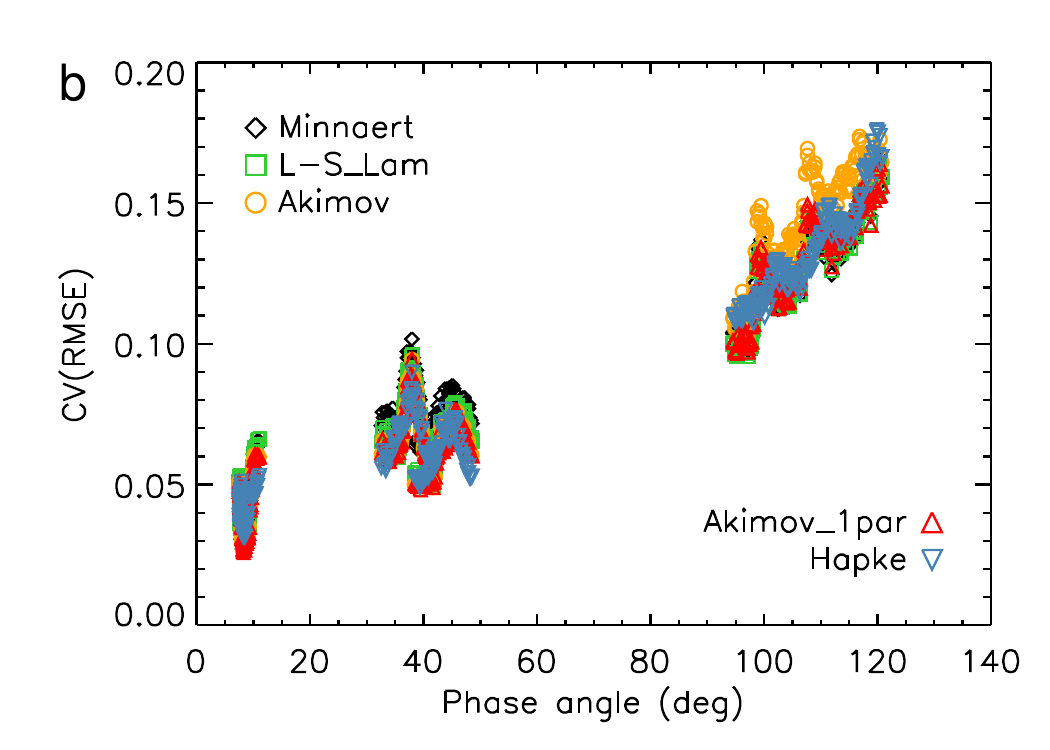}
\caption{The goodness-of-fit of different photometric models for 350 {\it RC3} clear filter images with $(\iota, \epsilon) <80^\circ$ as a function of ({\bf a}) image number and ({\bf b}) average image phase angle. ``Akimov\_1par'' denotes the parameterized Akimov model and ``L-S\_Lam'' the Lommel-Seeliger / Lambert combination. The two Akimov models align closely for phase angles below $60^\circ$.}
\label{fig:phot_mod_comp}
\end{figure}

\begin{figure}
\centering
\includegraphics[width=12cm,angle=0]{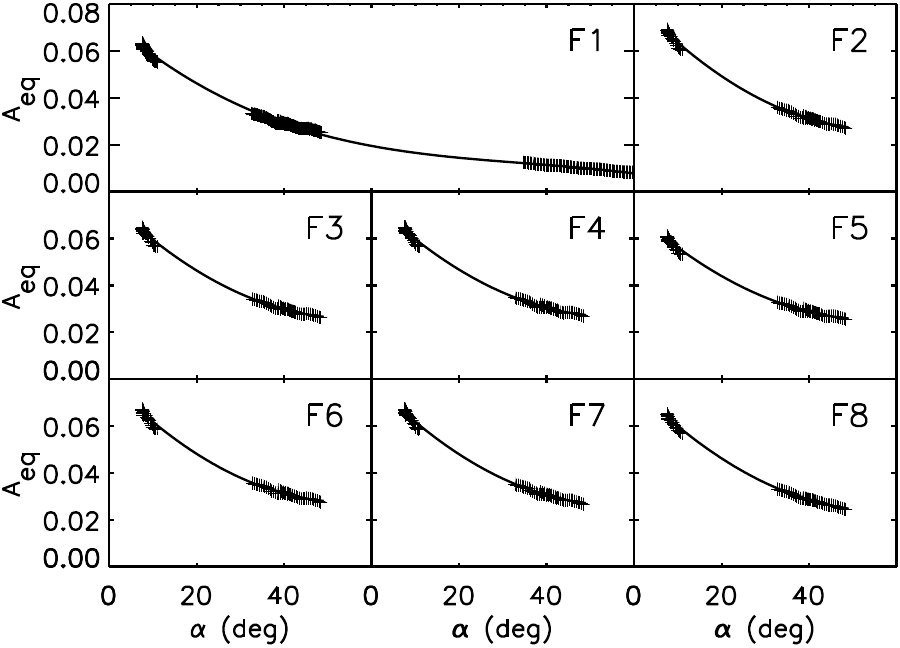}
\caption{Polynomial phase curves for all filters ($A_{\rm eq}$ in Eq.~\ref{eq:type_I}), valid in combination with the Akimov disk function (the parameterized version for F1). The plus symbols indicate the average equigonal albedo of the selected images. The phase angle ($\alpha$) range for F1 is ($0^\circ, 120^\circ$).}
\label{fig:phase_curves}
\end{figure}

\begin{figure}
\centering
\includegraphics[width=8cm,angle=0]{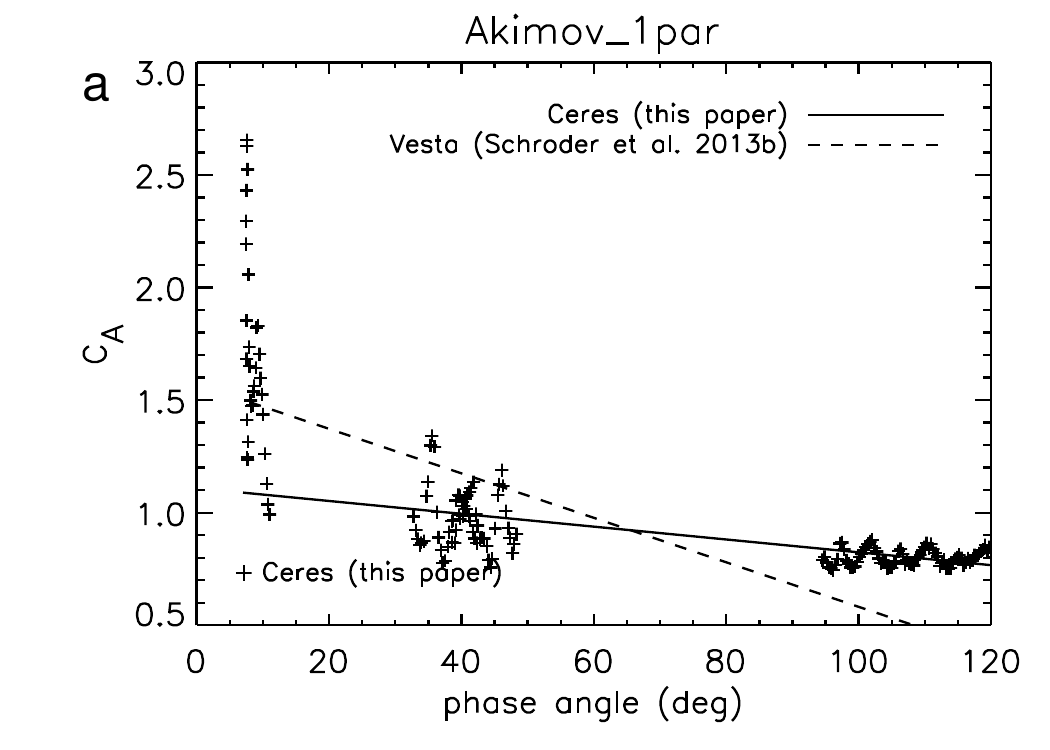}
\includegraphics[width=8cm,angle=0]{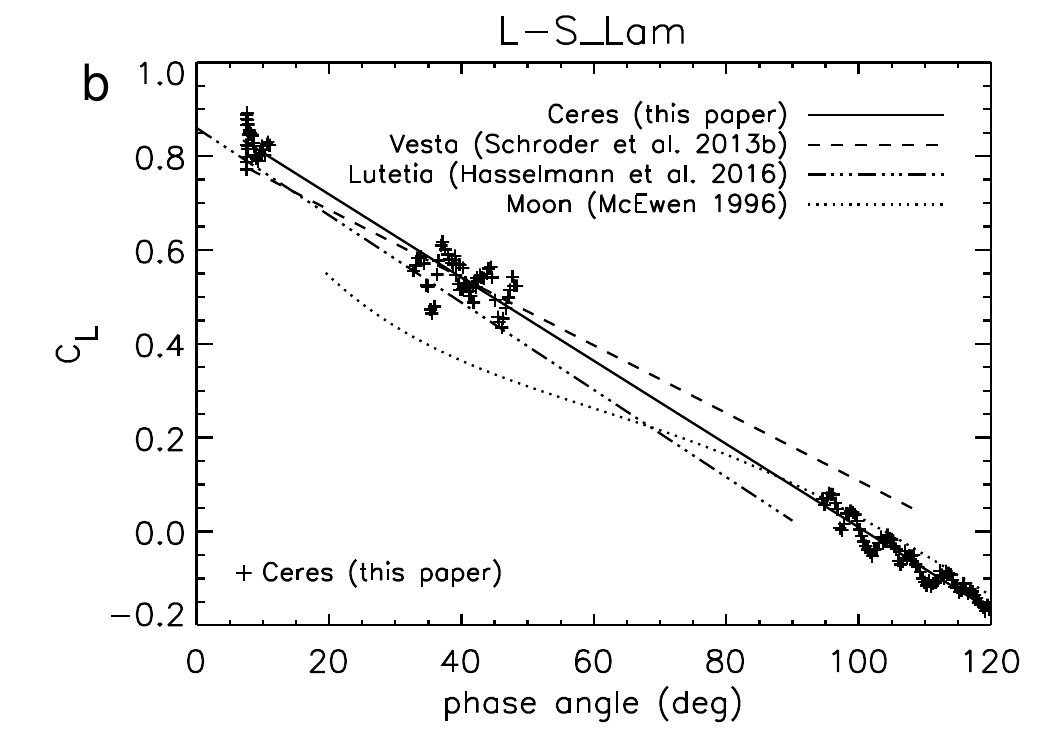}
\includegraphics[width=8cm,angle=0]{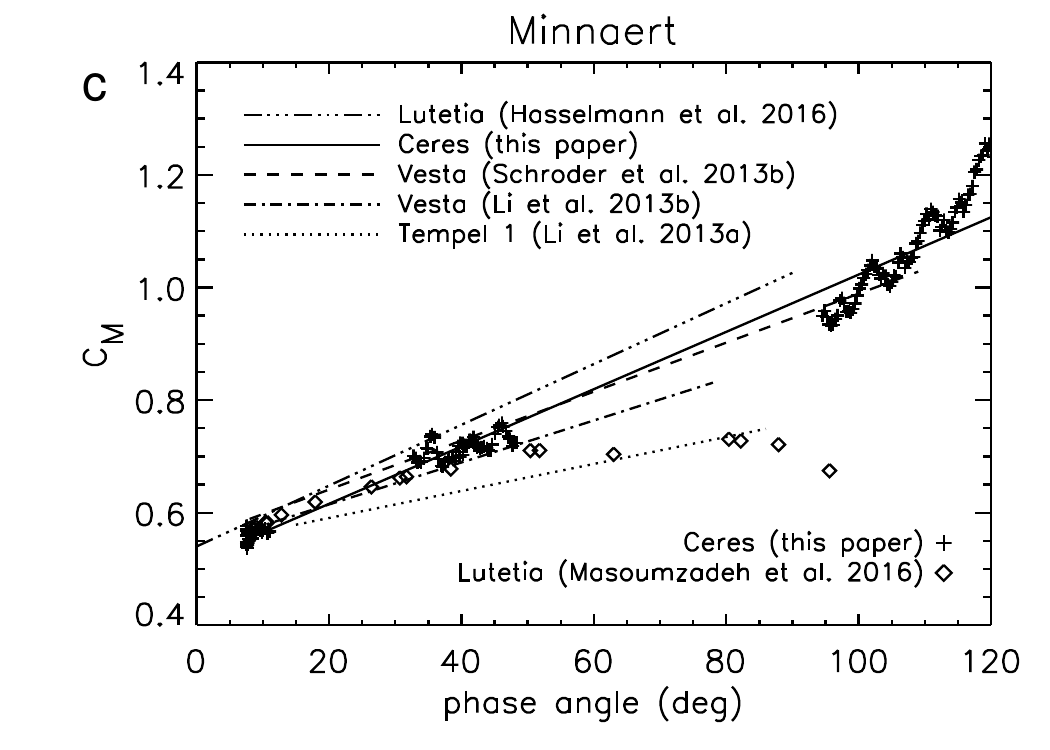}
\caption{Disk function model parameters as a function of average image phase angle with $(\iota, \epsilon) <80^\circ$. ({\bf a}) Parameterized Akimov. ({\bf b}) Lommel-Seeliger/Lambert. ({\bf c}) Minnaert.}
\label{fig:photmod_parameters}
\end{figure}

We derive the normal albedo distribution over the surface by creating a global map from a series of photometrically corrected images. We want the photometric correction to have the highest possible quality and therefore we select images at the lowest possible phase angle, which show least shadowing and promise the smallest projection artifacts. During one phase of {\it RC3} a series of images was acquired with the lowest phase angle range of the entire mission (Table~\ref{tab:images}). These images offer almost global coverage excluding only the polar extremes. We construct maps for the two most successful photometric models (parameterized Akimov and Hapke) for a qualitative assessment. As the photometrically corrected images show substantial artifacts at large incidence and emission angle for all models, we need to limit these angles. Fortunately, low phase angle images are particularly well corrected with few artifacts, and differences between the models are minor (Fig.~\ref{fig:artifacts}). We can therefore set the limit relatively high at $(\iota, \epsilon) < 80^\circ$. The maps constructed in this way offer only an approximation to the true normal albedo. The images were acquired at phase angles outside the range of the opposition effect, which typically occurs within a few degrees of zero phase. As we did not observe in this range, we cannot account for, or even assess, variations in the strength of the opposition effect over the surface.


The map-making procedure was the following. The registered reflectance images were projected to an equirectangular projection with a resolution of 5~pixels per degree using bi-linear interpolation. The photometric angles ($\iota$, $\epsilon$, $\alpha$) were calculated for the projected images for the purpose of photometric correction. The global albedo map was calculated as the median of all photometrically corrected projected images, including only illuminated pixels with $(\iota, \epsilon) < 80^\circ$. By using the median instead of the mean we avoided artifacts at crater rims. The median of the final map was set to the visual geometric albedo of Ceres, for which we adopt $0.10 \pm 0.01$ from \citet{T89}. The later estimate of $0.113 \pm 0.005$ by \citet{T02} may be more accurate but is not significantly different from the original. We can equate the normal albedo with the geometric albedo because the disk function at zero phase angle is very uniform over Ceres' disk. We also tested an alternative approach in which we photometrically corrected the registered images prior to projection, but found that the coverage of the global albedo map was, surprisingly, slightly worse. Figure~\ref{fig:albedo} shows the global albedo maps for the parameterized Akimov (a) and Hapke (b) models. The two maps appear similar at first glance, but differences exist. First, the albedo at latitudes above $60^\circ$ and below $-60^\circ$ is relatively high for Hapke but not Akimov. Given that the increase is pervasive over all longitudes this appears to be an artifact of the way the Hapke model handles the more extreme illumination conditions of local slopes toward the poles. Given the aforementioned increased brightness at the image poles (where $\iota$ and $\epsilon$ are high) for images of low to moderate phase angle (Fig.~\ref{fig:artifacts}), the increased brightness unlikely represents polar frost. Second, the large positive albedo feature centered on ($135^\circ$E, $0^\circ$) and the terrain to the west/southwest of it is of higher albedo for Akimov. It is unclear which map represents the ``truth'' in this respect, and we have to keep in mind that different photometric models can give different results. The maps show many discrete positive albedo features that appear to be associated with impacts and few negative ones. The bright areas inside Occator crater are the brightest objects on the surface, and their albedo is much higher than the map scale maximum (0.12). As these areas were overexposed for many images that make up this map we defer estimating their albedo to Sec.~\ref{sec:Occator}. The second brightest feature is Oxo crater with an albedo of 0.14-0.15. Then follow a number of craters: an unnamed crater northwest of Ahuna Mons (0.13-0.14), Kupalo (0.13), Haulani (0.12). Of the smaller bright areas it is worth to highlight the isolated mountain Ahuna Mons, with an albedo close to 0.12. All of these bright features have a very youthful appearance in high resolution images, which suggests that albedo is an indicator of age. Striking is the aforementioned large positive albedo feature centered on ($135^\circ$E, $0^\circ$) in a region called Vendimia Planitia, in the northern part of which we find Dantu crater. Dantu appears to be one of the youngest craters within this area and has uncovered material of relatively high albedo. Discrete low albedo features are rare, one example being the ejecta of Occator towards the northeast.

The maps in Fig.~\ref{fig:albedo} show good qualitative agreement with the VIR albedo maps at 0.55~\textmu m \citep{C16,Lo16}, which are of lower spatial resolution. There is also good agreement with the maps derived from HST images by \citet{L06}. It was reported by \citet{L16} that no albedo changes over the surface could be identified over the decade between the HST and FC observations. The authors concentrated their analysis on two bright areas on the surface (Haulani and Occator). We visually verify their assertion for the whole surface by comparing the HST map with maps constructed from FC images at HST resolution. One of the HST maps was acquired with the F555W filter, which has an effective wavelength of $533$~nm and a full-width-at-half-maximum (FWHM) of 123~nm. The most similar FC filter is F2, which has an effective wavelength of $555$~nm and a FWHM of 44~nm. Even though the FC color filters are affected by in-field stray light \citep{S14a}, the amount of stray light for F2 is relatively low and we can safely ignore it. We created two simulated HST albedos map from low-phase angle F2 images acquired in RC3 (selection in Table~\ref{tab:images}) using a method that differs from that employed for the clear filter map. The registered images were photometrically corrected prior to projection, both with the Akimov model (parameters in Sec.~\ref{sec:best_fit_color_model}) and a custom fit Hapke model with $w = 0.121$, $B_{\rm S0} = 4.3$, $h_{\rm S} = 0.02$ (fixed), $\bar{\theta} = 25^\circ$, $b = 0.30$ (fixed), and $c = -0.65$ (fixed). The photometric correction involved dividing the observed image by the model image, but prior to division both images were binned and subsequently convolved with the HST point spread function (PSF) associated with the F555W filter. We did not restrict the photometric angles in this step. Binning was done by averaging the signal in blocks of $24 \times 24$ pixels, giving the binned images a resolution of 30.4~km per pixel compared to the 30~km per pixel quoted for the HST images by \citet{L06}. The binned photometrically corrected images were then expanded to original size using nearest-neighbor interpolation, creating artificially large pixels in the process, and projected to equirectangular projection of 3~pixels per degree resolution assuming an ellipsoid shape and using nearest-neighbor interpolation. The final albedo map was constructed as the median of the projected images, including only pixels with $(\iota, \epsilon) < 50^\circ$ to be consistent with the HST map as calculated from the ellipsoid shape. The median of the map was set to 100\%.

The original HST map and the two FC maps (Akimov and Hapke) are compared in Fig.~\ref{fig:HST_comparison}. The agreement between the HST map and the FC maps is good, demonstrating the high quality of the \citet{L06} HST image reduction and map construction. The position of major features is slightly different because the HST map was produced with the \citet{T05} pole solution, which is different from the Dawn solution of \citet{Pr16}. There are striking differences between the two FC maps that use different models for the photometric correction. Like in the full-resolution albedo maps, the large feature in Vendimia Planitia is much brighter for Akimov than Hapke, as is the terrain to the west/southwest of it. On the other hand, the terrain in the top right of the map ($270^\circ$-$360^\circ$E, $40^\circ$N) is brighter for Hapke. The albedo distribution in the HST map shows several similarities with the FC Hapke map, such as the low albedo of the terrain west/southwest of the large feature and the high albedo around ($270^\circ$-$360^\circ$E, $40^\circ$N). This is not surprising, given that the Hapke model was also used for the HST map. The Hapke parameter settings were different, but this should not significantly affect the photometric correction for these low phase angle images. However, the albedo of the Vendimia Planitia feature, especially the area around Dantu ($137^\circ$E, $24^\circ$N) in the HST map is more like that in the Akimov map. Again, we cannot compare the brightness of Occator, as it was overexposed in the majority of the images used for making the map. Another difference between the Hapke and Akimov maps are the presence of obvious seams in the former, not only in the FC map, but also the original HST map. These may be artifacts particular to the Hapke model, but perhaps only associated with this particular data set and special way of processing. Given the large uncertainties associated with the photometric modeling we cannot conclude that any of the differences between the HST map and the FC maps are real; depending on whether we used the Akimov or Hapke model for photometrically correcting the images we might argue that the albedo of the Vendimia Planitia feature has either increased or decreased in time. We therefore agree with \citet{L16} that the albedo distribution over the surface of Ceres has most likely not changed over the last decade at the HST spatial resolution. We performed a dedicated HST simulation for one correctly exposed image and found no evidence for the area to have changed in brightness, in agreement with \citet{L16}.

\subsection{Color}
\label{sec:color_maps}

With the photometric models for the narrow-band filters derived in Sec.~\ref{sec:best_fit_color_model} we can construct global color maps. For this purpose we photometrically correct the {\it RC3} images that were acquired at the lowest phase angle of the mission (Table~\ref{tab:images}) using the Akimov disk function and the phase curves in Fig.~\ref{fig:phase_curves}. To avoid unduly influence of shadows, cosmic rays, and projection artifacts we calculate the albedo in each projected pixel as the median over all images that include that pixel. With all spectral information contained in 7 narrow band filters, we need to think of a way to convey as much of this information as possible using only the three image color bands (red, green, and blue, or RGB) for the purpose of visualization. First we note that the color variation on Ceres is relatively subdued compared to that on Vesta \citep{R12a}. The dominant variation is that of the spectral slope over the full FC wavelength range, which ranges from neutral (Ceres average) to negative; we refer to terrain that has a negative spectral slope compared to Ceres average as ``blue'' terrain. Other color variations exist, but seem to be mostly restricted to the Dantu area and the inside of Occator crater (see Sec.~\ref{sec:Occator}). After trying out different filter combinations, we found that virtually all spectral variation can be visualized in a color composite using only three filters: F8 (438~nm), F2 (555~nm), and F5 (965~nm). The F8 and F5 filters are at the extremes of the FC wavelength range and thereby are the best indicators for the blue spectral slope. Filter F2 was chosen to visualize the color variations around Dantu. A further advantage of F2 and F5 is that they are only weakly affected by stray light \citep{S14a}. However, F8 is more strongly affected by stray light and we must be aware of possible artifacts, since the images used to construct these maps were not corrected for this problem.

In Fig.~\ref{fig:color} we display several color maps in different representations. For reference we show the Akimov albedo map from in Fig.~\ref{fig:color}a. To limit distortions near the poles we choose the Mollweide projection. Ceres maps are commonly displayed with longitude $180^\circ$ in the center of the projection. In this case we put longitude $0^\circ$ in the center to allow the proper display of the Oxo crater, which is found exactly at this longitude and would otherwise be almost invisible at the edge of the projection. In Fig.~\ref{fig:color}b we assigned the (F8, F2, F5) filters to the RGB color channels, which conveys information about both color and albedo. The albedo of several locations is too high to display correctly with the chosen color stretch and thus appear white; these are Oxo crater, the bright areas in the Occator crater, and a small unnamed crater at ($309.9^\circ$E, $0.4^\circ$S). The Haulani and Kupalo craters stand out as bright blue, and Occator is mostly blue with dark ejecta in the northeast. Patches that appear beige in this color display dot the surface, especially around Dantu, and a single reddish patch is seen near the north pole next to Ernutet crater. To visualize only the color distribution over the surface we construct a false color map by taking the ratio of the aforementioned filters with F3 (749~nm), another filter low in stray light. The result is shown in Fig.~\ref{fig:color}c. Again, choices have to be made concerning the color saturation when displaying such a map. We do not only want to show the distribution of blue material, but also the intensity of the blue color, and we therefore saturate only the color of the bluest object on the surface, the ejecta of Haulani crater. In this way we can see, for example, that Occator is less blue than Haulani. The bluest objects are Haulani, Oxo, and Ahuna Mons, followed by Occator and Kupalo. By comparing with the map in Fig.~\ref{fig:color}b we note that blue color is positively correlated with albedo, but not all blue craters are bright. One example is an unnamed dark blue crater at ($237.1^\circ$E, $39.2^\circ$S) on the northwestern rim of Urvara crater. The false color map also clearly shows the small patch associated with Ernutet that is so unusually red for the surface of Ceres. We can visualize the distribution of blue material in a different way by displaying the F5/F8 ratio in gray scale (Fig.~\ref{fig:color}d). In this map dark terrain is blue, whereas white terrain is red. The single ratio map provides a particularly effective way to recognize subtle color features on the surface. For example, in Fig.~\ref{fig:color}d we see very long crater rays extending from Occator and Haulani that are not obvious in Figs.~\ref{fig:color}b and c. The Occator rays extend over almost an entire hemisphere. The maps in Fig.~\ref{fig:color} are the highest quality color maps available in terms of the clarity of the colors, due to the almost complete absence of artifacts related to the projection and photometric correction. Selecting only the lowest phase angle images ensured that craters are not recognizable other than through the color of their ejecta. Obvious artifacts are restricted to the poles, as they were not well illuminated. But these areas represent only a small fraction of the Ceres surface, something that can be appreciated best when using an equal area projection like in Fig.~\ref{fig:color}. Note the absence of color gradients towards the poles.

As said, the dominant color variation on Ceres is the negative spectral slope associated with blue material. What exactly this means in terms of the spectrum is shown in Fig.~\ref{fig:Haulani_spectra}, in which we focus on what appears to be the bluest terrain on the surface: Haulani. We calculate the average spectrum of two small areas shown in (a), one on the blue ejecta west of the crater, the other on a patch of surface that appears to be mostly free of ejecta and represents the background, or average Ceres (as verified in Fig.~\ref{fig:color}a). The absolute reflectance spectra in (b) are both relatively flat with slight depressions around 750~nm and towards 430~nm. The ratio of the two spectra shows a gradual drop of 20\% over the full FC wavelength range. All blue terrain on Ceres exhibits a similar spectral change as observed for Haulani, only weaker. The bluest features on Ceres, Haulani, Oxo, and Ahuna Mons, appear fresh, featuring sharp crater rims, high albedo contrasts in close vicinity, lack of secondary impact craters, and apparent flow features. Also, the presence of crater rays around Haulani and Occator support a young age for these craters. All this suggests that the intensity of the blue color is a measure for age. By inference, Haulani and Oxo are the youngest impact craters in their size class. A young age for Ahuna Mons is consistent with its status as a suspected cryovolcanic dome \citep{R16}. Occator crater, while still relatively young, is less blue than Haulani and may therefore be older. Crater counts confirm that bluish material is mainly associated with the youngest impact craters on Ceres \citep{SK16}.

Terrain that dominates the red end of the color spectrum is associated with Ernutet crater. It stands out like a sore thumb in Fig.~\ref{fig:color}c at high latitude. Figure~\ref{fig:Ernutet}a shows in detail how this red material is distributed in small patches inside and beyond the southeast corner of the crater. It is the reddest terrain on Ceres, with the exception of Cerealia Facula in Occator (see Sec.~\ref{sec:Occator}). Its spectrum relative to the surrounding terrain displays a featureless positive spectral slope (Figure~\ref{fig:Ernutet}b). \citet{dS17} found this terrain to be rich in organic material and argued for an endogenous origin. The FC images do not reveal an obvious source for the red material, such as a cryovolcanic vent. It is concentrated southeast of the crater and distributed in a patchy fashion. The color maps in Fig.~\ref{fig:color} suggest that its distribution is limited to the surroundings of Ernutet. If such red material is present in other locations on the surface covered by our maps, it should be distributed in units of small size ($< \sim 1$~km$^2$). \citet{P17} further explore the geologic context of the Ernutet red material.

Surprisingly, the large positive albedo feature in Vendimia Planitia that is prominent in the albedo map (Fig.~\ref{fig:albedo}) does not stand out in the color maps. Indeed, we verified that the spectrum of terrain inside this feature is not different from that of average Ceres, with the exception of the surroundings of Dantu, which are more blue. The major difference with the rest of Ceres is the albedo, which is about 10\% higher. Such a difference can be explained in two ways; the regolith inside this feature either is mixed with a bright material with a neutral visible spectrum, like certain salts, or has a smaller typical particle size. The latter also points at a different mineralogical composition, as the smaller particle size may indicate a different material strength and greater susceptibility to fracturing under the same space weathering conditions. It should be possible to distinguish between the two possibilities using VIR spectra. The lack of blue color supports an old age, with Dantu being a relatively recent addition.

\subsection{Phase curve slope}
\label{sec:exp_model}

For Vesta, \citet{S13b} identified variations in the slope of the phase curve that could be related to physical properties of the regolith. Can we find evidence for a similar phenomenon on Ceres? To answer this question we adopt the exponential phase function in Eq.~\ref{eq:exp_model}, in which the slope of the phase curve is given by the $\nu$ parameter, and model the reflectance with Eq.~\ref{eq:type_I} using the parameterized Akimov disk function (Eq.~\ref{eq:Akimov}). We fit the modeled reflectance to the observed reflectance in each pixel of a large number of RC3 clear filter images in equirectangular projection with a resolution of 5~pixels per degree. We applied the natural logarithm to the photometrically corrected images to give high and low reflectance values equal weight in the fitting procedure. That is, for each projected pixel we find $A_{\rm N}$ and $\nu$ in
\begin{equation}
\ln (r_{\rm F} / D_{\rm A}) = \ln A_{\rm N} - \nu \alpha,
\label{eq:log_exp_model}
\end{equation}
in which $(r_{\rm F} / D_{\rm A})$ is a vector of values, each element representing one pixel in one image with the restriction $(\iota, \epsilon) < 80^\circ$. We reject vectors that do not have at least 5 elements. The error adopted for $\ln (r_{\rm F} / D_{\rm A})$ in the fit procedure was 0.05. We would like to include as many images in the analysis as possible with a wide range of phase angles. Unfortunately, the phase angle coverage of the surface was very uneven during RC3. While the north pole of Ceres was observed at very high phase angles, the equator and south pole were only seen at phase angles smaller than $50^\circ$. To make a global evaluation of the phase curve variations on Ceres we are therefore restricted to 198 images with $\alpha < 50^\circ$ (Table~\ref{tab:images}).

\begin{figure}
\centering
\includegraphics[width=\textwidth,angle=0]{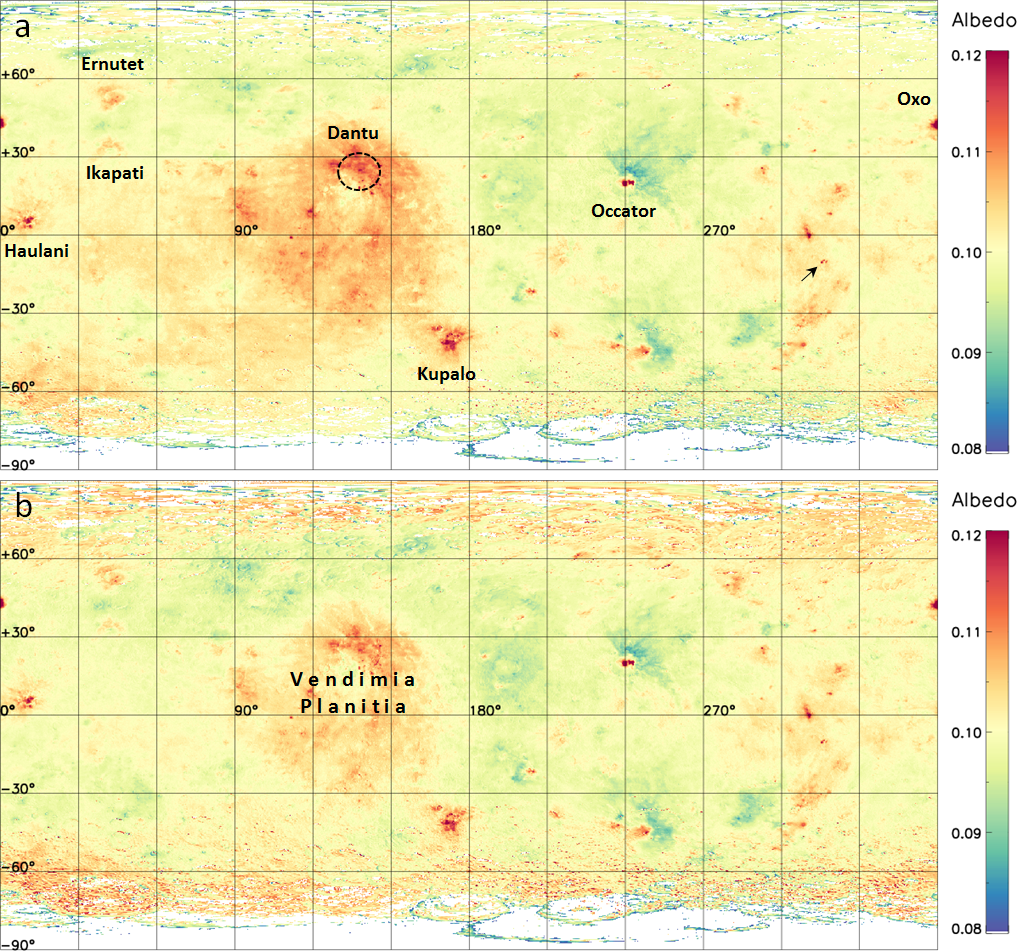}
\caption{Ceres global normal visual albedo distribution reconstructed from clear filter images acquired at low phase using the ({\bf a}) parameterized Akimov model and ({\bf b}) Hapke model. The projection is equirectangular and the median of the maps was set to 0.10 \citep{T89}. The arrow in (a) points at Ahuna Mons.}
\label{fig:albedo}
\end{figure}

\begin{figure}
\centering
\includegraphics[width=14cm,angle=0]{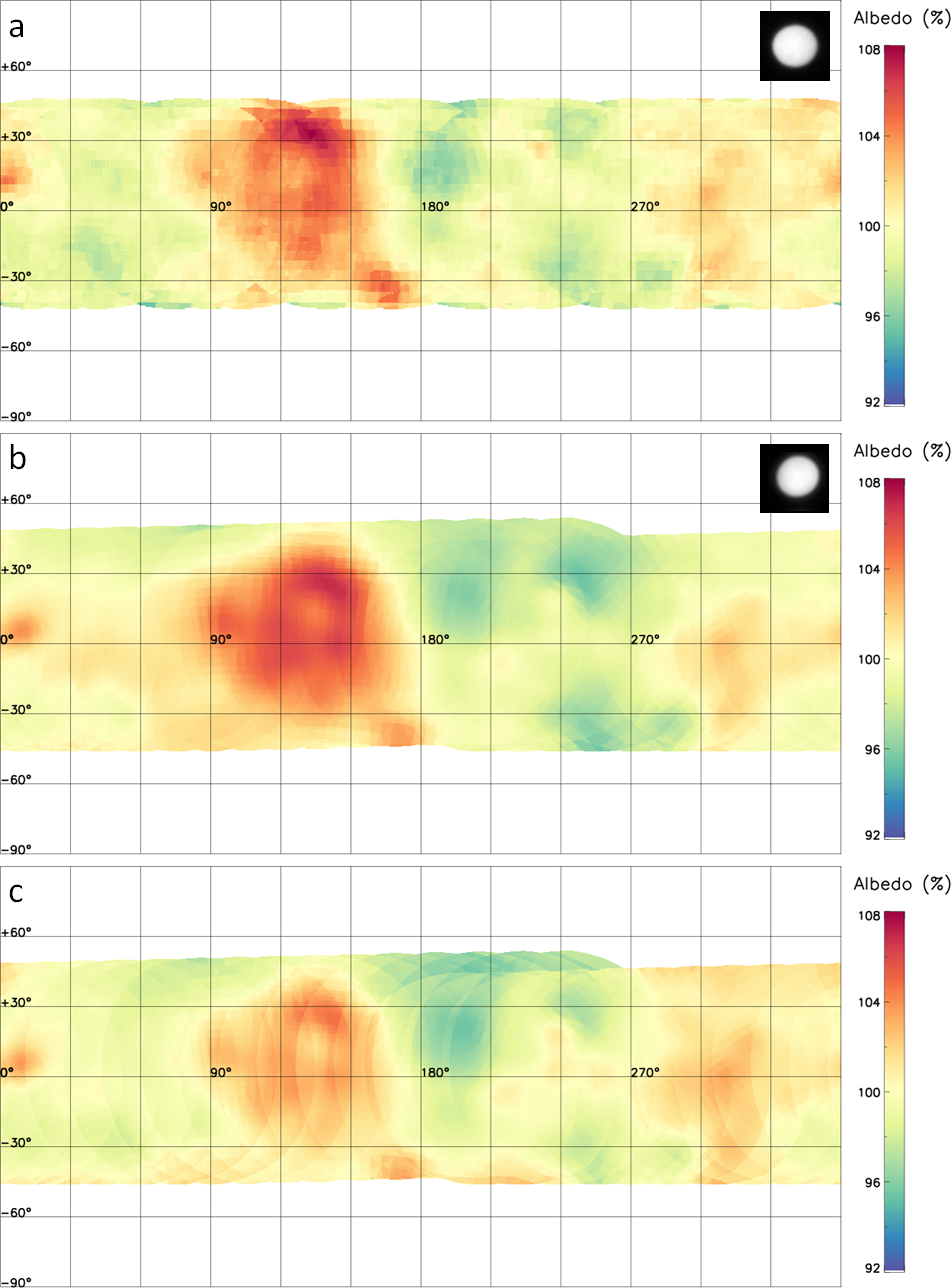}
\caption{Comparing Ceres surface albedo maps at HST resolution. ({\bf a}) HST albedo map (filter F555W) from \citet{L06} with Hapke photometric correction. ({\bf b}) FC2 albedo map (filter F2) reconstructed from images reduced to HST resolution with Akimov photometric correction. ({\bf c}) As (b), using Hapke photometric correction. The inset shows Ceres as it appeared to both instruments, with the resolution artificially reduced for the FC.}
\label{fig:HST_comparison}
\end{figure}

\begin{figure}
\centering
\includegraphics[width=10.5cm,angle=0]{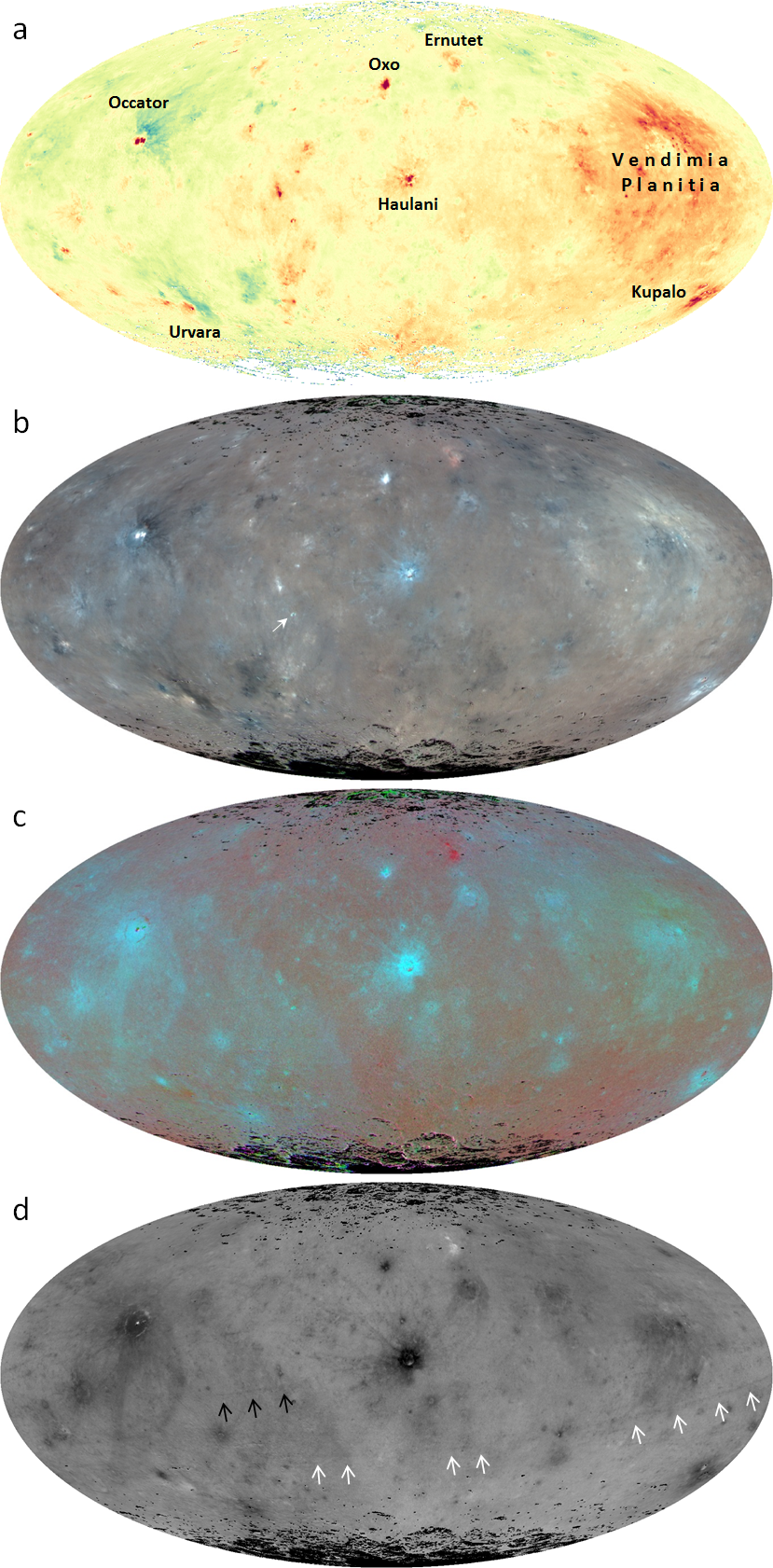}
\caption{Global maps in Mollweide projection with $0^\circ$ longitude at the center. ({\bf a}) Albedo map from Fig.~\ref{fig:albedo}a. ({\bf b}) Enhanced color composite with filters F5 (965~nm), F2 (555~nm), and F8 (438~nm) in the RGB color channels. The minimum and maximum value in each color channel is the median $\pm$ 25\%. The arrow points at Ahuna Mons ({\bf c}) False color composite with ratios F5/F3 (965~nm/749~nm), F2/F3 (555~nm/749~nm), and F8/F3 (438~nm/749~nm) in the RGB color channels, in which black and white are the map median $\pm$ ($7,7,10$)\%. ({\bf d}) F5/F8 color ratio (965~nm/438~nm) with white and black being the map median $\pm$ 15\%. Arrows indicate extended crater rays emanating from Occator (white) and Haulani (black).}
\label{fig:color}
\end{figure}

\begin{figure}
\centering
\includegraphics[width=\textwidth,angle=0]{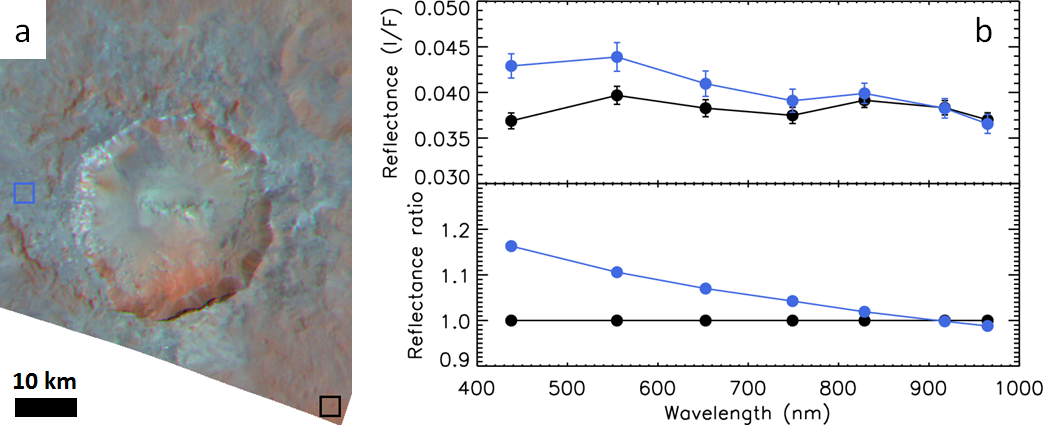}
\caption{Spectra of two selected regions around Haulani crater. ({\bf a}) Color composite of {\it HAMO} images (R, G, B) = (965, 555, 438)~nm = ({\bf 39655}, {\bf 39658}, {\bf 39652}) (colors saturated) showing the location of the two spectra shown in the plot. The blue and black squares represent ``blue terrain'' and the ``background''. ({\bf b}) Plots showing the reflectance spectra calculated as the average and standard deviation of the (projected) pixels in the square (top) and the average relative to the background
(bottom).}
\label{fig:Haulani_spectra}
\end{figure}

\begin{figure}
\centering
\includegraphics[width=\textwidth,angle=0]{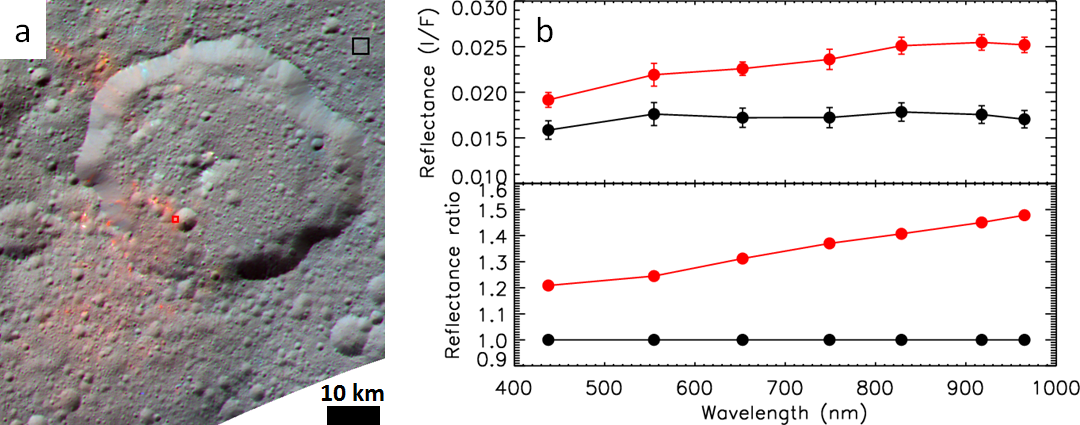}
\caption{Spectra of two selected regions around Ernutet crater. ({\bf a}) Color composite of {\it HAMO} images (R, G, B) = (965, 555, 438)~nm = ({\bf 39987}, {\bf 39990}, {\bf 39984}) (colors saturated) showing the location of the two spectra shown in the plot. The red and black squares representing ``red terrain'' and the ``background''. ({\bf b}) Plots showing the reflectance spectra calculated as the average and standard deviation of the (projected) pixels in the square (top) and the average relative to the background
(bottom).}
\label{fig:Ernutet}
\end{figure}

\begin{figure}
\centering
\includegraphics[width=\textwidth,angle=0]{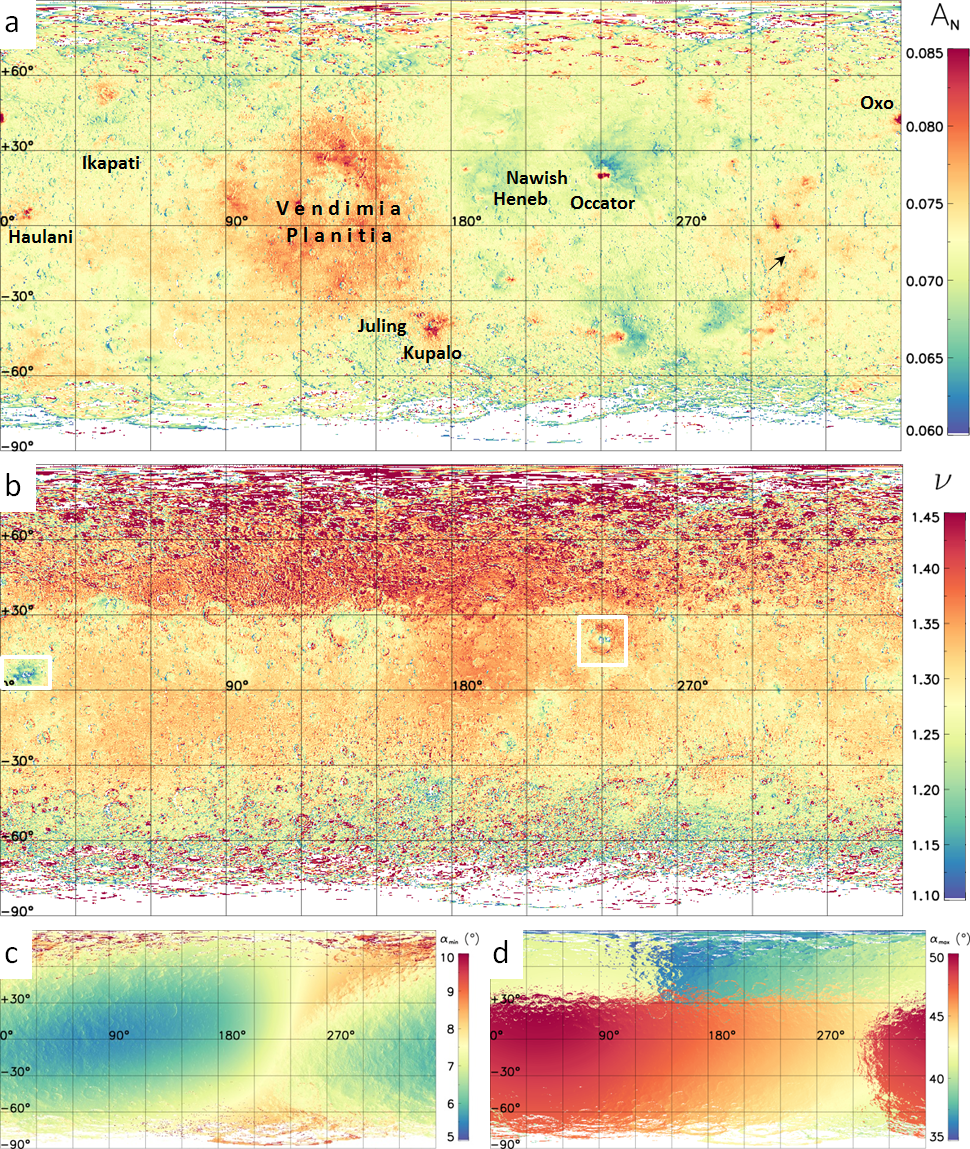}
\caption{Maps of the exponential model parameters amplitude $A_{\rm N}$ ({\bf a}) and slope $\nu$ ({\bf b}) as determined from {\it RC3} images, with the ({\bf c}) minimum and ({\bf d}) maximum phase angle of the observations. The white rectangles in (b) indicate areas selected for closer scrutiny in Figs.~\ref{fig:Haulani} and \ref{fig:Occator}.}
\label{fig:exp_model}
\end{figure}

The resulting global maps of the phase curve amplitude ($A_{\rm N}$) and slope ($\nu$) are shown in Fig.~\ref{fig:exp_model}. Note that the $\nu$ values shown are valid for $\alpha$ in radians. Before interpreting the maps we need to assess the reality of their features. The $A_{\rm N}$ map (Fig.~\ref{fig:exp_model}a) agrees well with the Akimov global albedo map in Fig.~\ref{fig:albedo}a. However, there is clear evidence for artifacts in the $\nu$ map (Fig.~\ref{fig:exp_model}b), with unusually high $\nu$ values prevalent above latitude $30^\circ$N. This boundary coincides with an observational limit. Despite the fact that the terrain above latitude $30^\circ$N was observed at very high phase angles early in RC3, Fig.~\ref{fig:exp_model}d shows that for the images included in the analysis, it was not observed in the $42^\circ$-$50^\circ$ phase angle range, skewing the exponential model fits. Whereas this does not seem to affect $A_{\rm N}$, the $\nu$ map can only be considered reliable for latitudes between $40^\circ$S and $30^\circ$N.

One would expect an inverse correlation between the phase curve amplitude $A_{\rm N}$ and slope $\nu$, given that brighter materials are often composed of more transparent particles. A terrain of bright material has weaker shadows, filled in by the lower particle opacity and the increased role of multiple scattering, and therefore its phase curve is shallower. At first glance, the phase curve amplitude and slope appear to be uncorrelated in Fig.~\ref{fig:exp_model}. For example, the large Vendimia Planitia feature is clearly recognizable in the $A_{\rm N}$ map but unremarkable in the $\nu$ map. Other instances of relatively bright terrain, like west and southwest of Vendimia Planitia and terrain in the $300^\circ$-$330^\circ$ longitude range, are not distinguished in the $\nu$ map. But then, for many other bright features an anti-correlation does exist: Small areas of relatively low $\nu$ (blue in Fig.~\ref{fig:exp_model}b) are bright craters like Oxo, Haulani, and Kupalo. Areas where $\nu$ is relatively high (red in Fig.~\ref{fig:exp_model}b) are around Heneb and Nawish craters and around Occator crater. Given the absence of a general anti-correlation between $A_{\rm N}$ and $\nu$, the physical condition of the surface probably plays a role in determining the phase curve slope. But it is not clear whether this is mostly the case for the regolith in Vendimia Planitia or that around Oxo, Haulani, and Kupalo. For the latter terrain, for which the (anti-)correlation does exist, it is also not clear whether the amplitude variations are large enough to explain the slope variations in the canonical way.

\begin{figure}
\centering
\includegraphics[width=\textwidth,angle=0]{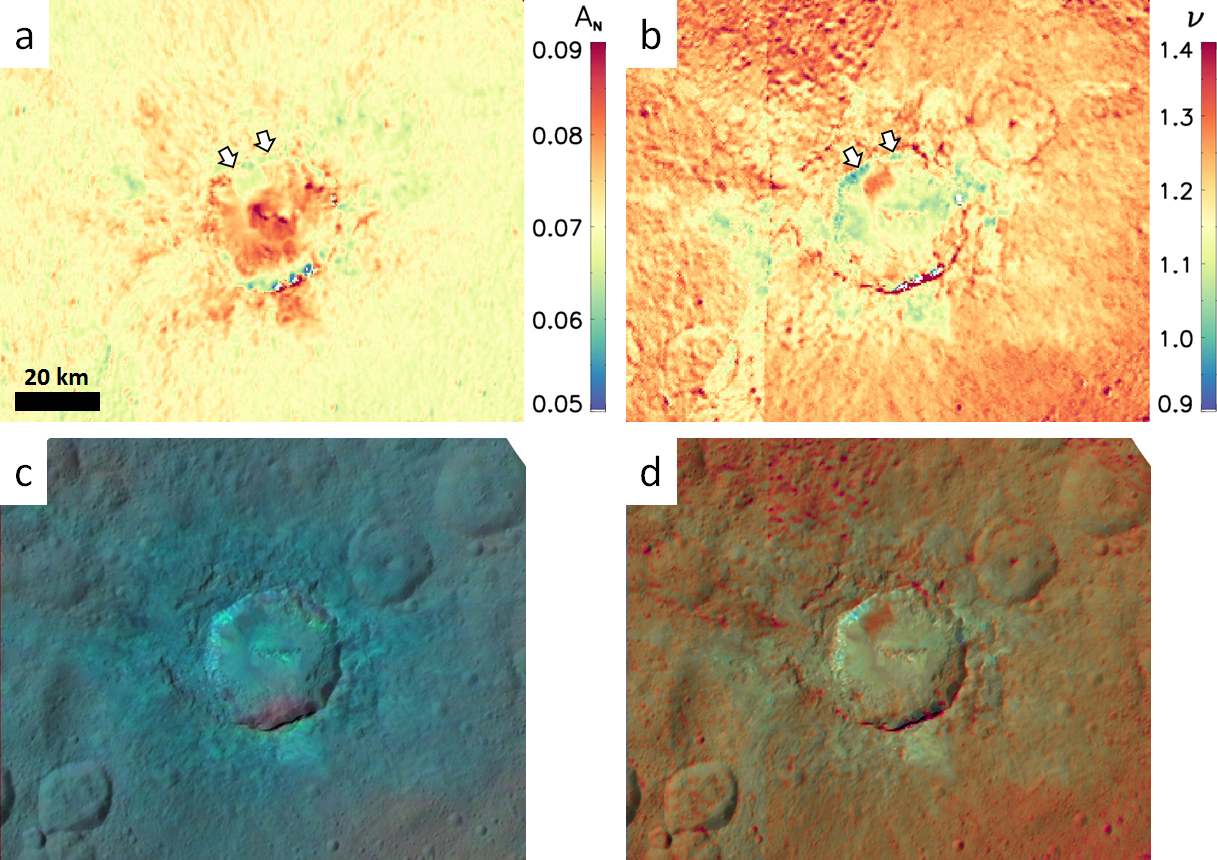}
\caption{Maps of Haulani crater derived from 14 {\it Survey} clear filter images showing the area from (lon, lat) = ($3^\circ, 0^\circ$) to ($18^\circ, 12^\circ$) in equirectangular projection. ({\bf a}) Exponential model amplitude parameter $A_{\rm N}$. ({\bf b}) Exponential model slope parameter $\nu$. Some artifacts related to phase angle coverage are visible. ({\bf c}) HAMO image {\bf 43383} with color from Fig.~\ref{fig:color}b with reduced color saturation. ({\bf d}) Same image with color from (b). The arrows in (a) and (b) point out crater walls with low $\nu$ and average $A_{\rm N}$.}
\label{fig:Haulani}
\end{figure}

To find clues at to what physical phenomena are responsible for the phase curve variations we look at several areas in detail, now fitting the exponential model to the higher resolution {\it Survey} images. The coverage of Haulani crater in {\it Survey} was restricted to only the $10^\circ$-$36^\circ$ range, so the derived phase curve amplitude and, especially, slope values may not directly be comparable to those in Fig.~\ref{fig:exp_model}. Figure~\ref{fig:Haulani} shows that Haulani has an interior and ejecta with relatively large $A_{\rm N}$ and low $\nu$. These terrains are the most bluish on Ceres. However, the (anti-)correlation between $A_{\rm N}$ and $\nu$ breaks down on the northwest crater walls, where we find low $\nu$ coupled to average $A_{\rm N}$ values. For asteroid Vesta, \citet{S13b} found a relation between phase curve slope and the physical properties of the regolith, with high $\nu$ values for the ejecta of young craters and low values for crater walls. They argued that this was related to the regolith roughness, most likely on very small scales: Crater walls of young craters are physically smooth whereas their ejecta are rough. For Haulani we also have evidence for smooth crater walls, but its ejecta are not rough. The low-$\nu$ terrain in and around the crater has a rather unusual morphology, exhibiting flow features \citep{K16}. Another low-$\nu$ area is Ikapati crater ($45^\circ$E, $33^\circ$N), which stands out as blue in the phase curve slope map (Fig.~\ref{fig:exp_model}b), but has only a slightly elevated normal albedo (Figs.~\ref{fig:albedo}a, \ref{fig:exp_model}a). Again, we point out that other terrains that have a normal albedo similar to that of Ikapati do not have low $\nu$ values. To the west and southwest of Ikapati we find what appears to be an outflow, either fluidized ejecta or of cryovolcanic origin \citep{K16}. The flow features appear physically smooth, even in the highest resolution {\it LAMO} images (35~m per pixel). The low $\nu$ associated with the Haulani and Ikapati flow features suggests that they are physically smooth below the {\it Survey} resolution of 0.4~km per pixel. As small scale roughness is thought to dominate the scattering behavior \citep{SH07}, this terrain may be physically smooth at the scale of the regolith.

\begin{figure}
\centering
\includegraphics[width=\textwidth,angle=0]{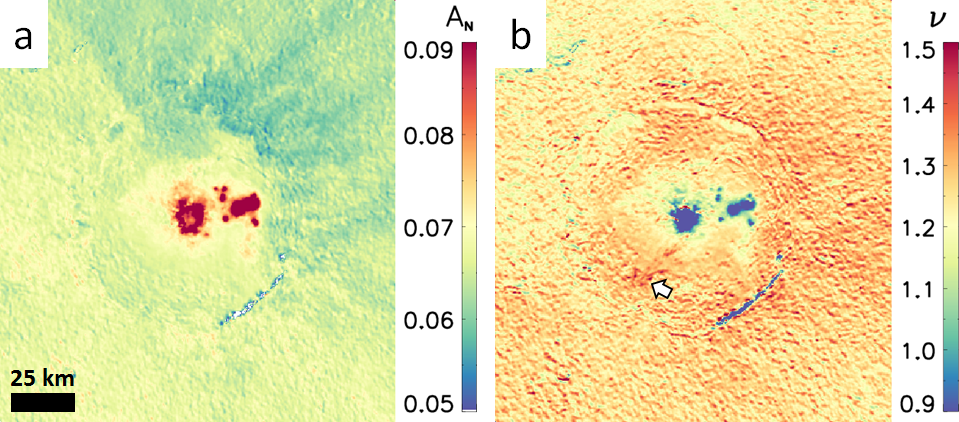}
\caption{Maps of Occator crater derived from 67 {\it Survey} clear filter images showing the area from (lon, lat) = ($230^\circ, 10^\circ$) to ($250^\circ, 30^\circ$) in equirectangular projection. ({\bf a}) Exponential model amplitude parameter $A_{\rm N}$. ({\bf b}) Exponential model slope parameter $\nu$. The interior of the three largest bright areas was overexposed ($A_{\rm N} \gg 0.09$). Parameter values on the southeast crater wall are not reliable due to the presence of shadows. The arrow points out the heavily cracked area mentioned in the text.}
\label{fig:Occator}
\end{figure}

The imaging coverage of Occator crater in {\it Survey} was good at a wide range of phase angles. We restrict the maximum phase angle to $50^\circ$ to allow a direct comparison with the maps in Fig.~\ref{fig:exp_model}. In the $A_{\rm N}$ and $\nu$ maps in Fig.~\ref{fig:Occator} the bright areas stand out. Due to their high albedo they have $A_{\rm N} \gg 0.09$ and $\nu \ll 0.9$, and are shown saturated in the maps. As they were routinely overexposed during {\it Survey}, the actual phase curve slope and amplitude are not known (we address the spectrophotometric properties of Cerealia Facula and the Vinalia Faculae in Sec.~\ref{sec:Occator}). Despite the ubiquitous presence of flow features \citep{K16}, the crater floor appears wrinkled and cracked in the images, which may explain the relatively high $\nu$. In particular, a small, heavily cracked area in the southwest corner of the crater, which may represent an emerging bright area, has elevated $\nu$ values that are likely artifacts due to the strong shadows of the crack network. The bright areas have high $A_{\rm N}$ and low $\nu$ values. The latter is to be expected if the bright areas are composed of particles that are more transparent than average (which is very likely). The ejecta northeast of the crater are relatively dark (low $A_{\rm N}$) and have increased $\nu$ values. It is not clear whether this is the consequence of physical roughness or the lower albedo of the material. We were able to find only a single instance of ejecta with an increased $\nu$ on the surface of Ceres: an area rich in boulders just beyond the southeast rim of the Juling crater ($167.5^\circ$E, $-36.4^\circ$S). Juling, a neighbor of Kupalo crater, is known for having an unusual red colored crater floor.

\citet{Ma16} proposed Vendimia Planitia to be a large, ancient impact basin on basis of the surface topography. The large, almost perfectly circular, albedo feature inside this area is clearly recognizable in Fig.~\ref{fig:albedo}, and was already seen in the HST maps of \citet{L06}. Its visible color and scattering properties are atypical for Ceres' surface (compare Figs.~\ref{fig:color} and \ref{fig:exp_model}). It is somewhat smaller than the basin outlined by \citet{Ma16}, and the albedo map may delineate the extent of the impact basin more clearly than the topography map. Two other basin candidates outlined by the authors (Planitia~B and C) are not recognized in our maps. If this feature is indeed an ancient impact basin, its color, albedo, and lack of a clear rim make it akin to the palimpsests on Ganymede and Callisto \citep{S79}. On asteroid Vesta, a large area with unusual spectral properties is associated with the large Veneneia impact basin. The dark, volatile-rich material making up this terrain is thought to derive from the impactor \citep{McC12,R12b}. It is worth investigating whether the Vendimia Planitia material is of exogenous or endogenous origin.

\subsection{Occator bright areas}
\label{sec:Occator}

\begin{figure}
\centering
\includegraphics[width=8cm,angle=0]{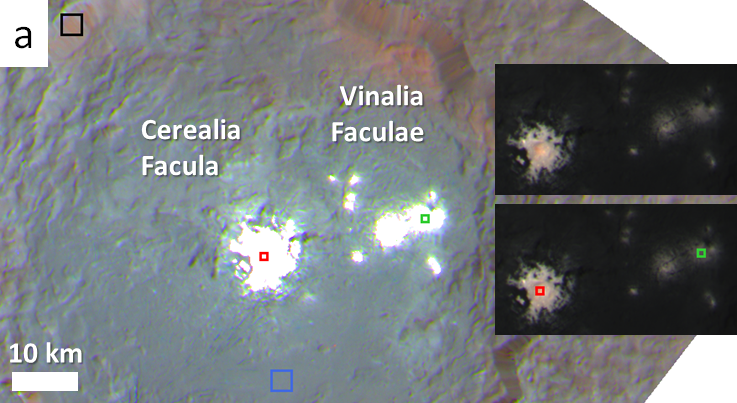}
\includegraphics[width=9cm,angle=0]{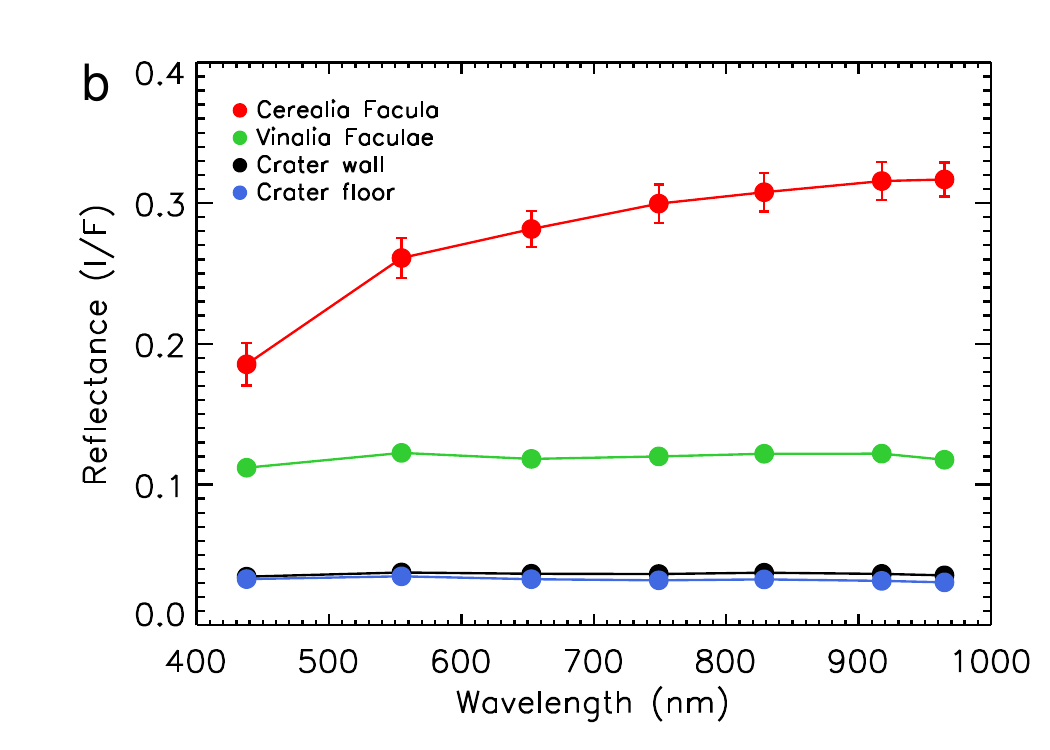}
\includegraphics[width=9cm,angle=0]{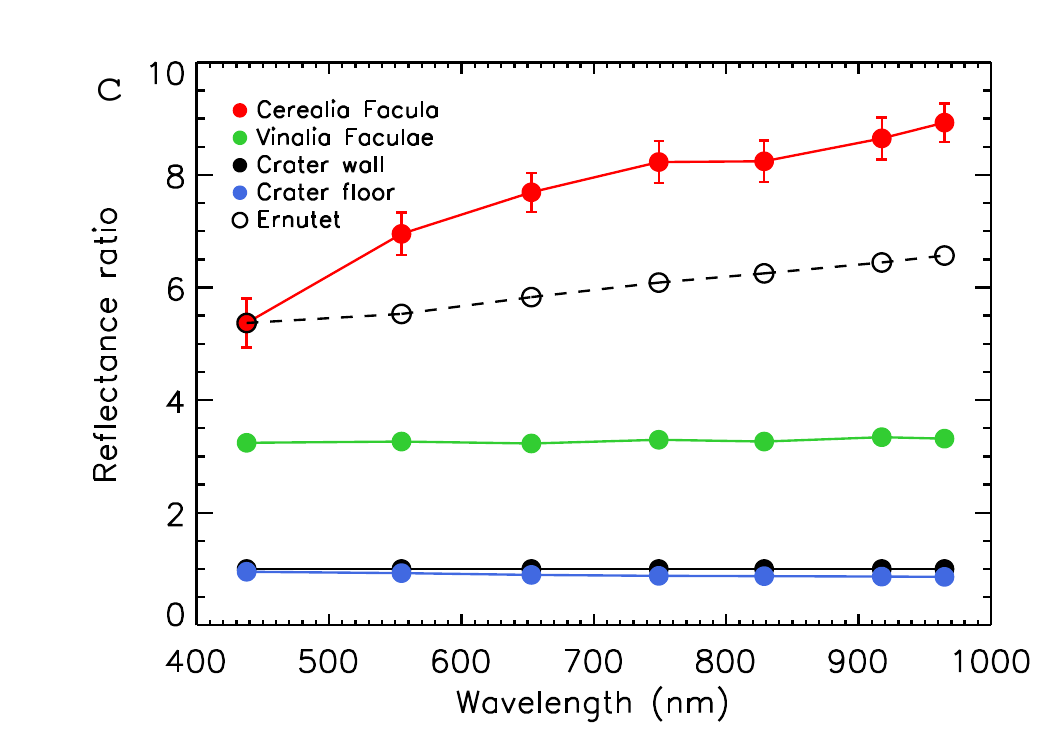}
\caption{Spectra of four selected regions inside the Occator crater. ({\bf a}) Location of the regions, marked as red, green, blue, and black squares, on a color composite of {\it HAMO} images (R, G, B) = (965, 555, 438)~nm = ({\bf 40748}, {\bf 40751}, {\bf 40745}) (colors saturated). The black square represents the ``background'' location on the crater wall. The large image shows the bright areas saturated. The two insets show the areas at their full brightness range, one with locations marked. ({\bf b}) Reflectance spectra calculated as the average and standard deviation of the projected pixels in the square. ({\bf c}) Spectra relative to the background spectrum. The Ernutet ratio spectrum from Fig.~\ref{fig:Ernutet} (dashed line) is included for reference, and is scaled to match the Cerealia Facula ratio spectrum at 430~nm.}
\label{fig:Occator_color}
\end{figure}

The spectrophotometric properties of the bright areas inside the Occator crater appear to be unique to Ceres. As these areas were often overexposed during {\it RC3}, they were not dealt with correctly in the previous sections and we need to take special care. To retrieve their spectral properties we analyze {\it HAMO} images {\bf 40745}-{\bf 40751} with phase angle around $30^\circ$, which we corrected for stray light according to \citet{S14a}. In Fig.~\ref{fig:Occator_color} we show how the reflectance spectra of four locations inside Occator compare: the central bright area (Cerealia Facula), an adjacent secondary bright area (one of the Vinalia Faculae), the crater floor, and a crater wall in the northwest corner. Of all locations, the reflectance spectrum of the crater wall is most like that of Ceres average, and we adopt it as our ``background'' spectrum. The crater floor is blue (see also Fig.~\ref{fig:color}), and compared to the background it has the typical negative spectral slope that we associate with Ceres blue material, the reflectance ratio dropping by 10\% over the full wavelength range. Cerealia Facula is the brightest area on Ceres and also the reddest. Its spectral slope is much larger than that of the red material near Ernutet crater. Its reflectance (Fig.~\ref{fig:Occator_color}b) increases towards larger wavelengths over the full FC wavelength range, most rapidly at the lower end of the range, confirming \citet{N15}. When compared to the background (ratio spectrum in Fig.~\ref{fig:Occator_color}c) a possible absorption feature around 830~nm becomes apparent. However, as Cerealia Facula is so much brighter than Ceres average, the \citet{S14a} stray light correction is not accurate for this terrain, and artifacts on the order of the depth of this feature are to be expected. Surprisingly, the visible spectrum of the neighboring Vinalia Faculae (green points) is very different from that of Cerealia Facula (red points); it is completely neutral with respect to the, much darker, background spectrum. If two terrains have the same spectral shape but different brightness, they may have the same composition but different average particle size, with the brighter terrain having smaller particles. But the Vinalia Faculae are so much brighter than Ceres average (3-4 times at this phase angle), that they most likely contain a significant fraction of more transparent particles that are spectrally neutral in the visible, perhaps a salt. The different spectral slope (red versus neutral) suggests a difference in composition between the bright areas. \citet{dS16} reported both Cerealia Facula and Vinalia Faculae to harbor sodium carbonate, interpreted as the solid crystallized residue of a brine. They did not address a possible compositional difference, but an investigation of the composition of the bright areas using VIR data is ongoing \citep{Pa16}.

\begin{figure}
\centering
\includegraphics[width=8cm,angle=0]{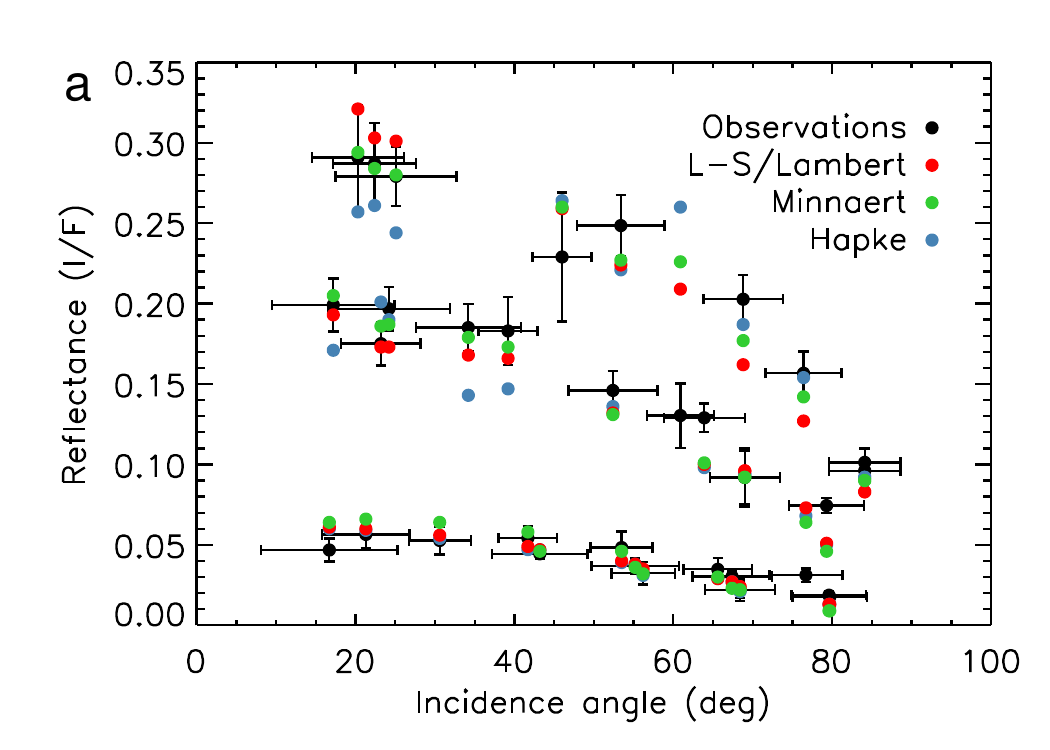}
\includegraphics[width=8cm,angle=0]{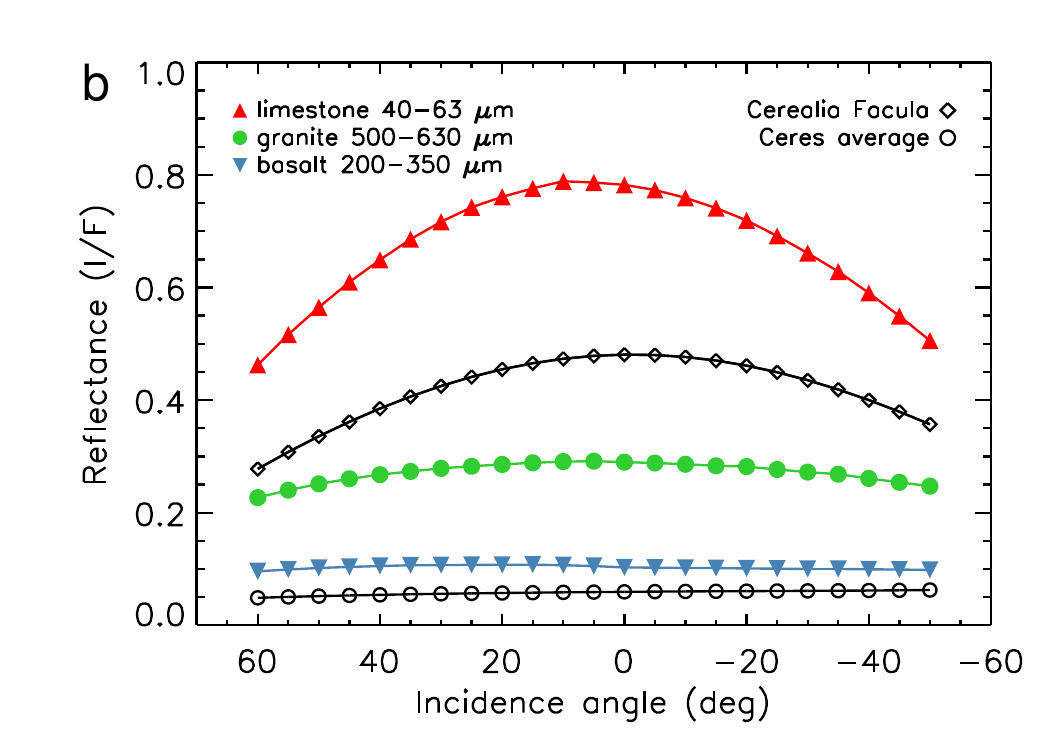}
\includegraphics[width=8cm,angle=0]{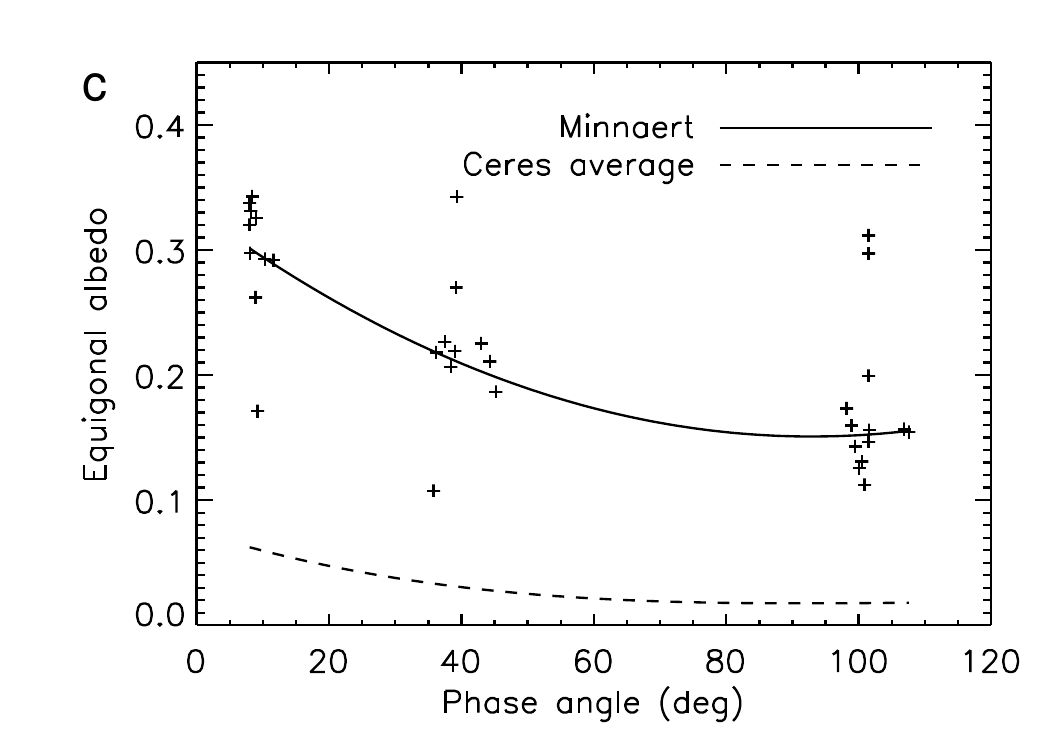}
\includegraphics[width=8cm,angle=0]{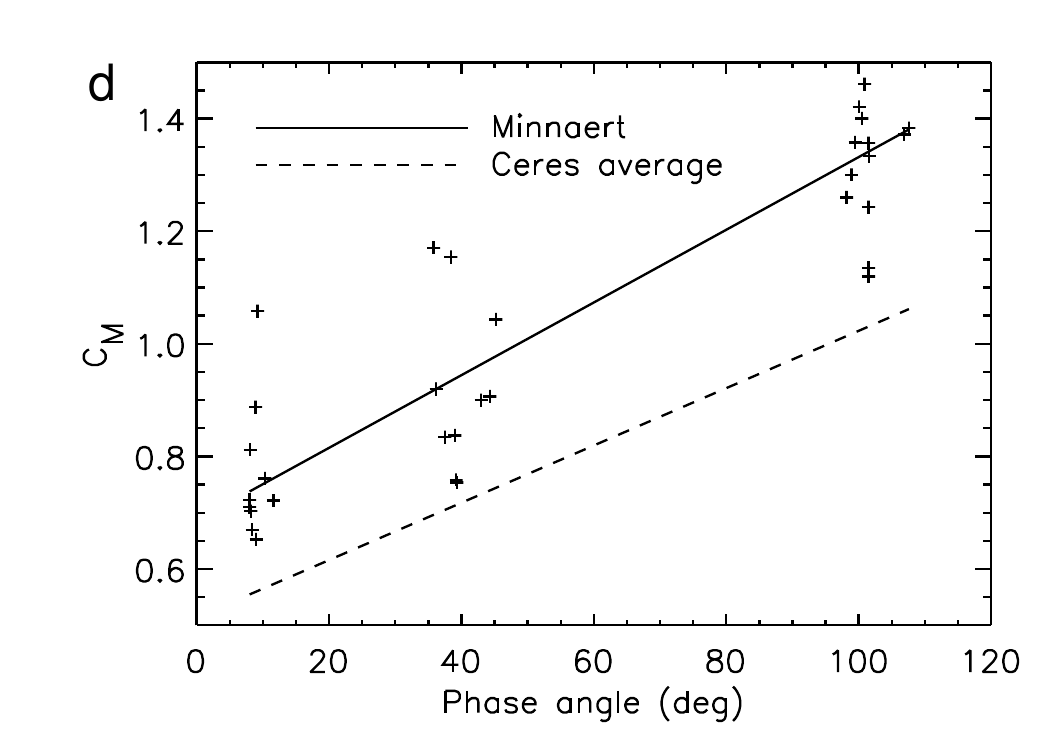}
\includegraphics[width=8cm,angle=0]{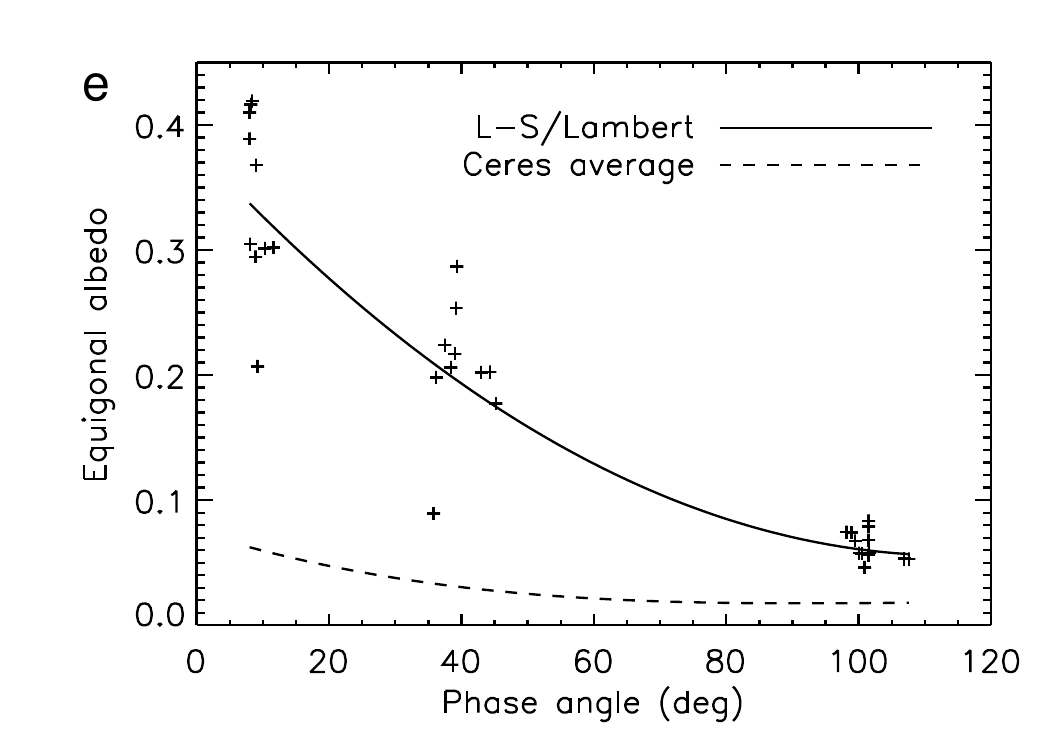}
\includegraphics[width=8cm,angle=0]{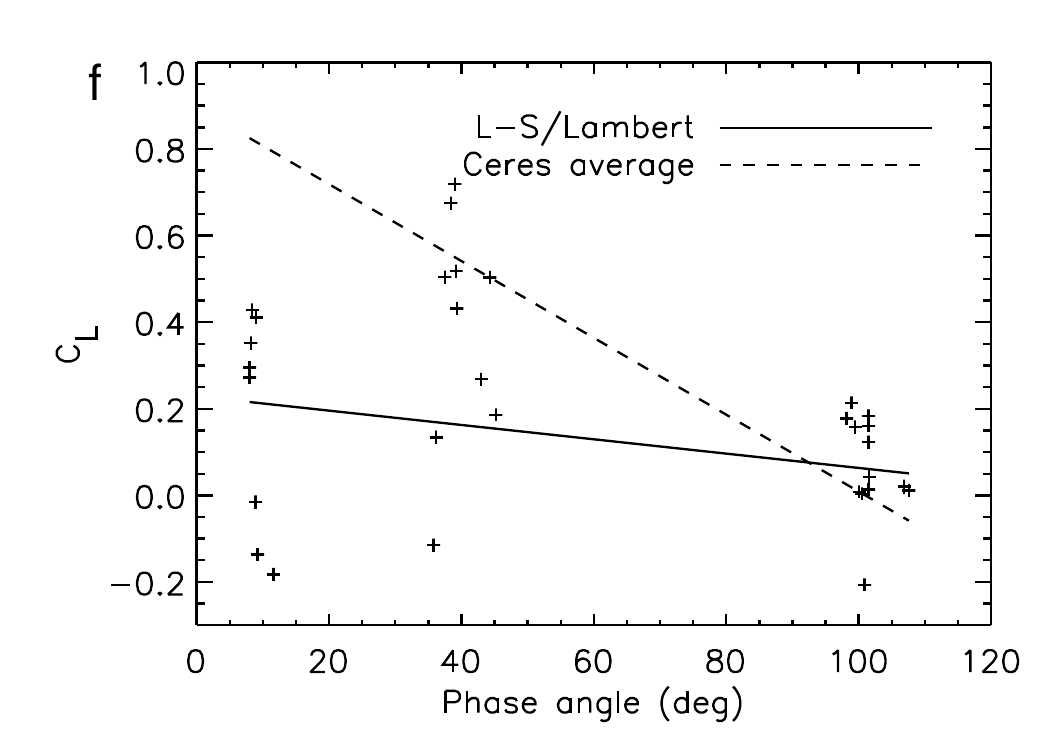}
\caption{Scattering properties of Cerealia Facula from {\it RC3} clear filter images with $(\iota, \epsilon) <85^\circ$. ({\bf a})~The observed reflectance (average of 6~pixels with standard deviation) compared to the Lommel-Seeliger/Lambert and Hapke model reflectances. ({\bf b})~Scattering properties at $10^\circ$ phase angle, calculated using the L-S/Lam model results, compared to experimental data from \citet{S14b}. ({\bf c})~The Minnaert model phase function ($A_{\rm eq}$) compared to that of Ceres average. The plus symbols are the $A_{\rm eq}$ of the data when adopting the $c_{\rm L}$ model in (d). ({\bf d})~The Minnaert disk function parameter compared to that for average Ceres. The plus symbols are the $c_{\rm M}$ of the data when adopting the $A_{\rm eq}$ model in (c). ({\bf e})~The L-S/Lam model phase function ($A_{\rm eq}$) of Cerealia Facula compared to that of Ceres average. The plus symbols are the $A_{\rm eq}$ of the data when adopting the $c_{\rm L}$ model in (f). ({\bf f})~The L-S/Lam disk function parameter compared to that for average Ceres. The plus symbols are the $c_{\rm L}$ of the data when adopting the $A_{\rm eq}$ model in (e).}
\label{fig:Occator_photometry}
\end{figure}

To retrieve the scattering properties of Cerealia Facula, we identified 111 {\it RC3} clear filter images of Occator crater of which 44 were correctly exposed, spanning a range of phase angles from $8^\circ$ to $109^\circ$. We projected the latter images to the same equirectangular projection and discarded 4~images for which the projection was grossly inaccurate. We restricted the photometric angles to $(\iota, \epsilon) <85^\circ$, leaving a total of 35~images. An analysis of {\it HAMO} images confirms that the terrain was resolved during {\it RC3}, albeit barely, with a size of about $3 \times 3$~pixels (the spatial resolution is 1.3~km per pixel). We model the reflectance as the average of 6 projected pixels that cover Cerealia Facula, using our suite of photometric models. When ranked according to phase angle the data can be divided into three relatively confined groups: $8^\circ < \alpha < 12^\circ$, $35^\circ < \alpha < 45^\circ$, and $95^\circ < \alpha < 110^\circ$. We compare the modeled reflectances to the observations in Fig.~\ref{fig:Occator_photometry}a. We show the reflectance as function of incidence angle to emphasize the strong drop in reflectance towards $\iota = 90^\circ$ (``limb darkening''). The three phase angle groups can be clearly distinguished. The Minnaert model provides the best fit to the data. The next best is the Lommel-Seeliger/Lambert combination, and then Hapke, but the differences are small. The Akimov models fail to reproduce the observations. The best fit Type~I model parameters are given in Tables~\ref{tab:Occator_c2_coef} and \ref{tab:Occator_phase_coef_clear}. For the phase function we employ a second-order polynomial, as the scatter in the data do not allow us to reliably determine higher order coefficients. The Minnaert phase function ($A_{\rm eq}$) and disk function parameter ($c_{\rm M}$) are shown in Figs.~\ref{fig:Occator_photometry}c and d, together with those of Ceres average. Despite the good fit to the reflectance data, the scatter of the data around these two model curves is quite large. As expected, the phase function is higher than that of Ceres average, but appears unremarkable otherwise. The disk function parameter is also larger. At low phase angle, $c_{\rm M}$ is about 0.70, larger than the 0.55 typical for asteroids (Fig.~\ref{fig:photmod_parameters}c.) but identical to the value derived by \citet{BV83} for Jupiter's moon Europa from Voyager images, a body of similarly high albedo. The L-S/Lambert phase function ($A_{\rm eq}$) and disk function parameter ($c_{\rm L}$) are shown in Figs.~\ref{fig:Occator_photometry}e and f. Again, we note that despite the good fit to the reflectance data, the scatter of the data around the two model curves is quite large. The phase function appears unusually steep for a surface of high albedo, in contrast to our finding in the previous section, that bright areas have shallow phase curves. The L-S/Lambert phase curve is steeper than the Minnaert phase curve because the Minnaert disk function has larger values than the L-S/Lambert disk function at high phase angles. The contribution of the Lambert term at small phase angles is much larger for the bright area than for Ceres average ($c_{\rm L}$ is small), consistent with the \citet{L16} finding that the bright areas scatter light more isotropically. The Hapke model fits the data about 10\% worse than the Minnaert model, and has best-fit parameters $w = 0.78$, $B_{\rm S0} = 0$, $\bar{\theta} = 59^\circ$, $b = 0.23$, and $c = -1$. None of these parameters were fixed; the fitting algorithm converged to $B_{\rm S0} = 0$ even though this parameter was non-zero for Ceres average with essentially the same phase angle coverage. The very high value of the photometric roughness ($\bar{\theta}$), is among the highest ever retrieved for a Solar System object, confirms that the phase curve is unusually steep. This is reinforced by the Henyey-Greenstein $c$-parameter value of $-1$, which, in the framework of the Hapke model, implies that the bright material particles are exclusively backscattering. The best-fit Hapke parameters are highly unusual, even for an object of high albedo. For example, \citet{D91} determined the following parameters for the leading side of Jupiter's moon Europa from a combination of telescopic observations and Voyager images: $w = 0.96$, $B_{\rm S0} = 0.5$, $h_{\rm S} = 0.0016$, $\bar{\theta} = 10^\circ$, $b = -0.43$, and $c = 0.77$\footnote{Note that the authors used a different definition of the $c$-parameter.}, where the phase angle range of the data was $0^\circ$-$120^\circ$. In fact, their photometric roughness value is one of the lowest determined for any Solar System body. \citet{K04} reported a Hapke roughness of $\bar{\theta} = 60^\circ$ for very dark terrain on comet Borrelly, and interpreted this large value in terms of physical roughness on a large scale (on the order of hundred meters).

If we analyze the scattering properties of Cerealia Facula from the perspective of the Type~I models, we conclude that the disk function is regular, but the phase function is unusual, at least at the relatively low spatial resolution of the {\it RC3} images (1.3~km per pixel). The contribution of the Lambertian scattering term to the disk function is larger, as expected for a particulate surface deposit of high albedo (see below for a quantitative estimate), for which the constituent particles are relatively transparent, leading to a preponderance of multiple scattering (e.g.\ \citealt{V78}). We demonstrate this by comparing the scattering behavior of the bright area and Ceres average to that of particulate surfaces composed of three common Earth rocks with different albedo \citep{S14b}. Figure~\ref{fig:Occator_photometry}d shows the reflectance for a fixed phase angle of $10^\circ$ as a function of incidence angle to emphasize the limb darkening behavior. The scattering behavior of the experimental surfaces are determined by the physical properties of the constituent particles, like the optical constants and average size. The albedo and limb darkening of Ceres average are closest to those of a surface of dark basalt particles, whereas the albedo and limb darkening of the bright area are intermediate to those of the granite and limestone particulate surfaces (the latter appears white to the eye). The phase curve of Cerealia Facula is very steep. This is unexpected for a high albedo surface, but confirmed by VIR measurements \citep{L17}. It is exactly the aforementioned preponderance of multiple scattering by and in the semi-transparent constituent particles that should lead to a combination of high albedo and shallow phase curve. Unresolved topography is probably at least partly responsible for the steep phase curve. Cerealia Facula consists of a fractured dome inside a depression \citep{SB16}, and at higher phase angles parts of the Facula floor may be obscured from view. Other factors that would contribute are an extremely rough nature of the surface and bright, yet opaque, constituent particles. The latter is consistent with a large size. As they are probably salt particles \citep{dS16}, their low hardness would allow the space environment to rapidly disintegrate them. This suggest Cerealia Facula is a young feature, consistent with its high albedo and the young age for Occator \citep{N15}. \citet{dS16} pointed out similarities with regions of outgassing on the south pole of Saturn's moon Enceladus. Indeed, a possible analog may be the young ``tiger stripes'' terrain, where the particles are comparatively large \citep{J08} and the Hapke photometric roughness parameter appears to be higher than in the surrounding plains \citep{A12}\footnote{Unfortunately, the authors did not report $\bar{\theta}$ values.}. Also the rough patches on Borrelly are suspected to be regions of outgassing \citep{K04}. We can conclude that the unusual photometric properties of Cerealia Facula are not inconsistent with it being a (former) site of volatile release. More insight into the nature of the bright terrain may be gained from spectrophotometric experiments with analog materials.

What is the normal visual albedo of Cerealia Facula? \citet{N15} quoted an absolute reflectance of 0.25, but did not define this quantity. \citet{L16} calculated the Bond albedo for the bright area to be 7.1 times higher than Ceres average, but did not provide an estimate for the normal visual albedo. The albedo of the area depends on the spatial resolution. The wide phase angle coverage of the {\it RC3} images allows for the most reliable estimate possible, but at the rather low resolution of 1.3~km per pixel. A number of assumptions are necessary. The visual reflectance is defined at a mean wavelength of 540~nm in the Johnson UVB photometric system, but the FC clear filter responsivity peaks at around 700~nm \citep{Si11}. Given that the bright material is very red (Fig.~\ref{fig:Occator_color}), we will overestimate the visual reflectance by about 10\% from the clear filter reflectance (factor 0.9). Furthermore, the area is hardly resolved in {\it RC3} images, and the point spread function (PSF) distributes signal from the pixel covering the very bright core of the area to neighboring pixels. To resolve this issue we apply the \citet{S08} method to deconvolve image {\bf 37114} with the clear filter PSF constructed from Saturn images\footnote{The diameter of Saturn in these images was 1.0 pixel.} acquired during the {\it Initial Checkout Operations} \citep{S13a}. We find that deconvolution consistently brightens the brightest pixel in the bright area by a factor 1.5. The exact value of this factor depends on the assumed noise characteristics of the image, but proves to be robust for settings that result in well-sharpened images without noticeable ringing. To account for the unknown strength of the opposition effect of the bright area we use the findings of \citet{BS00} concerning the strength of the opposition effect for asteroids of different type. Based on the albedo, the bright area should scatter most similarly to E-type asteroids, which have a ratio of intensity at phase angles $0.3^\circ$ and $5.0^\circ$ of about 1.3. Extrapolating Eq.~19 to $\alpha = 5^\circ$ gives 0.34. Our best estimate for the visual normal albedo of the bright area at a resolution of 1.3~km per pixel is then $0.9 \times 1.5 \times 1.3 \times 0.34 = 0.60$, or six times Ceres average. The largest uncertainty is associated with the unknown strength of the opposition effect, which should make our estimate accurate to about 20\%, or $0.6 \pm 0.1$.

\begin{figure}
\centering
\includegraphics[width=12cm,angle=0]{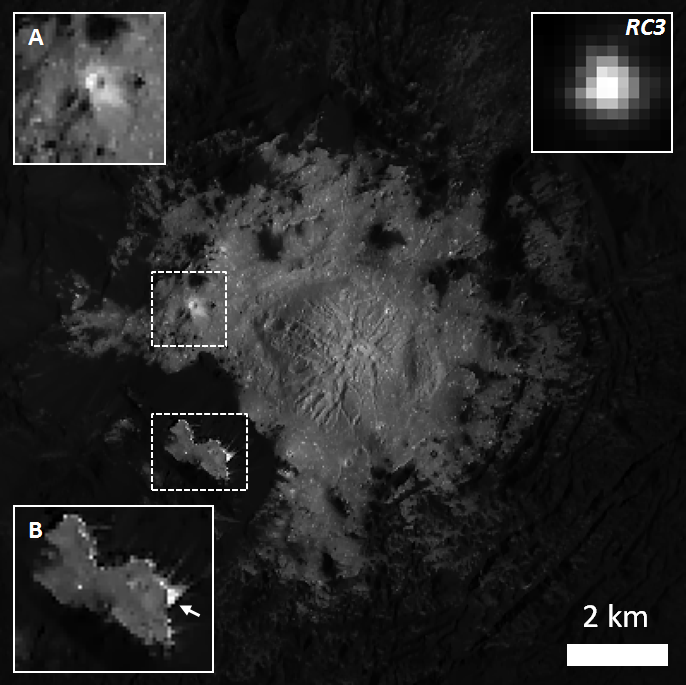}
\caption{Cerealia Facula in Occator in {\it LAMO} image {\bf 57273} ($\alpha = 47^\circ$), deconvolved with the clear filter PSF. The two terrains surrounded by a dashed line are shown enlarged in insets at the left top and bottom. Inset~A shows a small impact crater. Inset~B shows an area harboring terrain with the highest reflectance (arrow). The top right inset shows the central bright area as it appears in {\it RC3} image {\bf 36979} ($\alpha = 10^\circ$), displayed on a smaller scale and different orientation. For each image (inset), black and white are set to minimum and maximum reflectance, respectively.}
\label{fig:central_spot}
\end{figure}

The highest resolution images acquired in {\it LAMO} (35~m per pixel) reveal many small features on the surface of Cerealia Facula that are even brighter than their surroundings (Fig.~\ref{fig:central_spot}). The geology of this terrain is discussed by \citet{SM16,SB16}, and here we concentrate on its photometric properties. The brightest terrain, associated with the group of pixels indicated in the figure, has a reflectance that is about 2.5 times the average of the entire bright area at a phase angle of $47^\circ$. The presence of streaks in the vicinity of this terrain (Fig.~\ref{fig:central_spot}, inset~B) suggests that fresh material was exposed by mass wasting. Another group of bright pixels is associated with a small impact crater (Fig.~\ref{fig:central_spot}, inset~A). We propose that these very small, much brighter terrains represent fresh material, exposed by either impacts or mass wasting. The phase curve of these very bright terrains is expected be shallower than that of the average, and we therefore estimate the normal visual albedo to be on the order of unity, comparable to the geometric albedo of icy moons like Dione and Rhea \citep{V07}. The darkening of the bright material is a relatively rapid process, given that the age of Occator is estimated at 78~Ma \citep{N15}. It may be autonomous, for example through the release of volatiles, or a consequence of the exposure to the space environment (a form of space weathering). We also note large fractures in the center of Cerealia Facula in Fig.~\ref{fig:central_spot} described by \citet{SB16}, which may, to some degree, explain the high photometric roughness determined at a lower spatial resolution.

\section{The nature of blue material}
\label{sec:blue_material}

In Sec.~\ref{sec:color_maps} we presented global color maps that showed the wide-spread distribution of blue material over the surface, concentrated around geologically young features. What is the nature of this blue material? We can exclude water ice as an explanation, as it is exposed in only a few, isolated patches \citep{Co16}. \citet{J16} considered explanations for the blue color in terms of Rayleigh scattering and particle size but found them both wanting, and suggested compositional variations to be responsible. The blue spectral slope in the visible wavelength range is similar to that of several anhydrous salts \citep{Bi14,H14}. Below we suggest an alternative explanation that does not require an additional regolith component. \citet{St16} evaluate the different possibilities including data from the VIR spectrometer. The nature of blue material cannot be established with FC data alone, and VIR observations will provide additional constraints.

The blue spectral slope may be an indicator for desiccation. Ammoniated phyllosilicates are widespread \citep{dS15,A16}, and sub-surface water ice appears to be abundant \citep{PY16,SH16}. In a laboratory experiment, \citet{P16} mixed water ice and a smectite phyllosilicate (montmorillonite) into particles (``intra-mixture''). The spectrum of the smectite component has a positive visible spectral slope and several absorption bands in the near-IR. They subsequently sublimated the water ice under low temperature and pressure conditions ($<-70^\circ$C, $10^{-5}$ mbar). After sublimation, the smectite residue had a foam-like structure that retained the initial shape of the particles. Its reflectance spectrum had a negative slope over the full visible and near-IR wavelength range that was almost featureless in the near-IR; the smectite bands were reduced and no water ice bands were evident. The residual foam had a highly complex structure on the micrometer scale, whose scattering properties would be difficult to model. The authors speculated this structure to be responsible for the spectral bluing through a mechanism discussed by \citet{B14}, who attempted to provide a theoretical foundation to the idea that Rayleigh scattering can explain instances of bluing observed in the Solar System. Specifically, \citet{C08,C12} proposed that Rayleigh scattering by sub-micrometer particles embedded in a transparent matrix of water ice leads to a non-linear increase of the reflectance of the Saturnian moons Dione and Iapetus from about 0.7~\textmu m towards 0.5~\textmu m wavelength. Rayleigh scattering was also suggested to be responsible for the non-linear increase of the reflectance towards 0.5~\textmu m observed for Lunar regolith \citep{C10}. The model of sub-micrometer particles embedded in a transparent matrix may not apply for the Moon, but instead Rayleigh scattering may be associated with sub-micrometer particles on the surface of other, larger particles. The \citet{B14} modeling effort evaluates the consequences of particle size and particle size distribution for spectral bluing while assuming Mie and Rayleigh-type scattering. The author was forced to adopt several simplifications. For example, he assumed spherical particles and did not account for the close packing of particles in a regolith. As \citet{B14} writes, his work represents a first step towards understanding Rayleigh scattering in planetary regoliths. While it does provide the typical particle size responsible for the bluing, it is unclear whether the \citet{B14} model describes the scattering properties of the \citet{P16} phyllosilicate foam.

We note that bluing on Ceres is qualitatively different from that on Dione and Iapetus. It does not constitute a strong increase of the reflectance from 0.4~\textmu m towards 0.5~\textmu m wavelength, but rather a gradual increase towards the blue over the entire 0.4-1.0~\textmu m wavelength range of the Framing Camera. But the spectral slope of blue material on Ceres is similar to that of the desiccated phyllosilicate/ice particles seen by \citet{P16}. We therefore propose the following scenario to explain the blue color of fresh craters on Ceres. During a major impact, water ice from below the surface is mixed with silicates that are widespread on the surface, among which are phyllosilicates that result from extensive internal processes \citep{dS15,A16}. While the impact ejecta are initially rich in water ice, rapid sublimation leads to the formation of a foam-like residue. Water ice is very unstable on the surface, and a decimeter-sized desiccated layer would be expected to form on a time scale of hundreds of years \citep{F16}. The residue is blue like in the \citet{P16} experiment, perhaps as a consequence of Rayleigh-type scattering by sub-micrometer particles or structures (filaments, edges) in a complex foam-like structure. Over time, the ejecta color returns to the neutral background, possibly through gradual destruction of the foam in a process that can be regarded as a form of space weathering. The time scale of this process has been retrieved by age determination through crater counting \citep{SK16}. The validity of our hypothesis may be assessed with additional experiments under conditions tuned to the Ceres environment and a deeper investigation of spectral signatures in the near-infrared.

\section{No evidence for haze on Ceres}
\label{sec:haze}

A surprising finding was reported by \citet{N15}: a haze forms inside the Occator crater in diurnal fashion. FC images were shown in their Fig.~4 with stretched  contrast to show the haze. Their extended material Fig.~7 shows excess signal that peaks right in between Cerealia Facula and the Vinalia Faculae. The discovery of haze on such a small world would have important implications, but there are several red flags associated with this finding. First, its presence is inferred from an intensity profile and suggestive stretching of the image contrast; we do not see the haze unambiguously like we can see the layers around Titan and Pluto. Second, several of the involved images are considerably enlarged; as pointed out in Sec.~\ref{sec:Occator}, Cerealia Facula is only about 3~pixels across. The consequences of camera optical characteristics that become important for targets of such small size are not discussed. Third, the phase angle of the observation in their Fig.~7 ($43^\circ$) is not good at all for detecting a haze. Small particles, especially those much larger than the wavelength of light, are known to be strongly forward scattering\footnote{Corresponds to phase angles close to $180^\circ$.} (e.g.\ \citealt{W97}), as opposed to the more isotropic Rayleigh scattering by particles much smaller than the wavelength of light. Fourth, the authors did not provide a physical explanation of how an optically thick haze could form, and why it would be akin to a ground fog rather than expanding in plume-like fashion into the vacuum of space.

\begin{figure}
\centering
\includegraphics[width=6cm,angle=0]{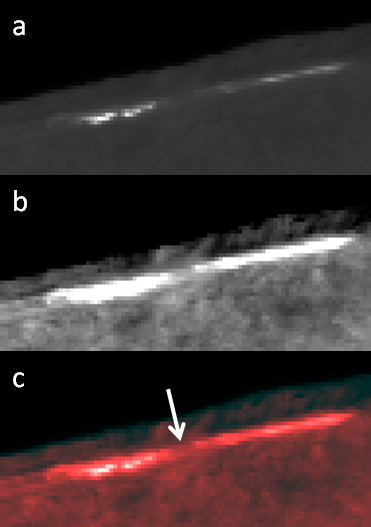}
\caption{Our assessment of {\it Survey} image {\bf 37570}, which purportedly shows the haze. {\bf a}: The Occator crater at the full brightness range. The bright areas are seen at an oblique angle. {\bf b}: The same view with the contrast enhanced as in Fig.~4c in \citet{N15}. {\bf (c)} Color composite with (b) in the red channel and (a) in both the blue and green channels, showing the bright areas located in the center of the ``haze''. Note that the area in between Cerealia Facula and the Vinalia Faculae (arrow), where \citet{N15} report the haze is most dense, has a similar brightness as the surroundings. If the white pixels in (b) are supposed to represent haze, then there is clearly no haze between the areas.}
\label{fig:haze_stretch}
\end{figure}

The evidence presented by \citet{N15} consists of two parts: (1)~several images in their Fig.~4 that purport to show the haze directly and (2)~the intensity profiles in their Fig.~7 of the extended material that aim to show that the haze density is maximum in between the bright areas. We list all images involved in Table~\ref{tab:haze_images} and calculated the photometric angles for the location in between the areas. We used an ellipsoid shape model for the calculation, as we are interested in these angles from the perspective of a free-floating haze and not the local surface. First, we evaluate the images in Fig.~4 of \citet{N15} that had their contrast stretched. It is not clear what these images represent. Are the white pixels haze? But the image in their Fig.~4a also has white pixels in places other than Occator crater. Their Fig.~4c shows a side view of the haze, showing it to be akin to a ground fog. We identified this image as {\it Survey C1} image {\bf 37570}. Figure~\ref{fig:haze_stretch} shows it at its full brightness range, revealing the bright areas at an oblique angle. We also stretched the image in a similar way as in their Fig.~4a. The composite image (c) shows that the brightest pixels are located exactly in the center of the white haze. This is what one would expect if the signal of those pixels were broadened by the PSF. It is not what we would expect if particles were erupting from the bright terrain, in which case we would expect a concentration of signal over the areas in the form of plumes. Earlier we showed that the surface surrounding the bright areas is slightly brighter than Ceres average (Fig.~\ref{fig:Occator}a). Such terrain shows up as white when applying a contrast stretch. Note that whereas in \citet{N15} the haze is most dense in between the areas (arrow in Fig.~\ref{fig:haze_stretch}), white pixels are absent in this location.

\begin{figure}
\centering
\includegraphics[height=5.5cm,angle=0]{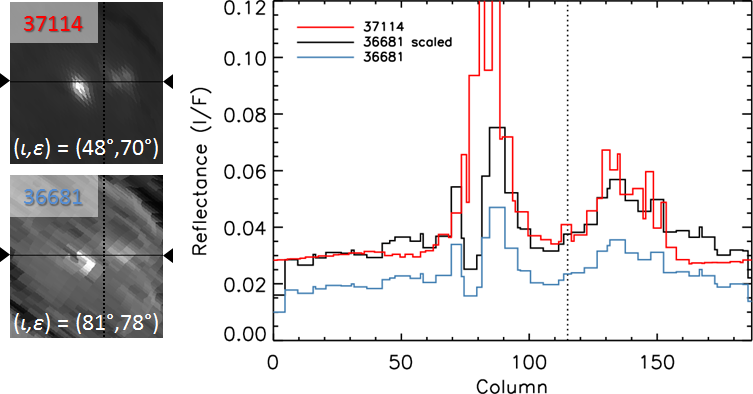}
\caption{Searching for haze in the {\it RC3} clear filter images {\bf 37114} and {\bf 36681} used in Fig.~7 of the extended material in \citet{N15}. {\bf Left}: Crops of the two images in identical projection that show the bright areas inside Occator, with photometric angles indicated. We investigate the reflectance profile along the drawn horizontal line. The intersection with the dotted vertical line identifies the location where \citet{N15} report the haze is most dense. {\bf Right}: Reflectance profiles for the two images. The black curve is the blue curve scaled to match the red curve for the terrain outside the bright areas (columns 0-40).}
\label{fig:haze_RC3}
\end{figure}

Now we evaluate the intensity profiles in Fig.~7 of the \citet{N15} extended material. The {\it noon} profile across the bright areas apparently shows haze in between them (``parabolic fit of haze''), while the {\it sunset} profile does not. This is explained as a diurnal variation, with the haze density peaking at noon when the insolation is maximum. The {\it noon} profile corresponds to {\it RC3} image {\bf 37113} and the {\it sunset} profile corresponds to image {\bf 36681}. As the bright areas in {\bf 37113} are overexposed we analyze the next acquired image ({\bf 37114}). We find that the {\it noon} profile does not, in fact, correspond to local noon, but rather to a time exactly in between sunrise and noon; the incidence angle we calculate for image {\bf 37114} (Table~\ref{tab:haze_images}) is $48^\circ$ rather than the $15^\circ$ that would correspond to noon. We projected both images to exactly the same geometry and plot identical linear profiles across the bright areas in Fig.~\ref{fig:haze_RC3}. Note that we plot the reflectance ($I/F$) rather than the uncalibrated data numbers shown by \citet{N15}. As the incidence angle is much higher for {\bf 36681} than for {\bf 37114} the overall reflectance is lower, and for the sake of comparison we scale the former profile to match the latter for a region far from the bright areas (columns 0-40). Now, if haze exists the {\bf 37114} profile (red) should be systematically higher (by about 20\%) than the {\bf 36681} profile (black) in between the areas (i.e.\ at the location of the dotted line). It is possible that we see a slight increase in Fig.~\ref{fig:haze_RC3}, but it is marginal at best. Would we be able to detect a 20\% increase with confidence? At several locations the {\bf 36681} profile is actually systematically higher than the {\bf 37114} profile (e.g.\ columns 40-60 and 155-180), opposite of what we hoped to see. An explanation in terms of unresolved local topography (which obscures terrain and casts shadows) is more reasonable. In other words, the spatial resolution of the {\it RC3} images is too low to detect intrinsic reflectance variability on the order of 20\%.

\begin{figure}
\centering
\includegraphics[height=5.5cm,angle=0]{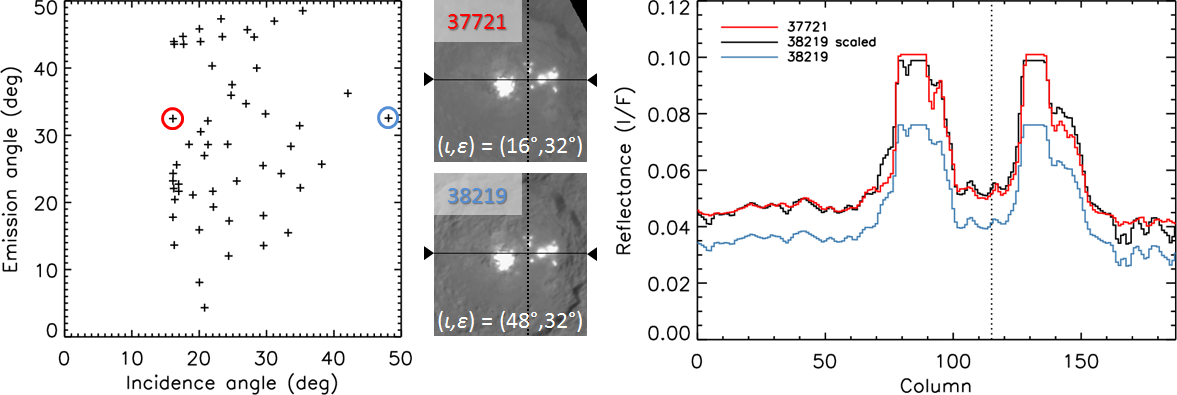}
\caption{Searching for haze in {\it Survey} clear filter images of Occator. {\bf Left}: The incidence and emission angle of an area between the bright areas in the center of Occator of all available {\it Survey} images. The images {\bf 37721} (red circle) and {\bf 38219} (blue circle) have identical emission angles but very different incidence angles. {\bf Middle}: The two $188 \times 200$ pixel sized image crops shown in identical projection, with photometric angles indicated. We investigate the reflectance profile along the drawn horizontal line. The intersection with the dotted vertical line identifies the area on the surface associated with the photometric angles in the left plot. The phase angle is $\alpha = 20^\circ$ for both images. {\bf Right}: Reflectance profiles for the two images. The black curve is the blue curve scaled to match the red curve for the terrain outside the bright areas (columns 0-40). If a diurnal haze were present between the bright areas as described by \citet{N15} then the red curve (reflectance at lower incidence angle) should be higher at the location of the dotted line than the black curve. In fact, the two curves overlap. The flat profile tops indicate where the bright areas were overexposed.}
\label{fig:haze_survey}
\end{figure}

While the diurnal haze hypothesis is a complex explanation of what may be a simple combination of instrumental and observational artifacts, it is testable. Occator is well resolved in {\it Survey} images, and we are less concerned with unresolved topography and contributions from the PSF wings. We search the location between the bright areas for the presence of haze, expecting its density to depend on the degree of insolation, i.e.\ the incidence angle. Our analysis is displayed in Fig.~\ref{fig:haze_survey}. First we identified two {\it Survey} images in which this location has different incidence angles but the same emission and phase angle (Fig.~\ref{fig:haze_survey}a). The low incidence angle image ({\bf 37721}, $\iota = 16^\circ$) is expected to have more haze than the high incidence angle image ({\bf 38219}, $\iota = 48^\circ$). Image {\bf 37721} can be classified as a noon image, whereas {\bf 38219} was acquired far from noon (the \citeauthor{N15} {\it noon} image also has $\iota = 48^\circ$). We projected the images to exactly the same geometry in (Fig.~\ref{fig:haze_survey}b). Note how the higher incidence angle in {\bf 38219} leads to increased shadows. The reflectance profiles in Fig.~\ref{fig:haze_survey}c are the same as in Fig.~\ref{fig:haze_RC3}. The {\bf 38219} profile (blue) is associated with a larger incidence angle and is lower than that of {\bf 37721} (red) because of the lower solar irradiance. We therefore scale the blue profile to match the red profile for the terrain west of the bright areas (black). The increase of signal in between the bright areas in image {\bf 37113} attributed to haze is about 20\%. At an emission angle of $70^\circ$ (Table~\ref{tab:haze_images}), the optical path length in {\bf 37113} is twice that in our images. But the solar irradiance is 50\% larger in {\bf 37721}. So, if a diurnal haze were present between Cerealia Facula and the Vinalia Faculae then the red profile in (c) should be roughly 15\% higher than the black profile. In fact, the two curves overlap.

How could a haze exist inside Occator? In the absence of a supporting atmosphere, haze particles would be on ballistic trajectories. If ejected from the bright terrain we would rather see plumes than a ground fog-type haze. But a plume would be very hard to observe. The Enceladus plumes, for example, are invisible in day-side images and can only be seen at high phase angles \citep{P06}. The \citet{N15} images were acquired at phase angles unfavorable for detecting forward scattering haze particles. If there were really a ground fog-type haze, it would be due to vigorous cometary-like activity over the entire crater floor. In that case we expect a correlation with insolation, which we do not observe. In fact, the scattering properties of Cerealia Facula are those of a particulate surface deposit of high albedo (see Sec.~\ref{sec:Occator}). We conclude that there is no evidence for haze inside Occator and thereby no evidence for present sublimation. That said, it may still be worthwhile to look for visible evidence for a faint, extended atmosphere around Ceres in light of the water detection by the Herschel space observatory \citep{K14}. \citet{M16} found large excursions in the radial velocity of Ceres and hinted at dramatic events being responsible, suggesting a link with Occator on basis of the haze claim. The suspected events happened between the Dawn {\it Survey} and {\it HAMO} observational phases. If a large expanding dust cloud was behind the radial velocity variations, it would have left characteristic signatures in the spectral profiles used for determining the radial velocity. Such an event may be expected to leave traces on the surface. We did not find any surface changes over that time in the FC data, in terms of neither reflectance nor morphology, although spectral changes may yet be identified with VIR.


\section{Summary}

We performed a global spectrophotometric characterization of Ceres using Dawn Framing Camera images. We evaluated a number of photometric models to identify the one that best describes the average scattering behavior of the surface. The models were used to apply a photometric correction to FC images for the production of global albedo and color maps. Of the two leading model candidates, Akimov and Hapke, we found that the Akimov disk function in combination with a polynomial phase function yielded the highest quality global maps. The Akimov model is also less complicated to use than the Hapke model, as also noted by \citet{S12}. The global albedo map is dominated by a large circular feature in Vendimia Planitia, which had earlier been seen by HST \citep{L06}. Smaller high albedo features are mostly associated with fresh-looking craters. The global color maps show that the dominant color variation over the surface relates to the presence of blue material in and around those same craters. Blue material has a negative spectral slope when compared to Ceres average. Next, we adopted a simple phase function, an exponential function, for which we determined the distribution of amplitude and slope over the surface. Using this technique, \citet{S13b} found variations of the slope associated with fresh craters on asteroid Vesta that could be related to physical roughness of the regolith: smooth crater walls and rough ejecta. On Ceres the situation is different: we found a few instances of smooth crater walls, but rough ejecta are very rare. The bright, blue ejecta of craters that show evidence for flows, like Haulani, may in fact be physically smooth. High albedo, blue color, and physical smoothness all appear to be indicators of youth. But whereas all blue terrain appears to be young, not all young terrain is necessarily blue.

Clues as to why ejecta of fresh craters on Ceres are blue are found in the experiments of \citet{P16}, who created particles by mixing phyllosilicates with water ice, applied low temperature and vacuum conditions to sublimate the ice, and found the phyllosilicate residue to have a negative spectral slope. The surface of Ceres is known to harbor ammoniated phyllosilicates \citep{dS15,A16} and the sub-surface may harbor water ice \citep{PY16,SH16}. We hypothesize that the ejecta of a fresh crater like Haulani were initially a liquid water/phyllosilicate mixture, perhaps as a consequence of impact melting of subsurface water ice, that turned blue after desiccation. The physical smoothness of the regolith is consistent with the initial liquid condition. Space weathering may be responsible for the fading of the blue color over time. Alternative explanations for the blue color are explored by \citet{St16}.

On the other end of the color spectrum we find terrain with a positive spectral slope in the visible wavelength range. The prime example of such red terrain is distributed in small patches near Ernutet crater. \citet{dS17} found the Ernutet red material to be rich in organic material. Our maps do not provide evidence for the presence of such red material in other locations. The large bright feature in Vendimia Planitia has spectrophotometric properties that are different from those of other bright terrains. Being prominent in the albedo map, it is not recognized in both color and phase curve slope maps. \citet{Ma16} proposed Vendimia Planitia to be a ancient impact basin. While slightly smaller than the proposed basin, the spectrophotometric properties of the large bright feature suggest it is a palimpsest, in agreement with the impact basin hypothesis.

The bright area in the center of Occator crater, Cerealia Facula, is the brightest feature on Ceres and is accompanied by a group of several secondary, less bright areas called the Vinalia Faculae. The scattering properties of Cerealia Facula are consistent with those of a high albedo surface deposit, albeit one with an unusually steep phase curve. Unresolved topography, high surface roughness, and large average particle size may contribute to the latter. The scattering properties of Cerealia Facula may be related to it being a (former) site of volatile release. We estimate its normal visual albedo to be $0.6 \pm 0.1$ at 1.3~km per pixel (six times higher than Ceres average) and confirm its red color \citep{N15}. Apparently recently exposed material seen at a resolution of 35~m per pixel may have an albedo around unity, suggesting the bright material to rapidly darken over time. The Vinalia Faculae have a visible spectrum that is not different from Ceres average, suggesting a different composition from that of the central area. There is no evidence for the existence of a diurnal ground fog-type haze inside the Occator crater as found by \citet{N15}. We are unable to replicate their findings using the same images, and we are unable to confirm their findings using higher-resolution images. Their suggestion that a haze is directly visible in certain images appears to result from stretching of the contrast of images of bright surface deposits. We conclude that these FC images show no evidence for present sublimation of water ice inside Occator.

\section*{Data availability}

Global maps of Ceres for all color filters as shown in Fig.~\ref{fig:color} are available at \url{https://doi.org/10.5281/zenodo.4251217}.

\section*{Acknowledgements}

The authors thank two anonymous referees for their valuable comments.


\section*{References}

\bibliography{Ceres_photometry}

\begin{thebibliography}{82}
\expandafter\ifx\csname natexlab\endcsname\relax\def\natexlab#1{#1}\fi
\expandafter\ifx\csname url\endcsname\relax
  \def\url#1{\texttt{#1}}\fi
\expandafter\ifx\csname urlprefix\endcsname\relax\def\urlprefix{URL }\fi

\bibitem[{{Ammannito} et~al.(2016){Ammannito}, {DeSanctis}, {Ciarniello},
  {Frigeri}, {Carrozzo}, {Combe}, {Ehlmann}, {Marchi}, {McSween}, {Raponi},
  {Toplis}, {Tosi}, {Castillo-Rogez}, {Capaccioni}, {Capria}, {Fonte},
  {Giardino}, {Jaumann}, {Longobardo}, {Joy}, {Magni}, {McCord}, {McFadden},
  {Palomba}, {Pieters}, {Polanskey}, {Rayman}, {Raymond}, {Schenk}, {Zambon},
  and {Russell}}]{A16}
{Ammannito}, E., {DeSanctis}, M.~C., {Ciarniello}, M., {Frigeri}, A.,
  {Carrozzo}, F.~G., {Combe}, J.-P., {Ehlmann}, B.~L., {Marchi}, S., {McSween},
  H.~Y., {Raponi}, A., {Toplis}, M.~J., {Tosi}, F., {Castillo-Rogez}, J.~C.,
  {Capaccioni}, F., {Capria}, M.~T., {Fonte}, S., {Giardino}, M., {Jaumann},
  R., {Longobardo}, A., {Joy}, S.~P., {Magni}, G., {McCord}, T.~B., {McFadden},
  L.~A., {Palomba}, E., {Pieters}, C.~M., {Polanskey}, C.~A., {Rayman}, M.~D.,
  {Raymond}, C.~A., {Schenk}, P.~M., {Zambon}, F., {Russell}, C.~T., Sep. 2016.
  {Distribution of phyllosilicates on the surface of Ceres}. Science 353,
  aaf4279.

\bibitem[{{Anderson} et~al.(2004){Anderson}, {Sides}, {Soltesz}, {Sucharski},
  and {Becker}}]{A04}
{Anderson}, J.~A., {Sides}, S.~C., {Soltesz}, D.~L., {Sucharski}, T.~L.,
  {Becker}, K.~J., Mar. 2004. {Modernization of the Integrated Software for
  Imagers and Spectrometers}. In: {Mackwell}, S., {Stansbery}, E. (Eds.), Lunar
  and Planetary Institute Science Conference Abstracts. Vol.~35 of Lunar and
  Planetary Institute Science Conference Abstracts. p. 2039.

\bibitem[{{Annex} et~al.(2012){Annex}, {Verbiscer}, and {Helfenstein}}]{A12}
{Annex}, A.~M., {Verbiscer}, A.~J., {Helfenstein}, P., Mar. 2012. {Photometric
  Properties of Enceladus' South Polar Terrain}. In: Lunar and Planetary
  Science Conference. Vol.~43 of Lunar and Planetary Science Conference. p.
  2698.

\bibitem[{{Becker} et~al.(2012){Becker}, {Anderson}, {Barrett}, {Sides}, and
  {Titus}}]{B12}
{Becker}, K.~J., {Anderson}, J.~A., {Barrett}, J.~M., {Sides}, S.~C., {Titus},
  T.~N., Mar. 2012. {ISIS Support for Dawn Instruments}. In: Lunar and
  Planetary Institute Science Conference Abstracts. Vol.~43 of Lunar and
  Planetary Institute Science Conference Abstracts. p. 2892.

\bibitem[{{Belskaya} and {Shevchenko}(2000)}]{BS00}
{Belskaya}, I.~N., {Shevchenko}, V.~G., Sep. 2000. {Opposition Effect of
  Asteroids}. \icarus 147, 94--105.

\bibitem[{{Bishop} et~al.(2014){Bishop}, {Quinn}, and {Dyar}}]{Bi14}
{Bishop}, J.~L., {Quinn}, R., {Dyar}, M.~D., Aug. 2014. {Spectral and thermal
  properties of perchlorate salts and implications for Mars}. American
  Mineralogist 99, 1580--1592.

\bibitem[{{Brown}(2014)}]{B14}
{Brown}, A.~J., Sep. 2014. {Spectral bluing induced by small particles under
  the Mie and Rayleigh regimes}. \icarus 239, 85--95.

\bibitem[{{Buratti} and {Veverka}(1983)}]{BV83}
{Buratti}, B., {Veverka}, J., Jul. 1983. {Voyager photometry of Europa}.
  \icarus 55, 93--110.

\bibitem[{{Ciarniello} et~al.(2016){Ciarniello}, {De Sanctis}, {Ammannito},
  {Raponi}, {Longobardo}, {Palomba}, {Carrozzo}, {Tosi}, {Li}, {Schr\"oder},
  {Zambon}, {Frigeri}, {Fonte}, {Giardino}, {Pieters}, {Raymond}, and
  {Russell}}]{C16}
{Ciarniello}, M., {De Sanctis}, M.~C., {Ammannito}, E., {Raponi}, A.,
  {Longobardo}, A., {Palomba}, E., {Carrozzo}, F.~G., {Tosi}, F., {Li}, J.-Y.,
  {Schr\"oder}, S., {Zambon}, F., {Frigeri}, A., {Fonte}, S., {Giardino}, M.,
  {Pieters}, C., {Raymond}, C.~A., {Russell}, C.~T., 2016. {Spectrophotometric
  properties of dwarf planet Ceres from the VIR spectrometer on board the Dawn
  mission}, {A\&A, in press}.

\bibitem[{{Clark} et~al.(2012){Clark}, {Cruikshank}, {Jaumann}, {Brown},
  {Stephan}, {Dalle Ore}, {Eric Livo}, {Pearson}, {Curchin}, {Hoefen},
  {Buratti}, {Filacchione}, {Baines}, and {Nicholson}}]{C12}
{Clark}, R.~N., {Cruikshank}, D.~P., {Jaumann}, R., {Brown}, R.~H., {Stephan},
  K., {Dalle Ore}, C.~M., {Eric Livo}, K., {Pearson}, N., {Curchin}, J.~M.,
  {Hoefen}, T.~M., {Buratti}, B.~J., {Filacchione}, G., {Baines}, K.~H.,
  {Nicholson}, P.~D., Apr. 2012. {The surface composition of Iapetus: Mapping
  results from Cassini VIMS}. \icarus 218, 831--860.

\bibitem[{{Clark} et~al.(2008){Clark}, {Curchin}, {Jaumann}, {Cruikshank},
  {Brown}, {Hoefen}, {Stephan}, {Moore}, {Buratti}, {Baines}, {Nicholson}, and
  {Nelson}}]{C08}
{Clark}, R.~N., {Curchin}, J.~M., {Jaumann}, R., {Cruikshank}, D.~P., {Brown},
  R.~H., {Hoefen}, T.~M., {Stephan}, K., {Moore}, J.~M., {Buratti}, B.~J.,
  {Baines}, K.~H., {Nicholson}, P.~D., {Nelson}, R.~M., Feb. 2008.
  {Compositional mapping of Saturn's satellite Dione with Cassini VIMS and
  implications of dark material in the Saturn system}. \icarus 193, 372--386.

\bibitem[{{Clark} et~al.(2010){Clark}, {Pieters}, {Taylor}, {Petro},
  {Isaacson}, {Nettles}, {Combe}, and {M3 Team}}]{C10}
{Clark}, R.~N., {Pieters}, C.~M., {Taylor}, L.~A., {Petro}, N.~E., {Isaacson},
  P.~J., {Nettles}, J.~W., {Combe}, J.~P., {M3 Team}, Mar. 2010. {Rayleigh
  Scattering in Reflectance Spectra of the Moon}. In: Lunar and Planetary
  Science Conference. Vol.~41 of Lunar and Planetary Science Conference. p.
  2337.

\bibitem[{{Combe} et~al.(2016){Combe}, {McCord}, {Tosi}, {Ammannito},
  {Carrozzo}, {De Sanctis}, {Raponi}, {Byrne}, {Landis}, {Hughson}, {Raymond},
  and {Russell}}]{Co16}
{Combe}, J.-P., {McCord}, T.~B., {Tosi}, F., {Ammannito}, E., {Carrozzo},
  F.~G., {De Sanctis}, M.~C., {Raponi}, A., {Byrne}, S., {Landis}, M.~E.,
  {Hughson}, K.~H.~G., {Raymond}, C.~A., {Russell}, C.~T., Sep. 2016.
  {Detection of local H$_{2}$O exposed at the surface of Ceres}. Science 353,
  aaf3010.

\bibitem[{{De Sanctis} et~al.(2017){De Sanctis}, {Ammannito}, {McSween},
  {Raponi}, {Marchi}, {Capaccioni}, {Capria}, {Carrozzo}, {Ciarniello},
  {Fonte}, {Formisano}, {Frigeri}, {Giardino}, {Longobardo}, {Magni},
  {McFadden}, {Palomba}, {Pieters}, {Tosi}, {Zambon}, {Raymond}, and
  {Russell}}]{dS17}
{De Sanctis}, M.~C., {Ammannito}, E., {McSween}, H.~Y., {Raponi}, A., {Marchi},
  S., {Capaccioni}, F., {Capria}, M.~T., {Carrozzo}, F.~G., {Ciarniello}, M.,
  {Fonte}, S., {Formisano}, M., {Frigeri}, A., {Giardino}, M., {Longobardo},
  A., {Magni}, G., {McFadden}, L.~A., {Palomba}, E., {Pieters}, C.~M., {Tosi},
  F., {Zambon}, F., {Raymond}, C.~A., {Russell}, C.~T., 2017. {Localized
  aliphatic organic material on the surface of Ceres}, {Science, in press}.

\bibitem[{{De Sanctis} et~al.(2015){De Sanctis}, {Ammannito}, {Raponi},
  {Marchi}, {McCord}, {McSween}, {Capaccioni}, {Capria}, {Carrozzo},
  {Ciarniello}, {Longobardo}, {Tosi}, {Fonte}, {Formisano}, {Frigeri},
  {Giardino}, {Magni}, {Palomba}, {Turrini}, {Zambon}, {Combe}, {Feldman},
  {Jaumann}, {McFadden}, {Pieters}, {Prettyman}, {Toplis}, {Raymond}, and
  {Russell}}]{dS15}
{De Sanctis}, M.~C., {Ammannito}, E., {Raponi}, A., {Marchi}, S., {McCord},
  T.~B., {McSween}, H.~Y., {Capaccioni}, F., {Capria}, M.~T., {Carrozzo},
  F.~G., {Ciarniello}, M., {Longobardo}, A., {Tosi}, F., {Fonte}, S.,
  {Formisano}, M., {Frigeri}, A., {Giardino}, M., {Magni}, G., {Palomba}, E.,
  {Turrini}, D., {Zambon}, F., {Combe}, J.-P., {Feldman}, W., {Jaumann}, R.,
  {McFadden}, L.~A., {Pieters}, C.~M., {Prettyman}, T., {Toplis}, M.,
  {Raymond}, C.~A., {Russell}, C.~T., Dec. 2015. {Ammoniated phyllosilicates
  with a likely outer Solar System origin on (1) Ceres}. \nat 528, 241--244.

\bibitem[{{De Sanctis} et~al.(2016){De Sanctis}, {Raponi}, {Ammannito},
  {Ciarniello}, {Toplis}, {McSween}, {Castillo-Rogez}, {Ehlmann}, {Carrozzo},
  {Marchi}, {Tosi}, {Zambon}, {Capaccioni}, {Capria}, {Fonte}, {Formisano},
  {Frigeri}, {Giardino}, {Longobardo}, {Magni}, {Palomba}, {McFadden},
  {Pieters}, {Jaumann}, {Schenk}, {Mugnuolo}, {Raymond}, and {Russell}}]{dS16}
{De Sanctis}, M.~C., {Raponi}, A., {Ammannito}, E., {Ciarniello}, M., {Toplis},
  M.~J., {McSween}, H.~Y., {Castillo-Rogez}, J.~C., {Ehlmann}, B.~L.,
  {Carrozzo}, F.~G., {Marchi}, S., {Tosi}, F., {Zambon}, F., {Capaccioni}, F.,
  {Capria}, M.~T., {Fonte}, S., {Formisano}, M., {Frigeri}, A., {Giardino}, M.,
  {Longobardo}, A., {Magni}, G., {Palomba}, E., {McFadden}, L.~A., {Pieters},
  C.~M., {Jaumann}, R., {Schenk}, P., {Mugnuolo}, R., {Raymond}, C.~A.,
  {Russell}, C.~T., Aug. 2016. {Bright carbonate deposits as evidence of
  aqueous alteration on (1) Ceres}. Nature 536~(7614), 54--57.

\bibitem[{{Domingue} et~al.(1991){Domingue}, {Hapke}, {Lockwood}, and
  {Thompson}}]{D91}
{Domingue}, D.~L., {Hapke}, B.~W., {Lockwood}, G.~W., {Thompson}, D.~T., Mar.
  1991. {Europa's phase curve - Implications for surface structure}. \icarus
  90, 30--42.

\bibitem[{{Formisano} et~al.(2016){Formisano}, {De Sanctis}, {Magni},
  {Federico}, and {Capria}}]{F16}
{Formisano}, M., {De Sanctis}, M.~C., {Magni}, G., {Federico}, C., {Capria},
  M.~T., Jan. 2016. {Ceres water regime: surface temperature, water sublimation
  and transient exo(atmo)sphere}. \mnras 455, 1892--1904.

\bibitem[{{Hanley} et~al.(2014){Hanley}, {Dalton}, {Chevrier}, {Jamieson}, and
  {Barrows}}]{H14}
{Hanley}, J., {Dalton}, J.~B., {Chevrier}, V.~F., {Jamieson}, C.~S., {Barrows},
  R.~S., Nov. 2014. {Reflectance spectra of hydrated chlorine salts: The effect
  of temperature with implications for Europa}. \jgre 119, 2370--2377.

\bibitem[{{Hapke}(1981)}]{H81}
{Hapke}, B., Apr. 1981. {Bidirectional reflectance spectroscopy. I - Theory}.
  \jgr 86, 3039--3054.

\bibitem[{{Hapke}(1984)}]{H84}
{Hapke}, B., Jul. 1984. {Bidirectional reflectance spectroscopy. III -
  Correction for macroscopic roughness}. \icarus 59, 41--59.

\bibitem[{{Hapke}(1986)}]{H86}
{Hapke}, B., Aug. 1986. {Bidirectional reflectance spectroscopy. IV - The
  extinction coefficient and the opposition effect}. \icarus 67, 264--280.

\bibitem[{{Hapke}(2002)}]{H02}
{Hapke}, B., Jun. 2002. {Bidirectional Reflectance Spectroscopy. 5. The
  Coherent Backscatter Opposition Effect and Anisotropic Scattering}. \icarus
  157, 523--534.

\bibitem[{{Hasselmann} et~al.(2016){Hasselmann}, {Barucci}, {Fornasier},
  {Leyrat}, {Carvano}, {Lazzaro}, and {Sierks}}]{H16}
{Hasselmann}, P.~H., {Barucci}, M.~A., {Fornasier}, S., {Leyrat}, C.,
  {Carvano}, J.~M., {Lazzaro}, D., {Sierks}, H., Mar. 2016. {Asteroid (21)
  Lutetia: Disk-resolved photometric analysis of Baetica region}. \icarus 267,
  135--153.

\bibitem[{{Helfenstein} and {Veverka}(1989)}]{SH89}
{Helfenstein}, P., {Veverka}, J., 1989. Physical characterization of asteroid
  surfaces from photometric analysis. In: {Binzel}, R.~P., {Gehrels}, T.,
  {Matthews}, M.~S. (Eds.), Asteroids II. Univ. of Arizona Press, Tucson, pp.
  557--593.

\bibitem[{{Jaumann} et~al.(2008){Jaumann}, {Stephan}, {Hansen}, {Clark},
  {Buratti}, {Brown}, {Baines}, {Newman}, {Bellucci}, {Filacchione},
  {Coradini}, {Cruikshank}, {Griffith}, {Hibbitts}, {McCord}, {Nelson},
  {Nicholson}, {Sotin}, and {Wagner}}]{J08}
{Jaumann}, R., {Stephan}, K., {Hansen}, G.~B., {Clark}, R.~N., {Buratti},
  B.~J., {Brown}, R.~H., {Baines}, K.~H., {Newman}, S.~F., {Bellucci}, G.,
  {Filacchione}, G., {Coradini}, A., {Cruikshank}, D.~P., {Griffith}, C.~A.,
  {Hibbitts}, C.~A., {McCord}, T.~B., {Nelson}, R.~M., {Nicholson}, P.~D.,
  {Sotin}, C., {Wagner}, R., Feb. 2008. {Distribution of icy particles across
  Enceladus' surface as derived from Cassini-VIMS measurements}. \icarus 193,
  407--419.

\bibitem[{{Jaumann} et~al.(2016){Jaumann}, {Stephan}, {Krohn}, {Matz}, {Otto},
  {Neumann}, {Kneissl}, {Schmedemann}, {Schroeder}, {Tosi}, {De Sanctis},
  {Preusker}, {Buczkowski}, {Capaccioni}, {Carsenty}, {Elgner}, {von der
  Gathen}, {Gieber}, {Hiesinger}, {Hoffmann}, {Kersten}, {Li}, {McCord},
  {McFadden}, {Mottola}, {Nathues}, {Neesemann}, {Raymond}, {Roatsch},
  {Russell}, {Schmidt}, {Schulzeck}, {Wagner}, and {Williams}}]{J16}
{Jaumann}, R., {Stephan}, K., {Krohn}, K., {Matz}, K.-D., {Otto}, K.,
  {Neumann}, W., {Kneissl}, T., {Schmedemann}, N., {Schroeder}, S., {Tosi}, F.,
  {De Sanctis}, M.~C., {Preusker}, F., {Buczkowski}, D., {Capaccioni}, F.,
  {Carsenty}, U., {Elgner}, S., {von der Gathen}, I., {Gieber}, T.,
  {Hiesinger}, H., {Hoffmann}, M., {Kersten}, E., {Li}, J.-Y., {McCord}, T.~B.,
  {McFadden}, L., {Mottola}, S., {Nathues}, A., {Neesemann}, A., {Raymond}, C.,
  {Roatsch}, T., {Russell}, C.~T., {Schmidt}, B., {Schulzeck}, F., {Wagner},
  R., {Williams}, D.~A., Mar. 2016. {Age-Dependent Morphological and
  Compositional Variations on Ceres}. In: Lunar and Planetary Science
  Conference. Vol.~47 of Lunar and Planetary Science Conference. p. 1455.

\bibitem[{{Kaasalainen} et~al.(2001){Kaasalainen}, {Torppa}, and
  {Muinonen}}]{K01}
{Kaasalainen}, M., {Torppa}, J., {Muinonen}, K., Sep. 2001. {Optimization
  Methods for Asteroid Lightcurve Inversion. II. The Complete Inverse Problem}.
  \icarus 153, 37--51.

\bibitem[{{Kirk} et~al.(2004){Kirk}, {Howington-Kraus}, {Soderblom}, {Giese},
  and {Oberst}}]{K04}
{Kirk}, R.~L., {Howington-Kraus}, E., {Soderblom}, L.~A., {Giese}, B.,
  {Oberst}, J., Jan. 2004. {Comparison of USGS and DLR topographic models of
  Comet Borrelly and photometric applications}. \icarus 167, 54--69.

\bibitem[{{Krohn} et~al.(2016){Krohn}, {Jaumann}, {Stephan}, {Otto},
  {Schmedemann}, {Wagner}, {Matz}, {Tosi}, {Zambon}, {von der Gathen},
  {Schulzeck}, {Schr\"oder}, {Buczkowski}, {Hiesinger}, {McSween}, {Pieters},
  {Preusker}, {Roatsch}, {Raymond}, {Russell}, and {Williams}}]{K16}
{Krohn}, K., {Jaumann}, R., {Stephan}, K., {Otto}, K.~A., {Schmedemann}, N.,
  {Wagner}, R.~J., {Matz}, K.-D., {Tosi}, F., {Zambon}, F., {von der Gathen},
  I., {Schulzeck}, F., {Schr\"oder}, S.~E., {Buczkowski}, D.~L., {Hiesinger},
  H., {McSween}, H.~Y., {Pieters}, C.~M., {Preusker}, F., {Roatsch}, T.,
  {Raymond}, C.~A., {Russell}, C.~T., {Williams}, D.~A., 2016. Cryogenic flow
  features on ceres: Implications for crater-related cryovolcanism. Geophysical
  Research Letters 43~(23), 11,994--12,003.

\bibitem[{{K{\"u}ppers} et~al.(2014){K{\"u}ppers}, {O'Rourke},
  {Bockel{\'e}e-Morvan}, {Zakharov}, {Lee}, {von Allmen}, {Carry}, {Teyssier},
  {Marston}, {M{\"u}ller}, {Crovisier}, {Barucci}, and {Moreno}}]{K14}
{K{\"u}ppers}, M., {O'Rourke}, L., {Bockel{\'e}e-Morvan}, D., {Zakharov}, V.,
  {Lee}, S., {von Allmen}, P., {Carry}, B., {Teyssier}, D., {Marston}, A.,
  {M{\"u}ller}, T., {Crovisier}, J., {Barucci}, M.~A., {Moreno}, R., Jan. 2014.
  {Localized sources of water vapour on the dwarf planet (1)Ceres}. \nat 505,
  525--527.

\bibitem[{{Li} et~al.(2013{\natexlab{a}}){Li}, {A'Hearn}, {Belton}, {Farnham},
  {Klaasen}, {Sunshine}, {Thomas}, and {Veverka}}]{Li13}
{Li}, J.-Y., {A'Hearn}, M.~F., {Belton}, M.~J.~S., {Farnham}, T.~L., {Klaasen},
  K.~P., {Sunshine}, J.~M., {Thomas}, P.~C., {Veverka}, J., Feb.
  2013{\natexlab{a}}. {Photometry of the nucleus of Comet 9P/Tempel 1 from
  Stardust-NExT flyby and the implications}. \icarus 222, 467--476.

\bibitem[{{Li} et~al.(2016{\natexlab{a}}){Li}, {Le Corre}, {Reddy}, {Nathues},
  {Hoffmann}, {Schaefer}, {Ciarniello}, {Mottola}, {Schr{\"o}der}, {Raymond},
  and {Russell}}]{L16b}
{Li}, J.-Y., {Le Corre}, L., {Reddy}, V., {Nathues}, A., {Hoffmann}, M.,
  {Schaefer}, M., {Ciarniello}, M., {Mottola}, S., {Schr{\"o}der}, S.~E.,
  {Raymond}, C.~A., {Russell}, C.~T., Apr. 2016{\natexlab{a}}.
  {Spectrophotometric Modeling and Mapping of Ceres}. In: EGU General Assembly
  Conference Abstracts. Vol.~18 of EGU General Assembly Conference Abstracts.
  p. 17302.

\bibitem[{{Li} et~al.(2013{\natexlab{b}}){Li}, {Le Corre}, {Schr{\"o}der},
  {Reddy}, {Denevi}, {Buratti}, {Mottola}, {Hoffmann}, {Gutierrez-Marques},
  {Nathues}, {Russell}, and {Raymond}}]{L13}
{Li}, J.-Y., {Le Corre}, L., {Schr{\"o}der}, S.~E., {Reddy}, V., {Denevi},
  B.~W., {Buratti}, B.~J., {Mottola}, S., {Hoffmann}, M., {Gutierrez-Marques},
  P., {Nathues}, A., {Russell}, C.~T., {Raymond}, C.~A., Nov.
  2013{\natexlab{b}}. {Global photometric properties of Asteroid (4) Vesta
  observed with Dawn Framing Camera}. \icarus 226, 1252--1274.

\bibitem[{{Li} et~al.(2006){Li}, {McFadden}, {Parker}, {Young}, {Stern},
  {Thomas}, {Russell}, and {Sykes}}]{L06}
{Li}, J.-Y., {McFadden}, L.~A., {Parker}, J.~W., {Young}, E.~F., {Stern},
  S.~A., {Thomas}, P.~C., {Russell}, C.~T., {Sykes}, M.~V., May 2006.
  {Photometric analysis of 1 Ceres and surface mapping from HST observations}.
  \icarus 182, 143--160.

\bibitem[{{Li} et~al.(2015){Li}, {Nathues}, {Mottola}, {Sykes}, {Polanskey},
  {Joy}, {Mastrodemos}, {McFadden}, {Skillman}, {Memarsadeghi}, {Hoffmann},
  {Schr{\"o}der}, {Carsenty}, {Raymond}, and {Russell}}]{Li15}
{Li}, J.-Y., {Nathues}, A., {Mottola}, S., {Sykes}, M.~V., {Polanskey}, C.~A.,
  {Joy}, S., {Mastrodemos}, N., {McFadden}, L.~A., {Skillman}, D.,
  {Memarsadeghi}, N., {Hoffmann}, M., {Schr{\"o}der}, S.~E., {Carsenty}, U.,
  {Raymond}, C.~A., {Russell}, C.~T., Oct. 2015. {Search for Dust Around
  Ceres}. European Planetary Science Congress 10, EPSC2015--385.

\bibitem[{{Li} et~al.(2016{\natexlab{b}}){Li}, {Reddy}, {Nathues}, {Le Corre},
  {Izawa}, {Cloutis}, {Sykes}, {Carsenty}, {Castillo-Rogez}, {Hoffmann},
  {Jaumann}, {Krohn}, {Mottola}, {Prettyman}, {Schaefer}, {Schenk},
  {Schr{\"o}der}, {Williams}, {Smith}, {Zuber}, {Konopliv}, {Park}, {Raymond},
  and {Russell}}]{L16}
{Li}, J.-Y., {Reddy}, V., {Nathues}, A., {Le Corre}, L., {Izawa}, M.~R.~M.,
  {Cloutis}, E.~A., {Sykes}, M.~V., {Carsenty}, U., {Castillo-Rogez}, J.~C.,
  {Hoffmann}, M., {Jaumann}, R., {Krohn}, K., {Mottola}, S., {Prettyman},
  T.~H., {Schaefer}, M., {Schenk}, P., {Schr{\"o}der}, S.~E., {Williams},
  D.~A., {Smith}, D.~E., {Zuber}, M.~T., {Konopliv}, A.~S., {Park}, R.~S.,
  {Raymond}, C.~A., {Russell}, C.~T., Feb. 2016{\natexlab{b}}. {Surface Albedo
  and Spectral Variability of Ceres}. \apjl 817, L22.

\bibitem[{{Longobardo} et~al.(2017){Longobardo}, {Palomba}, {De Sanctis},
  {Ciarniello}, {Galiano}, {Tosi}, {Carrozzo}, {Capria}, {Zambon}, {Raponi},
  {Ammannito}, {Zinzi}, {Raymond}, {Russell}, and {Dawn/VIR Team}}]{L17}
{Longobardo}, A., {Palomba}, E., {De Sanctis}, M.~C., {Ciarniello}, M.,
  {Galiano}, A., {Tosi}, F., {Carrozzo}, F.~G., {Capria}, M.~T., {Zambon}, F.,
  {Raponi}, A., {Ammannito}, E., {Zinzi}, A., {Raymond}, C., {Russell}, C.~T.,
  {Dawn/VIR Team}, 2017. {Photometric properties of the Occator bright spot}.
  In: Asteroids, Comets, Meteors 2017 Conference.

\bibitem[{{Longobardo} et~al.(2016){Longobardo}, {Palomba}, {De Sanctis},
  {Ciarniello}, {Tosi}, {Carrozzo}, {Raponi}, {Zambon}, {Ammannito}, {Li},
  {Raymond}, and {Russell}}]{Lo16}
{Longobardo}, A., {Palomba}, E., {De Sanctis}, M.~C., {Ciarniello}, M., {Tosi},
  F., {Carrozzo}, F.~G., {Raponi}, A., {Zambon}, F., {Ammannito}, E., {Li},
  J.-Y., {Raymond}, C.~A., {Russell}, C.~T., Mar. 2016. {Average Photometric
  Properties of Ceres Spectral Parameters}. In: Lunar and Planetary Science
  Conference. Vol.~47 of Lunar and Planetary Science Conference. p. 2239.

\bibitem[{{Marchi} et~al.(2016){Marchi}, {Ermakov}, {Raymond}, {Fu}, {O'Brien},
  {Bland}, {Ammannito}, {De Sanctis}, {Bowling}, {Schenk}, {Scully},
  {Buczkowski}, {Williams}, {Hiesinger}, and {Russell}}]{Ma16}
{Marchi}, S., {Ermakov}, A.~I., {Raymond}, C.~A., {Fu}, R.~R., {O'Brien},
  D.~P., {Bland}, M.~T., {Ammannito}, E., {De Sanctis}, M.~C., {Bowling}, T.,
  {Schenk}, P., {Scully}, J.~E.~C., {Buczkowski}, D.~L., {Williams}, D.~A.,
  {Hiesinger}, H., {Russell}, C.~T., Jul. 2016. {The missing large impact
  craters on Ceres}. Nature Communications 7, 12257.

\bibitem[{{Markwardt}(2009)}]{M09}
{Markwardt}, C.~B., Sep. 2009. {Non-linear Least-squares Fitting in IDL with
  MPFIT}. In: {Bohlender}, D.~A., {Durand}, D., {Dowler}, P. (Eds.),
  Astronomical Data Analysis Software and Systems XVIII. Vol. 411 of
  Astronomical Society of the Pacific Conference Series. p. 251.

\bibitem[{{Masoumzadeh} et~al.(2015){Masoumzadeh}, {Boehnhardt}, {Li}, and
  {Vincent}}]{M15}
{Masoumzadeh}, N., {Boehnhardt}, H., {Li}, J.-Y., {Vincent}, J.-B., Sep. 2015.
  {Photometric analysis of Asteroid (21) Lutetia from Rosetta-OSIRIS images}.
  \icarus 257, 239--250.

\bibitem[{{McCord} et~al.(2012){McCord}, {Li}, {Combe}, {McSween}, {Jaumann},
  {Reddy}, {Tosi}, {Williams}, {Blewett}, {Turrini}, {Palomba}, {Pieters}, {de
  Sanctis}, {Ammannito}, {Capria}, {Le Corre}, {Longobardo}, {Nathues},
  {Mittlefehldt}, {Schr{\"o}der}, {Hiesinger}, {Beck}, {Capaccioni},
  {Carsenty}, {Keller}, {Denevi}, {Sunshine}, {Raymond}, and {Russell}}]{McC12}
{McCord}, T.~B., {Li}, J.-Y., {Combe}, J.-P., {McSween}, H.~Y., {Jaumann}, R.,
  {Reddy}, V., {Tosi}, F., {Williams}, D.~A., {Blewett}, D.~T., {Turrini}, D.,
  {Palomba}, E., {Pieters}, C.~M., {de Sanctis}, M.~C., {Ammannito}, E.,
  {Capria}, M.~T., {Le Corre}, L., {Longobardo}, A., {Nathues}, A.,
  {Mittlefehldt}, D.~W., {Schr{\"o}der}, S.~E., {Hiesinger}, H., {Beck}, A.~W.,
  {Capaccioni}, F., {Carsenty}, U., {Keller}, H.~U., {Denevi}, B.~W.,
  {Sunshine}, J.~M., {Raymond}, C.~A., {Russell}, C.~T., Nov. 2012. {Dark
  material on Vesta from the infall of carbonaceous volatile-rich material}.
  \nat 491, 83--86.

\bibitem[{{Minnaert}(1941)}]{M41}
{Minnaert}, M., May 1941. {The reciprocity principle in lunar photometry}. \apj
  93, 403--410.

\bibitem[{{Molaro} et~al.(2016){Molaro}, {Lanza}, {Monaco}, {Tosi}, {Lo Curto},
  {Fulle}, and {Pasquini}}]{M16}
{Molaro}, P., {Lanza}, A.~F., {Monaco}, L., {Tosi}, F., {Lo Curto}, G.,
  {Fulle}, M., {Pasquini}, L., May 2016. {Daily variability of Ceres' albedo
  detected by means of radial velocities changes of the reflected sunlight}.
  \mnras 458, L54--L58.

\bibitem[{Mor{\'e}(1978)}]{M78}
Mor{\'e}, J.~J., 1978. {The Levenberg-Marquardt algorithm: Implementation and
  theory}. In: Watson, G.~A. (Ed.), Numerical Analysis: Proceedings of the
  Biennial Conference Held at Dundee, June 28--July 1, 1977. Springer, Berlin,
  Heidelberg, pp. 105--116.

\bibitem[{{Nathues} et~al.(2016){Nathues}, {Hoffmann}, {Platz}, {Thangjam},
  {Cloutis}, {Reddy}, {Le Corre}, {Li}, {Mengel}, {Rivkin}, {Applin},
  {Schaefer}, {Christensen}, {Sierks}, {Ripken}, {Schmidt}, {Hiesinger},
  {Sykes}, {Sizemore}, {Preusker}, and {Russell}}]{N16}
{Nathues}, A., {Hoffmann}, M., {Platz}, T., {Thangjam}, G.~S., {Cloutis},
  E.~A., {Reddy}, V., {Le Corre}, L., {Li}, J.-Y., {Mengel}, K., {Rivkin}, A.,
  {Applin}, D.~M., {Schaefer}, M., {Christensen}, U., {Sierks}, H., {Ripken},
  J., {Schmidt}, B.~E., {Hiesinger}, H., {Sykes}, M.~V., {Sizemore}, H.~G.,
  {Preusker}, F., {Russell}, C.~T., Dec. 2016. {FC colour images of dwarf
  planet Ceres reveal a complicated geological history}. \planss 134, 122--127.

\bibitem[{{Nathues} et~al.(2015){Nathues}, {Hoffmann}, {Schaefer}, {Le Corre},
  {Reddy}, {Platz}, {Cloutis}, {Christensen}, {Kneissl}, {Li}, {Mengel},
  {Schmedemann}, {Schaefer}, {Russell}, {Applin}, {Buczkowski}, {Izawa},
  {Keller}, {O'Brien}, {Pieters}, {Raymond}, {Ripken}, {Schenk}, {Schmidt},
  {Sierks}, {Sykes}, {Thangjam}, and {Vincent}}]{N15}
{Nathues}, A., {Hoffmann}, M., {Schaefer}, M., {Le Corre}, L., {Reddy}, V.,
  {Platz}, T., {Cloutis}, E.~A., {Christensen}, U., {Kneissl}, T., {Li}, J.-Y.,
  {Mengel}, K., {Schmedemann}, N., {Schaefer}, T., {Russell}, C.~T., {Applin},
  D.~M., {Buczkowski}, D.~L., {Izawa}, M.~R.~M., {Keller}, H.~U., {O'Brien},
  D.~P., {Pieters}, C.~M., {Raymond}, C.~A., {Ripken}, J., {Schenk}, P.~M.,
  {Schmidt}, B.~E., {Sierks}, H., {Sykes}, M.~V., {Thangjam}, G.~S., {Vincent},
  J.-B., Dec. 2015. {Sublimation in bright spots on (1) Ceres}. \nat 528,
  237--240.

\bibitem[{{Palomba} et~al.(2016){Palomba}, {Longobardo}, {De Sanctis}, {Stein},
  {Ehlmann}, {Ammannito}, {Giacomo Carrozzo}, {Raponi}, {Ciarniello},
  {Frigeri}, {Capria}, {Tosi}, {Zambon}, {Fonte}, {Giardino}, {Capaccioni},
  {Raymond}, {Russell}, and {VIR-Dawn Team}}]{Pa16}
{Palomba}, E., {Longobardo}, A., {De Sanctis}, M.~C., {Stein}, N., {Ehlmann},
  B., {Ammannito}, E., {Giacomo Carrozzo}, F., {Raponi}, A., {Ciarniello}, M.,
  {Frigeri}, A., {Capria}, M.~T., {Tosi}, F., {Zambon}, F., {Fonte}, S.,
  {Giardino}, M., {Capaccioni}, F., {Raymond}, C., {Russell}, C.~T., {VIR-Dawn
  Team}, Oct. 2016. {Compositional differences among Bright Spots on the Ceres
  surface}. In: AAS/Division for Planetary Sciences Meeting Abstracts. Vol.~48
  of AAS/Division for Planetary Sciences Meeting Abstracts. p. 511.02.

\bibitem[{{Pieters} et~al.(2017){Pieters}, {Nathues}, {Thangiam}, {Hoffman},
  {De Sanctis}, {Ammannito}, {Hiesinger}, {Pasckert}, {O'Brien},
  {Castillo-Rogez}, {Ruesch}, {McFadden}, {Tosi}, and {Zambon}}]{P17}
{Pieters}, C.~M., {Nathues}, A., {Thangiam}, G., {Hoffman}, H., {De Sanctis},
  C., {Ammannito}, E., {Hiesinger}, H., {Pasckert}, J.~H., {O'Brien}, D.~P.,
  {Castillo-Rogez}, J.~C., {Ruesch}, O., {McFadden}, L.~A., {Tosi}, F.,
  {Zambon}, F.~{Raymond}, C.~A., 2017. {Context of unusual red organic-rich
  areas on Ceres and geologic constraints for their origin}. In: Lunar and
  Planetary Science Conference. Lunar and Planetary Science Conference.

\bibitem[{{Poch} et~al.(2016){Poch}, {Pommerol}, {Jost}, {Carrasco}, {Szopa},
  and {Thomas}}]{P16}
{Poch}, O., {Pommerol}, A., {Jost}, B., {Carrasco}, N., {Szopa}, C., {Thomas},
  N., Mar. 2016. {Sublimation of water ice mixed with silicates and tholins:
  Evolution of surface texture and reflectance spectra, with implications for
  comets}. \icarus 267, 154--173.

\bibitem[{{Porco} et~al.(2006){Porco}, {Helfenstein}, {Thomas}, {Ingersoll},
  {Wisdom}, {West}, {Neukum}, {Denk}, {Wagner}, {Roatsch}, {Kieffer}, {Turtle},
  {McEwen}, {Johnson}, {Rathbun}, {Veverka}, {Wilson}, {Perry}, {Spitale},
  {Brahic}, {Burns}, {Del Genio}, {Dones}, {Murray}, and {Squyres}}]{P06}
{Porco}, C.~C., {Helfenstein}, P., {Thomas}, P.~C., {Ingersoll}, A.~P.,
  {Wisdom}, J., {West}, R., {Neukum}, G., {Denk}, T., {Wagner}, R., {Roatsch},
  T., {Kieffer}, S., {Turtle}, E., {McEwen}, A., {Johnson}, T.~V., {Rathbun},
  J., {Veverka}, J., {Wilson}, D., {Perry}, J., {Spitale}, J., {Brahic}, A.,
  {Burns}, J.~A., {Del Genio}, A.~D., {Dones}, L., {Murray}, C.~D., {Squyres},
  S., Mar. 2006. {Cassini Observes the Active South Pole of Enceladus}. Science
  311, 1393--1401.

\bibitem[{{Prettyman} et~al.(2016){Prettyman}, {Yamashita}, {Castillo-Rogez},
  {Feldman}, {Lawrence}, {McSween}, {Schorghofer}, {Toplis}, {Forni}, {Joy},
  {Marchi}, {Platz}, {Polanskey}, {De Sanctis}, {Rayman}, {Raymond}, and
  {Russell}}]{PY16}
{Prettyman}, T.~H., {Yamashita}, N., {Castillo-Rogez}, J.~C., {Feldman}, W.~C.,
  {Lawrence}, D.~J., {McSween}, H.~Y., {Schorghofer}, N., {Toplis}, M.~J.,
  {Forni}, O., {Joy}, S.~P., {Marchi}, S., {Platz}, T., {Polanskey}, C.~A., {De
  Sanctis}, M.~C., {Rayman}, M.~D., {Raymond}, C.~A., {Russell}, C.~T., Mar.
  2016. {Elemental Composition of Ceres by Dawn's Gamma Ray and Neutron
  Detector}. In: Lunar and Planetary Science Conference. Vol.~47 of Lunar and
  Planetary Science Conference. p. 2228.

\bibitem[{{Preusker} et~al.(2016){Preusker}, {Scholten}, {Matz}, {Elgner},
  {Jaumann}, {Roatsch}, {Joy}, {Polanskey}, {Raymond}, and {Russell}}]{Pr16}
{Preusker}, F., {Scholten}, F., {Matz}, K.-D., {Elgner}, S., {Jaumann}, R.,
  {Roatsch}, T., {Joy}, S.~P., {Polanskey}, C.~A., {Raymond}, C.~A., {Russell},
  C.~T., Mar. 2016. {Dawn at Ceres - Shape Model and Rotational State}. In:
  Lunar and Planetary Science Conference. Vol.~47 of Lunar and Planetary
  Science Conference. p. 1954.

\bibitem[{{Reddy} et~al.(2012{\natexlab{a}}){Reddy}, {Le Corre}, {O'Brien},
  {Nathues}, {Cloutis}, {Durda}, {Bottke}, {Bhatt}, {Nesvorny}, {Buczkowski},
  {Scully}, {Palmer}, {Sierks}, {Mann}, {Becker}, {Beck}, {Mittlefehldt}, {Li},
  {Gaskell}, {Russell}, {Gaffey}, {McSween}, {McCord}, {Combe}, and
  {Blewett}}]{R12b}
{Reddy}, V., {Le Corre}, L., {O'Brien}, D.~P., {Nathues}, A., {Cloutis}, E.~A.,
  {Durda}, D.~D., {Bottke}, W.~F., {Bhatt}, M.~U., {Nesvorny}, D.,
  {Buczkowski}, D., {Scully}, J.~E.~C., {Palmer}, E.~M., {Sierks}, H., {Mann},
  P.~J., {Becker}, K.~J., {Beck}, A.~W., {Mittlefehldt}, D., {Li}, J.-Y.,
  {Gaskell}, R., {Russell}, C.~T., {Gaffey}, M.~J., {McSween}, H.~Y., {McCord},
  T.~B., {Combe}, J.-P., {Blewett}, D., Nov. 2012{\natexlab{a}}. {Delivery of
  dark material to Vesta via carbonaceous chondritic impacts}. \icarus 221,
  544--559.

\bibitem[{{Reddy} et~al.(2015){Reddy}, {Li}, {Gary}, {Sanchez}, {Stephens},
  {Megna}, {Coley}, {Nathues}, {Le Corre}, and {Hoffmann}}]{R15}
{Reddy}, V., {Li}, J.-Y., {Gary}, B.~L., {Sanchez}, J.~A., {Stephens}, R.~D.,
  {Megna}, R., {Coley}, D., {Nathues}, A., {Le Corre}, L., {Hoffmann}, M., Nov.
  2015. {Photometric properties of Ceres from telescopic observations using
  Dawn Framing Camera color filters}. \icarus 260, 332--345.

\bibitem[{{Reddy} et~al.(2012{\natexlab{b}}){Reddy}, {Nathues}, {Le Corre},
  {Sierks}, {Li}, {Gaskell}, {McCoy}, {Beck}, {Schr{\"o}der}, {Pieters},
  {Becker}, {Buratti}, {Denevi}, {Blewett}, {Christensen}, {Gaffey},
  {Gutierrez-Marques}, {Hicks}, {Keller}, {Maue}, {Mottola}, {McFadden},
  {McSween}, {Mittlefehldt}, {O'Brien}, {Raymond}, and {Russell}}]{R12a}
{Reddy}, V., {Nathues}, A., {Le Corre}, L., {Sierks}, H., {Li}, J.-Y.,
  {Gaskell}, R., {McCoy}, T., {Beck}, A.~W., {Schr{\"o}der}, S.~E., {Pieters},
  C.~M., {Becker}, K.~J., {Buratti}, B.~J., {Denevi}, B., {Blewett}, D.~T.,
  {Christensen}, U., {Gaffey}, M.~J., {Gutierrez-Marques}, P., {Hicks}, M.,
  {Keller}, H.~U., {Maue}, T., {Mottola}, S., {McFadden}, L.~A., {McSween},
  H.~Y., {Mittlefehldt}, D., {O'Brien}, D.~P., {Raymond}, C., {Russell}, C.,
  May 2012{\natexlab{b}}. {Color and Albedo Heterogeneity of Vesta from Dawn}.
  Science 336, 700.

\bibitem[{{Roatsch} et~al.(2016){Roatsch}, {Kersten}, {Matz}, {Preusker},
  {Scholten}, {Jaumann}, {Raymond}, and {Russell}}]{Ro16}
{Roatsch}, T., {Kersten}, E., {Matz}, K.-D., {Preusker}, F., {Scholten}, F.,
  {Jaumann}, R., {Raymond}, C.~A., {Russell}, C.~T., Feb. 2016. {Ceres Survey
  Atlas derived from Dawn Framing Camera images}. \planss 121, 115--120.

\bibitem[{{Ruesch} et~al.(2016){Ruesch}, {Platz}, {Schenk}, {McFadden},
  {Castillo-Rogez}, {Quick}, {Byrne}, {Preusker}, {O'Brien}, {Schmedemann},
  {Williams}, {Li}, {Bland}, {Hiesinger}, {Kneissl}, {Neesemann}, {Schaefer},
  {Pasckert}, {Schmidt}, {Buczkowski}, {Sykes}, {Nathues}, {Roatsch},
  {Hoffmann}, {Raymond}, and {Russell}}]{R16}
{Ruesch}, O., {Platz}, T., {Schenk}, P., {McFadden}, L.~A., {Castillo-Rogez},
  J.~C., {Quick}, L.~C., {Byrne}, S., {Preusker}, F., {O'Brien}, D.~P.,
  {Schmedemann}, N., {Williams}, D.~A., {Li}, J.-Y., {Bland}, M.~T.,
  {Hiesinger}, H., {Kneissl}, T., {Neesemann}, A., {Schaefer}, M., {Pasckert},
  J.~H., {Schmidt}, B.~E., {Buczkowski}, D.~L., {Sykes}, M.~V., {Nathues}, A.,
  {Roatsch}, T., {Hoffmann}, M., {Raymond}, C.~A., {Russell}, C.~T., Sep. 2016.
  {Cryovolcanism on Ceres}. Science 353, aaf4286.

\bibitem[{{Schenk} et~al.(2016{\natexlab{a}}){Schenk}, {Marchi}, {O'Brien},
  {Bland}, {Platz}, {Hoogenboom}, {Kramer}, {Schroeder}, {De Sanctis},
  {Buczkowski}, {Sykes}, {McFadden}, {Ruesch}, {Le Corre}, {Schmidt},
  {Hughson}, {Russell}, {Scully}, and {Raymond}}]{SM16}
{Schenk}, P., {Marchi}, S., {O'Brien}, D., {Bland}, M., {Platz}, T.,
  {Hoogenboom}, T., {Kramer}, G., {Schroeder}, S., {De Sanctis}, M.,
  {Buczkowski}, D., {Sykes}, M., {McFadden}, L., {Ruesch}, O., {Le Corre}, L.,
  {Schmidt}, B., {Hughson}, K., {Russell}, C.~T., {Scully}, J., {Raymond}, C.,
  Mar. 2016{\natexlab{a}}. {Impact Cratering on the Small Planets Ceres and
  Vesta: S-C Transitions, Central Pits, and the Origin of Bright Spots}. In:
  Lunar and Planetary Science Conference. Vol.~47 of Lunar and Planetary
  Science Conference. p. 2697.

\bibitem[{{Schenk} et~al.(2016{\natexlab{b}}){Schenk}, {Buczkowski}, {Scully},
  {De Sanctis}, {Schmidt}, {O'Brien}, {Hiesinger}, {Sizemore}, {Ammannito},
  {Raymond}, {Russell}, and {Dawn Science Team}}]{SB16}
{Schenk}, P.~M., {Buczkowski}, D., {Scully}, J.~E.~C., {De Sanctis}, M.~C.,
  {Schmidt}, B.~E., {O'Brien}, D.~P., {Hiesinger}, H., {Sizemore}, H.~G.,
  {Ammannito}, E., {Raymond}, C., {Russell}, C.~T., {Dawn Science Team}, Oct.
  2016{\natexlab{b}}. {Central Pit and Dome Formation as Seen in Occator
  Crater, Ceres}. In: AAS/Division for Planetary Sciences Meeting Abstracts.
  Vol.~48 of AAS/Division for Planetary Sciences Meeting Abstracts. p. 506.04.

\bibitem[{Schmedemann et~al.(2016)Schmedemann, Kneissl, Neesemann, Stephan,
  Jaumann, Krohn, Michael, Matz, Otto, Raymond, and Russell}]{SK16}
Schmedemann, N., Kneissl, T., Neesemann, A., Stephan, K., Jaumann, R., Krohn,
  K., Michael, G.~G., Matz, K.~D., Otto, K.~A., Raymond, C.~A., Russell, C.~T.,
  2016. Timing of optical maturation of recently exposed material on ceres.
  Geophysical Research Letters 43~(23), 11,987--11,993.

\bibitem[{{Schmidt} et~al.(2016){Schmidt}, {Hughson}, {Chilton}, {Scully},
  {Platz}, {Nathues}, {Sizemore}, {Bland}, {Byrne}, {Marchi}, {O'Brien},
  {Schorghofer}, {Hiesinger}, {Jaumann}, {Lawrence}, {Buczkowski}, {Castillo},
  {Schenk}, {Sykes}, {De Sanctis}, {Mitri}, {Formisano}, {Li}, {Reddy},
  {LeCorre}, {Russell}, {Raymond}, {Dawn Science}, and {Operations
  Team}}]{SH16}
{Schmidt}, B.~E., {Hughson}, K.~G., {Chilton}, H.~T., {Scully}, J.~E.~C.,
  {Platz}, T., {Nathues}, A., {Sizemore}, H., {Bland}, M.~T., {Byrne}, S.,
  {Marchi}, S., {O'Brien}, D.~P., {Schorghofer}, N., {Hiesinger}, H.,
  {Jaumann}, R., {Lawrence}, J., {Buczkowski}, D., {Castillo}, J.~C., {Schenk},
  P.~M., {Sykes}, M.~V., {De Sanctis}, M.~C., {Mitri}, G., {Formisano}, M.,
  {Li}, J.-Y., {Reddy}, V., {LeCorre}, L., {Russell}, C.~T., {Raymond}, C.~A.,
  {Dawn Science}, {Operations Team}, Mar. 2016. {Ground Ice on Ceres?} In:
  Lunar and Planetary Science Conference. Vol.~47 of Lunar and Planetary
  Science Conference. p. 2677.

\bibitem[{{Schorghofer} et~al.(2016){Schorghofer}, {Mazarico}, {Platz},
  {Preusker}, {Schr{\"o}der}, {Raymond}, and {Russell}}]{S16}
{Schorghofer}, N., {Mazarico}, E., {Platz}, T., {Preusker}, F., {Schr{\"o}der},
  S.~E., {Raymond}, C.~A., {Russell}, C.~T., Jul. 2016. {The permanently
  shadowed regions of dwarf planet Ceres}. \grl 43, 6783--6789.

\bibitem[{{Schr{\"o}der} et~al.(2014{\natexlab{a}}){Schr{\"o}der}, {Grynko},
  {Pommerol}, {Keller}, {Thomas}, and {Roush}}]{S14b}
{Schr{\"o}der}, S.~E., {Grynko}, Y., {Pommerol}, A., {Keller}, H.~U., {Thomas},
  N., {Roush}, T.~L., Sep. 2014{\natexlab{a}}. {Laboratory observations and
  simulations of phase reddening}. \icarus 239, 201--216.

\bibitem[{{Schr{\"o}der} et~al.(2013{\natexlab{a}}){Schr{\"o}der}, {Maue},
  {Guti{\'e}rrez Marqu{\'e}s}, {Mottola}, {Aye}, {Sierks}, {Keller}, and
  {Nathues}}]{S13a}
{Schr{\"o}der}, S.~E., {Maue}, T., {Guti{\'e}rrez Marqu{\'e}s}, P., {Mottola},
  S., {Aye}, K.~M., {Sierks}, H., {Keller}, H.~U., {Nathues}, A., Nov.
  2013{\natexlab{a}}. {In-flight Calibration of the Dawn Framing Camera}.
  \icarus 226, 1304--1317.

\bibitem[{{Schr{\"o}der} et~al.(2013{\natexlab{b}}){Schr{\"o}der}, {Mottola},
  {Keller}, {Raymond}, and {Russell}}]{S13b}
{Schr{\"o}der}, S.~E., {Mottola}, S., {Keller}, H.~U., {Raymond}, C.~A.,
  {Russell}, C.~T., Sep. 2013{\natexlab{b}}. {Resolved photometry of Vesta
  reveals physical properties of crater regolith}. \planss 85, 198--213.

\bibitem[{{Schr{\"o}der} et~al.(2014{\natexlab{b}}){Schr{\"o}der}, {Mottola},
  {Matz}, and {Roatsch}}]{S14a}
{Schr{\"o}der}, S.~E., {Mottola}, S., {Matz}, K.-D., {Roatsch}, T., May
  2014{\natexlab{b}}. {In-flight calibration of the Dawn Framing Camera II:
  Flat fields and stray light correction}. \icarus 234, 99--108.

\bibitem[{{Shan} et~al.(2008){Shan}, {Jia}, and {Agarwala}}]{S08}
{Shan}, Q., {Jia}, J., {Agarwala}, A., 2008. High-quality motion deblurring
  from a single image. ACM Transactions on Graphics 27~(3), Article 73.

\bibitem[{{Shepard} and {Helfenstein}(2007)}]{SH07}
{Shepard}, M.~K., {Helfenstein}, P., Mar. 2007. {A test of the Hapke
  photometric model}. Journal of Geophysical Research (Planets) 112, E03001.

\bibitem[{{Shkuratov} et~al.(2011){Shkuratov}, {Kaydash}, {Korokhin},
  {Velikodsky}, {Opanasenko}, and {Videen}}]{S11}
{Shkuratov}, Y., {Kaydash}, V., {Korokhin}, V., {Velikodsky}, Y., {Opanasenko},
  N., {Videen}, G., Oct. 2011. {Optical measurements of the Moon as a tool to
  study its surface}. \planss 59, 1326--1371.

\bibitem[{{Shkuratov} et~al.(2012){Shkuratov}, {Kaydash}, {Korokhin},
  {Velikodsky}, {Petrov}, {Zubko}, {Stankevich}, and {Videen}}]{S12}
{Shkuratov}, Y., {Kaydash}, V., {Korokhin}, V., {Velikodsky}, Y., {Petrov}, D.,
  {Zubko}, E., {Stankevich}, D., {Videen}, G., Dec. 2012. {A critical
  assessment of the Hapke photometric model}. \jqsrt 113, 2431--2456.

\bibitem[{{Sierks} et~al.(2011){Sierks}, {Keller}, {Jaumann}, {Michalik},
  {Behnke}, {Bubenhagen}, {B{\"u}ttner}, {Carsenty}, {Christensen}, {Enge},
  {Fiethe}, {Guti{\'e}rrez Marqu{\'e}s}, {Hartwig}, {Kr{\"u}ger}, {K{\"u}hne},
  {Maue}, {Mottola}, {Nathues}, {Reiche}, {Richards}, {Roatsch},
  {Schr{\"o}der}, {Szemerey}, and {Tschentscher}}]{Si11}
{Sierks}, H., {Keller}, H.~U., {Jaumann}, R., {Michalik}, H., {Behnke}, T.,
  {Bubenhagen}, F., {B{\"u}ttner}, I., {Carsenty}, U., {Christensen}, U.,
  {Enge}, R., {Fiethe}, B., {Guti{\'e}rrez Marqu{\'e}s}, P., {Hartwig}, H.,
  {Kr{\"u}ger}, H., {K{\"u}hne}, W., {Maue}, T., {Mottola}, S., {Nathues}, A.,
  {Reiche}, K.-U., {Richards}, M.~L., {Roatsch}, T., {Schr{\"o}der}, S.~E.,
  {Szemerey}, I., {Tschentscher}, M., Dec. 2011. {The Dawn Framing Camera}.
  \ssr 163, 263--327.

\bibitem[{{Smith} et~al.(1979){Smith}, {Soderblom}, {Beebe}, {Boyce}, {Briggs},
  {Carr}, {Collins}, {Johnson}, {Cook}, {Danielson}, and {Morrison}}]{S79}
{Smith}, B.~A., {Soderblom}, L.~A., {Beebe}, R., {Boyce}, J., {Briggs}, G.,
  {Carr}, M., {Collins}, S.~A., {Johnson}, T.~V., {Cook}, II, A.~F.,
  {Danielson}, G.~E., {Morrison}, D., Nov. 1979. {The Galilean satellites and
  Jupiter - Voyager 2 imaging science results}. Science 206, 927--950.

\bibitem[{{Stephan} et~al.(2016){Stephan}, {Jaumann}, {Krohn}, {Schmedemann},
  {Zambon}, {Tosi}, {Carrozzo}, {McFadden}, {Otto}, {De Sanctis}, {Ammannito},
  {Matz}, {Roatsch}, {Preusker}, {Raymond}, and {Russell}}]{St16}
{Stephan}, K., {Jaumann}, R., {Krohn}, K., {Schmedemann}, N., {Zambon}, F.,
  {Tosi}, F., {Carrozzo}, F.~G., {McFadden}, L.~A., {Otto}, K.~A., {De
  Sanctis}, M.~C., {Ammannito}, E., {Matz}, K.-D., {Roatsch}, T., {Preusker},
  F., {Raymond}, C.~A., {Russell}, C.~T., 2016. {The investigation of spectral
  nature of the bluish material on Ceres}, submitted to Geophys. Res. Letters.

\bibitem[{{Tedesco}(1989)}]{T89}
{Tedesco}, E.~F., 1989. Asteroid magnitudes, ubv colors, and iras albedos and
  diameters. In: {Binzel}, R.~P., {Gehrels}, T., {Matthews}, M.~S. (Eds.),
  Asteroids II. Univ. of Arizona Press, Tucson, pp. 1090--1161.

\bibitem[{{Tedesco} et~al.(2002){Tedesco}, {Noah}, {Noah}, and {Price}}]{T02}
{Tedesco}, E.~F., {Noah}, P.~V., {Noah}, M., {Price}, S.~D., Feb. 2002. {The
  Supplemental IRAS Minor Planet Survey}. \aj 123, 1056--1085.

\bibitem[{{Tedesco} et~al.(1983){Tedesco}, {Taylor}, {Drummond}, {Harwood},
  {Nickoloff}, {Scaltriti}, {Schober}, and {Zappala}}]{T83}
{Tedesco}, E.~F., {Taylor}, R.~C., {Drummond}, J., {Harwood}, D., {Nickoloff},
  I., {Scaltriti}, F., {Schober}, H.~J., {Zappala}, V., Apr. 1983. {Worldwide
  photometry and lightcurve observations of 1 Ceres during the 1975-1976
  apparition}. \icarus 54, 23--29.

\bibitem[{{Thomas} et~al.(2005){Thomas}, {Parker}, {McFadden}, {Russell},
  {Stern}, {Sykes}, and {Young}}]{T05}
{Thomas}, P.~C., {Parker}, J.~W., {McFadden}, L.~A., {Russell}, C.~T., {Stern},
  S.~A., {Sykes}, M.~V., {Young}, E.~F., Sep. 2005. {Differentiation of the
  asteroid Ceres as revealed by its shape}. \nat 437, 224--226.

\bibitem[{{Verbiscer} et~al.(2007){Verbiscer}, {French}, {Showalter}, and
  {Helfenstein}}]{V07}
{Verbiscer}, A., {French}, R., {Showalter}, M., {Helfenstein}, P., Feb. 2007.
  {Enceladus: Cosmic Graffiti Artist Caught in the Act}. Science 315, 815.

\bibitem[{{Veverka} et~al.(1978){Veverka}, {Goguen}, {Yang}, and
  {Elliot}}]{V78}
{Veverka}, J., {Goguen}, J., {Yang}, S., {Elliot}, J., May 1978. {Scattering of
  light from particulate surfaces. I - A laboratory assessment of
  multiple-scattering effects}. \icarus 34, 406--414.

\bibitem[{{West} et~al.(1997){West}, {Doose}, {Eibl}, {Tomasko}, and
  {Mishchenko}}]{W97}
{West}, R.~A., {Doose}, L.~R., {Eibl}, A.~M., {Tomasko}, M.~G., {Mishchenko},
  M.~I., Jul. 1997. {Laboratory measurements of mineral dust scattering phase
  function and linear polarization}. \jgr 102, 16.

\end{thebibliography}

\newpage
\clearpage

\begin{table}
\centering
\caption{Coordinates of selected features on Ceres.}
\vspace{5mm}
\begin{tabular}{lll}
\hline
Name & Longitude & Latitude \\
\hline
Ahuna Mons & $316^\circ$E & $10^\circ$S \\
Dantu & $137^\circ$E & $24^\circ$N \\
Ernutet & $46^\circ$E & $53^\circ$N \\
Haulani & $10^\circ$E & $5^\circ$N \\
Heneb & $191^\circ$E & $11^\circ$N \\
Ikapati & $45^\circ$E & $34^\circ$N \\
Juling & $168^\circ$E & $37^\circ$S \\
Kupalo & $172^\circ$E & $40^\circ$S \\
Nawish & $194^\circ$E & $18^\circ$N \\
Occator & $239^\circ$E & $20^\circ$N \\
Oxo & $0^\circ$E & $41^\circ$N \\
Urvara & $249^\circ$E & $45^\circ$S \\
\hline
\end{tabular}
\label{tab:craters}
\end{table}

\begin{table}
\centering
\caption{RC3 images selected for analysis in different sections of this paper. $\bar{\alpha}$ is the average image phase angle.}
\vspace{5mm}
\begin{tabular}{lllll}
\hline
Section & ID range & $\bar{\alpha}$ range & $n$ & Filter \\
\hline
\ref{sec:best_fit_clear_model} & {\bf 36357}-{\bf 37264} & $7.5^\circ$-$121^\circ$ & 350 & F1 \\
\ref{sec:best_fit_color_model} & {\bf 36531}-{\bf 37263} & $7.5^\circ$-$48.5^\circ$ & 525 & F2-F8 \\
\ref{sec:albedo} & {\bf 36778}-{\bf 37015} & $7.5^\circ$-$11.0^\circ$ & 63 & F1 \\
\ref{sec:albedo} & {\bf 36780}-{\bf 37008} & $7.7^\circ$-$10.9^\circ$ & 25 & F2 \\
\ref{sec:color_maps} & {\bf 36780}-{\bf 37014} & $7.7^\circ$-$10.9^\circ$ & 175 & F2-F8 \\
\ref{sec:exp_model} & {\bf 36529}-{\bf 37264} & $7.5^\circ$-$48.6^\circ$ & 198 & F1 \\
\hline
\end{tabular}
\label{tab:images}
\end{table}

\begin{table}
\centering
\caption{Linear coefficients for the clear filter disk function parameter $c = a + b \alpha$, valid for phase angles $7^\circ < \alpha < 121^\circ$.}
\vspace{5mm}
\begin{tabular}{lllll}
\hline
Model & Eq. & $c$ & $a$ & $b$ \\
\hline
L-S/Lambert & \ref{eq:L-S+Lam} & $c_{\rm L}$ & $0.896$ & $-8.87 \cdot 10^{-3}$ \\
Minnaert & \ref{eq:Minnaert} & $c_{\rm M}$ & $0.514$ & $+5.09 \cdot 10^{-3}$ \\
Akimov & \ref{eq:Akimov} & $c_{\rm A}$ & $1.109$ & $-2.85 \cdot 10^{-3}$ \\
\hline
\end{tabular}
\label{tab:c2_coef}
\end{table}

\begin{table}
\centering
\caption{Coefficients for the polynomial phase functions for clear filter photometric models with phase and disk function separated and $(\iota, \epsilon) < 80^\circ$. Parameters are valid for phase angles $7^\circ < \alpha < 121^\circ$.}
\vspace{5mm}
\begin{tabular}{lllll}
\hline
Model & $C_0$ & $C_1$ & $C_2$ & $C_3$ \\
\hline
L-S/Lambert & $0.0746$ & $-1.65 \cdot 10^{-3}$ & $1.56 \cdot 10^{-5}$ & $-4.79 \cdot 10^{-8}$ \\
Minnaert & $0.0740$ & $-1.67 \cdot 10^{-3}$ & $1.65 \cdot 10^{-5}$ & $-5.41 \cdot 10^{-8}$ \\
Akimov & $0.0731$ & $-1.64 \cdot 10^{-3}$ & $1.59 \cdot 10^{-5}$ & $-5.56 \cdot 10^{-8}$ \\
Akimov ($c_{\rm A}$) & $0.0731$ & $-1.64 \cdot 10^{-3}$ & $1.57 \cdot 10^{-5}$ & $-5.49 \cdot 10^{-8}$ \\
\hline
\end{tabular}
\label{tab:phase_coef_clear}
\end{table}

\begin{table}
\centering
\caption{Coefficients for the polynomial phase functions for narrow band filter photometric model with Akimov disk function and $(\iota, \epsilon) < 80^\circ$. Parameters are valid for phase angles $7^\circ < \alpha < 49^\circ$. The effective wavelength $\lambda_{\rm eff}$ (in nm) of each filter is indicated.}
\vspace{5mm}
\begin{tabular}{lllll}
\hline
Filter & $\lambda_{\rm eff}$ & $C_0$ & $C_1$ & $C_2$ \\
\hline
F2 & $555^{+15}_{-28}$ & $0.0813$ & $-1.95 \cdot 10^{-3}$ & $1.74 \cdot 10^{-5}$ \\
F3 & $749^{+22}_{-22}$ & $0.0759$ & $-1.80 \cdot 10^{-3}$ & $1.63 \cdot 10^{-5}$ \\
F4 & $917^{+24}_{-21}$ & $0.0759$ & $-1.79 \cdot 10^{-3}$ & $1.62 \cdot 10^{-5}$ \\
F5 & $965^{+56}_{-29}$ & $0.0713$ & $-1.67 \cdot 10^{-3}$ & $1.51 \cdot 10^{-5}$ \\
F6 & $829^{+18}_{-18}$ & $0.0789$ & $-1.88 \cdot 10^{-3}$ & $1.72 \cdot 10^{-5}$ \\
F7 & $653^{+18}_{-24}$ & $0.0790$ & $-1.89 \cdot 10^{-3}$ & $1.70 \cdot 10^{-5}$ \\
F8 & $438^{+10}_{-30}$ & $0.0769$ & $-1.86 \cdot 10^{-3}$ & $1.62 \cdot 10^{-5}$ \\
\hline
\end{tabular}
\label{tab:phase_coef_color}
\end{table}

\begin{table}
\centering
\caption{Acquisition parameters of the images used in the Occator haze analysis. The photometric angles ($\iota$, $\epsilon$, $\alpha$) are calculated with an ellipsoid shape model and valid for a location between the bright areas in Occator (intersection of drawn and dotted lines in the projections in Fig.~\ref{fig:haze_survey}). All images are clear filter (F1). Date and time are defined at the start of acquisition. Exposure time ($t_{\rm exp}$) is given in msec.}
\vspace{5mm}
\begin{tabular}{llllrlll}
\hline
Image & Date & UTC & Phase & $t_{\rm exp}$ & $\iota$ & $\epsilon$ & $\alpha$ \\
\hline
{\bf 36681} & 4 May 2015 & 09:50:04 & {\it RC3} & 66 & $81^\circ$ & $78^\circ$ & $36^\circ$ \\
{\bf 37113} & 7 May 2015 & 16:17:33 & {\it RC3} & 82 & $48^\circ$ & $70^\circ$ & $43^\circ$ \\
{\bf 37114} & 7 May 2015 & 16:17:37 & {\it RC3} & 36 & $48^\circ$ & $70^\circ$ & $43^\circ$ \\
{\bf 37570} & 6 Jun 2015 & 23:10:03 & {\it Survey C1} & 32 & $22^\circ$ & $84^\circ$ & $69^\circ$ \\
{\bf 37721} & 9 Jun 2015 & 15:03:59 & {\it Survey C2} & 91 & $16^\circ$ & $32^\circ$ & $20^\circ$ \\
{\bf 38219} & 15 Jun 2015 & 15:06:51 & {\it Survey C4} & 118 & $48^\circ$ & $32^\circ$ & $20^\circ$ \\
\hline
\end{tabular}
\label{tab:haze_images}
\end{table}

\begin{table}
\centering
\caption{Coefficients for the clear filter disk function parameter $c = a + b \alpha$ for Cerealia Facula at a resolution of 1.3~km per pixel, $8^\circ < \alpha < 108^\circ$, and $(\iota, \epsilon) < 85^\circ$.}
\vspace{5mm}
\begin{tabular}{lllll}
\hline
Model & Eq. & $c$ & $a$ & $b$ \\
\hline
L-S/Lambert & \ref{eq:L-S+Lam} & $c_{\rm L}$ & $0.229$ & $-1.65 \cdot 10^{-3}$ \\
Minnaert & \ref{eq:Minnaert} & $c_{\rm M}$ & $0.686$ & $+6.46 \cdot 10^{-3}$ \\
\hline
\end{tabular}
\label{tab:Occator_c2_coef}
\end{table}

\begin{table}
\centering
\caption{Coefficients for the polynomial phase functions for clear filter Type~I photometric models for Cerealia Facula at a resolution of 1.3~km per pixel, $8^\circ < \alpha < 108^\circ$, and $(\iota, \epsilon) < 85^\circ$.}
\vspace{5mm}
\begin{tabular}{llll}
\hline
Model & $C_0$ & $C_1$ & $C_2$ \\
\hline
L-S/Lambert & $0.381$ & $-5.70 \cdot 10^{-3}$ & $2.49 \cdot 10^{-5}$ \\
Minnaert & $0.331$ & $-3.88 \cdot 10^{-3}$ & $2.09 \cdot 10^{-5}$ \\
\hline
\end{tabular}
\label{tab:Occator_phase_coef_clear}
\end{table}

\newpage
\clearpage

\newpage
\clearpage
\thispagestyle{empty}

\newpage
\clearpage

\end{document}